\documentclass[11pt]{amsart}
\usepackage{bxpdfver}
\setpdfversion{1.4}

\usepackage{latexsym, amssymb, amsmath, amsthm, natbib}
\usepackage[leqno]{amsmath}
\usepackage{xy}
\xyoption{all}
\usepackage{verbatim}

\usepackage{caption,subcaption}

\usepackage[left=1in,top=1in,right=1in,bottom=1in]{geometry}

\usepackage{mathtools}

\usepackage{tgpagella}
\usepackage{setspace}
\usepackage{enumerate}
\usepackage[multiple]{footmisc}
\usepackage{hyperref}
\usepackage{xcolor}

\doublespace

\usepackage{tikz}
\usetikzlibrary{positioning}

\usepackage{pgfplots}
\pgfplotsset{compat=1.14}
\usepackage{mathrsfs}
\usetikzlibrary{arrows}

\tikzset{
  every node/.style    = {
    text centered,
    line width = .5,
    anchor = center,
  },
  every label/.style   = {
    fill = white, anchor = mid,
  },
  every path/.style   = {
    > = stealth
  },
  point/.style args   = {(#1)#2}{
    rounded corners,
    fill = white,
    minimum height = 20,
    minimum width = 10,
    label = { [name = #1] above:#2 },
  },
  point a/.style args   = {(#1)#2}{
    rounded corners,
    fill = white,
    minimum height = 10,
    minimum width = 20,
    label = { [name = #1] right:#2 },
  },
  point b/.style args   = {(#1)#2}{
    rounded corners,
    fill = white,
    minimum height = 10,
    minimum width = 75,
    label = { [name = #1] right:#2 },
  },
}

\usepackage{epigraph}
\let\originalepigraph\epigraph
\renewcommand\epigraph[2]{\originalepigraph{\textit{#1}}{\textsc{#2}}}

\usepackage{etoolbox}
\patchcmd{\section}{\scshape}{\bfseries}{}{}
\makeatletter
\renewcommand{\@secnumfont}{\bfseries}
\makeatother

\newcommand{\corigin}{c'}
\newtheorem{theorem}{Theorem}
\newtheorem{corollary}{Corollary}
\newtheorem{definition}{Definition}
\newtheorem{lemma}{Lemma}
\newtheorem{proposition}{Proposition}
\newtheorem{claim}{Claim}

\theoremstyle{remark}
\newtheorem{remark}{Remark}
\newtheorem{example}{Example}

   \def\calx{\mathcal{X}} \def\calt{\mathcal{T}}
 \def\cals{\mathcal{S}}  \def\calf{\mathcal{F}}
 \def\calc{\mathcal{C}}

\def\oconcave{ordinally concave} 

\newcommand{\df}[1]{\textbf{\textit{#1}}}

\newcommand{\norm}[1]{|| #1 ||}
\newcommand{\abs}[1]{\left| #1 \right|}

\newcommand{\ih}[1]{{\color{purple} IH: #1}}
\newcommand{\fk}[1]{{\color{red} FK: #1 }}
\newcommand{\mby}[1]{{\color{blue} MBY: #1 }}
\newcommand{\ky}[1]{{\color{violet} KY: #1 }}

\begin{document}

\title[Market Design with Distributional Objectives]{Market Design with Distributional Objectives$^{\dagger}$}


\author[Hafalir, Kojima, Yenmez, and Yokote] {Isa E. Hafalir \and Fuhito Kojima  \and M. Bumin Yenmez \and Koji Yokote$^{*}$}

\thanks{
An earlier version of this paper has been circulated under the title ``Design on Matroids: Diversity vs. Meritocracy." We thank the seminar participants at Brown University, Durham University, Michigan State University, Pennsylvania State University, and
Washington University in St. Louis; and the attendees at the International Conference on Applied Economic Theory Innovation, Iowa University Market Design Workshop, and SAET Conference.
We are grateful to Masanori Kobayashi, Leo Nonaka, Ryosuke Sato, and Ryo Shirakawa for their excellent research assistance.
Fuhito Kojima is supported by the JSPS KAKENHI Grant-In-Aid 21H04979 and JST ERATO Grant Number JPMJER2301, Japan.
Hafalir is affiliated with the UTS Business School, University of Technology Sydney, Sydney, Australia; Kojima is with the Department of Economics, the University of Tokyo, Tokyo, Japan;
Yenmez is with the Department of Economics, Washington University, St. Louis, 
MO, USA and Durham University Business School, Durham, United Kingdom;
Yokote is with the Graduate School of Economics, the University of Tokyo, Tokyo, Japan.
Emails: \texttt{isa.hafalir@uts.edu.au}, \texttt{fuhitokojima1979@gmail.com}, \texttt{bumin@wustl.edu}, \texttt{koji.yokote@gmail.com}.}

\begin{abstract}
We provide optimal solutions to an institution that has distributional objectives 
when choosing from a set of applications based on merit (or priority). For example, in college
admissions, administrators may want to admit a diverse class
in addition to choosing students with the highest qualifications. We provide a family
of choice rules that maximize merit subject to attaining a level of the distributional objective. 
We study the desirable properties of choice rules in this family
and use them to find all subsets of applications on the Pareto frontier of the distributional objective and merit.  
In addition, we provide two novel characterizations of matroids. 
\end{abstract}




\maketitle




\section{Introduction}
\epigraph{\textit{To see high merit and be unable to raise it to office, to raise it but not to give such promotion precedence, is just destiny.}}{-Confucius}


Meritocratic systems in which goods and political power are given to
people based on qualifications rather than their wealth or social status
have been idealized since ancient times. The Chinese philosopher
Confucius argued that those who govern should do so because of merit, not because of inherited status. The Han dynasty adopted Confucianism and
implemented civil service examinations to select and promote government
officials \citep{dien2001}.
The Greek philosopher Plato, in his book
\emph{The Republic}, stated that the wisest should rule, and hence rulers
shall be philosopher kings. A system based on meritocracy, however, may
increase economic inequality and social and political dysfunction,
the so-called \emph{meritocracy trap} \citep{markovits2019}. To decrease the
inequities that exist between different groups in society, affirmative action
and diversity policies have been implemented worldwide \citep{sowell04}.
Therefore, in practice, it is crucial to find a balance between meritocracy
and diversity.

In this paper, we find optimal subsets of applications to an institution that
is not only interested in choosing applications with the most merit but also satisfying a distributional objective 
such as having a diverse group. The institution ranks applications according to merit (or priority).  
For example, applicants may take an exam to determine how qualified they are. American universities rank students using SAT scores and other criteria. Meanwhile, the distributional objective is represented by a function of traits that applicants have.
The type of a student specifies the student's traits and may include information 
about gender, race, ethnicity, socioeconomic status, and disability status.

Our focus is on \emph{choice rules} that select a subset of each possible set of applications.\footnote{Choice rules
are one of the most basic primitives of economics (e.g., \cite{masco95} and \cite{kreps2023}).}
We study a family of choice rules that maximize the merit of the selected group
subject to attaining a level of the distributional objective. To do so, we start with an extreme member
of this family that maximizes the distributional objective first and then merit among the groups maximizing the distributional objective. Even though this rule can be defined in very general
environments, it may lack basic desirable properties,
which may make its implementation infeasible.
Indeed, some institutions, such as universities, get
thousands of applications every year. For example, in fall 2020, the average number
of applications for the ten colleges in the US that received the
most applications was 84,865.\footnote{See \url{https://tinyurl.com/uv6h3jsh} for
statistics from the U.S. News \& World Report.} Therefore, the choice rule
must be implementable in a computationally efficient way, and its outcome should not
depend on the order in which applications are evaluated, which is the \emph{path-independence}
property of a choice rule \citep{plott1973path}.\footnote{Path independence is equivalent to the conjunction of the \textit{substitutes condition}
and a mild consistency axiom standard in matching theory. See the proof of Theorem \ref{thm:pi} in Appendix \ref{app:proofs}.}
To this end, we define the distribution-conscious choice rule as follows. In the first step, we find
distributions of applicant types that maximize the distributional objective.
In the second step, we choose applications one by one using the merit ranking as long as the set
of chosen applications has a distribution smaller than an optimal distribution found in the first step.\footnote{A distribution $\xi$ is smaller than
distribution $\xi'$ if every coordinate of $\xi'$ is greater than or equal to the same coordinate of $\xi$.}
By construction, the distribution-conscious choice rule always finds a set of applications that
maximizes the distributional objective. However, because it
is myopic in the second step, it does not necessarily maximize the merit of the
chosen set among groups maximizing the distributional objectives.
To address this problem, we consider a
restriction on the distributional objective under which it lexicographically maximizes the distributional objective and merit, and is computationally fast.

We use a concept of concavity on functions with discrete domains introduced by \citet{murotashioura2003}, and we call it
\emph{ordinal concavity}.\footnote{\citet{murotashioura2003} introduce SSQM (semi-strict quasi M-concavity) as an ordinal implication of M-concavity. M-concavity is a cardinal notion, and it has a weaker variant called M$^\natural$-concavity. Analogous to weakening M-concavity to M$^\natural$-concavity,
SSQM$^\natural$-concavity is the natural counterpart of SSQM-concavity. Our ordinal concavity is equivalent to SSQM$^\natural$-concavity.\label{footnote}} 
Roughly, ordinal concavity requires that, when two different distributions
are made closer to each other, either the value of the distributional objective strictly increases for at least one of the two distributions or the value of the distributional objective
 remains the same for both.  
Ordinal concavity is weaker than the two standard concavity notions used in a field of mathematics known as discrete convex analysis: 
M-concavity and M$^{\natural}$-concavity.\footnote{See Appendix \ref{app:compare} for the definitions
of M-concavity and M$^{\natural}$-concavity.} While these two are cardinal, ordinal concavity is an ordinal notion, as it only depends on comparisons of values that the distributional objective takes.

When the distributional objective is ordinally concave, the distribution-conscious choice rule 
maximizes merit among all sets of applications that maximize the distributional objective,
and its outcome can be constructed in polynomial time (Theorem \ref{thm:diversitychoice}).
To prove the first part, we show that the collection of maximal distributions among optimal ones identified in the first step is well-behaved: Specifically,
it satisfies a notion of discrete convexity called \emph{M-convexity} (Lemma \ref{lem:pareto}).\footnote{See Section \ref{sec:convexity} for the definition of M-convexity.} Furthermore,
the myopic addition of contracts in the second step is equivalent to the outcome of a procedure in the combinatorial optimization literature known as the
\emph{greedy algorithm} on a \emph{matroid} that we construct
(Lemmas \ref{lem:matroid} and \ref{lem:equal}).\footnote{See Section \ref{sec:matroid}
for the definitions of the
greedy algorithm and matroids.}
The computational efficiency proof has two main parts.
In the first part, we establish the \emph{maximizer-cut theorem}, which allows us
to dissect the domain of feasible distributions in the search for an optimal distribution
(Theorem \ref{thm:maximizer-cut}). Using this result, we construct the
\emph{domain-reduction algorithm} that
allows us to find an optimal distribution efficiently.\footnote{The domain-reduction algorithm and the maximizer-cut theorem are originally established for minimizing M-convex functions. This optimization problem is equivalent to maximization of M-concave functions, a subclass of ordinally-concave functions; see Remark \ref{rem:maximizer-cut} in Section \ref{sec:proofsketch}.
} 
In the second part, we construct
a modified version of the distribution-conscious choice rule, which is more computationally
tractable than our original definition, and show that finding an outcome of the distribution-conscious choice rule takes quadratic time in the number of applications.

A desirable property of choice rules is path independence. Path independence states that applications can be viewed
in batches in any order without changing the final outcome, an appealing property to
institutions that receive many applications. Furthermore, it
guarantees the existence of a desirable matching in two-sided matching markets.\footnote{See, for example, \cite{chayen17} who study
two-sided matching markets with path-independent choice rules.}
We show that the distribution-conscious choice rule is path independent when the distributional objective is ordinally concave (Theorem \ref{thm:pi}).
In most matching clearinghouses, the deferred-acceptance algorithm of \cite{gale62} is used to assign applicants to institutions.
This algorithm produces a desirable matching when institution choice rules satisfy path independence, and it
is strategy-proof for applicants when institution choice rules further satisfy the
\emph{law of aggregate demand} \citep{hatmil05}. The law of aggregate demand
requires that the number of applications that are chosen weakly increases when
there are more applications (in the sense of set inclusion) to choose from.
The distribution-conscious choice rule does not necessarily satisfy the law of aggregate demand
even when the distributional objective is ordinally concave. However, if the distributional objective is also \textit{size-restricted concave},
then the distribution-conscious choice rule satisfies the law of aggregate demand (Theorem \ref{prop:structure}).
Size-restricted concavity was recently introduced by \cite{hakoyeyo24}; we provide a generalization of it
to more general domains.

Next, we consider the family of choice rules that maximize merit subject to achieving
a certain (exogenously given) level of the distributional objective.
We first observe that when the distributional objective
is capped at a level, the distribution-conscious choice rule for
the modified objective maximizes merit subject to attaining that level. 
Therefore, every choice rule in this family has the same desirable properties like the
distribution-conscious choice rule when the modified objectives are ordinally concave. 
We provide a characterization of distributional objectives such that the modified distributional objective for every 
level is ordinally concave (Proposition \ref{prop:ordinal-equivalence}). 
Using this family, we provide the \emph{trace algorithm} that
finds all subsets of applications on the Pareto frontier of the distributional objective and merit, 
and show that the trace algorithm is \emph{pseudo polynomial}, which
means that the time complexity is polynomial in the largest integer present in the
input data (Theorem \ref{thm:trace}).\footnote{For this result, we assume that the distributional objective takes integer values. Any ordinally concave distributional objective can be replaced with another distributional objective that takes integer values and is ordinally concave without changing the distribution-conscious choice rule.}  The trace algorithm is useful for an
institution that has the dual goals of maximizing distributional objective and merit, as
it presents the institution with all alternatives on the Pareto frontier.

One special case of our model is when the university has a utility function
over sets of contracts.\footnote{This can be modeled as a special case of our
model as follows: All agents have different types, and
the distributional objective takes distinct values on different distributions. Therefore, the
distribution-conscious choice rule is uniquely determined by the first step that maximizes the distributional objective.}
In this particular case, the distribution-conscious choice rule maximizes distributional objective on subsets
of a set of available applications.
An immediate corollary of Theorem \ref{thm:pi} is that when the utility function
over sets of applications satisfies ordinal concavity, the choice rule
constructed by maximizing the utility function satisfies path independence.\footnote{See \cite{murota:metr:2013} for a similar result under a set of different assumptions.}
In \cite{hakoyeyo24}, we show the reverse direction that if a choice rule is path independent, then there exists an ordinally concave
utility function such that the choice from any set of contracts is equal to
the subset that maximizes the utility function among all subsets. In other words, every path-independent choice rule can be rationalized by an ordinally concave
utility function. Therefore, there is a sense in which ordinal concavity is necessary for the path-independence property.

Our results apply to markets where institutions have two
distinct goals that may conflict with each other. We state the model in terms of
the main application of college admissions, where universities admit students
to maximize the merit of the incoming class as well as its distributional objective.
Other applications include school choice, hiring by public institutions or private firms, and auctions
with distributional goals (e.g., procurement auctions and spectrum license auctions).

Our paper is part of the recent literature on market-design problems
with distributional objectives. In practice, distributional objectives are
typically implemented by reserving a number of positions for target groups.
In the market-design literature, reserves are introduced and
analyzed by \cite{hayeyi13}, \cite{ehayeyi14}, and \cite{echyen12}.
Distributional objectives play an important role in matching problems with regional constraints \citep{kamakoji-basic,kamakoji-concepts,kamakoji-iff,kamada2020accommodating}.
Another matching market with distributional constraints is interdistrict school choice
\citep{hafalir2022interdistrict}. In another recent work, \cite{kuku22} study matching markets with
adjustable capacities and introduce a new concept of stability. Unlike these papers, we do not focus on a particular policy but
model it as a function on distributions that satisfies ordinal concavity. In a recent work, \cite{hakoye2022} study the
implementation of distributional objectives in a constrained efficient mechanism and introduce pseudo
M$^{\natural}$-concavity.
Like us, they also represent the distributional objective as a function, but their research question
is the existence of constrained efficient mechanisms, whereas we focus on the desirable
properties of institutional choice rules.

The most closely related papers in terms of
motivation to the current work are \cite{imamura2020} and \citet{grigoryan2021priority}, who introduce axioms
to compare meritocracy and diversity
 of choice rules and use
these axioms to characterize choice rules with reserves and quotas. Another related paper is
\cite{kojima-tamura-yokoo}, who study two-sided matching markets with agents
that have M$^{\natural}$-concave utility functions and show the existence of
stable matchings in a variety of matching problems with constraints based on properties
of M$^{\natural}$-concave utility functions. Choice rules with reserves and quotas
can be modeled as special cases of our distribution-conscious choice rule by choosing the appropriate distributional objective (see Example \ref{ex:saturated} in Section \ref{sec:choice}.)
We also provide two novel characterizations
of matroids (Lemma \ref{lem:matchar} and Proposition \ref{prop:matroid})
that may be helpful in other work.

In operations research, \cite{chenli2020} study parametric maximization problems using ordinal concavity.\footnote{Following
\citet{murotashioura2003}, they refer to the condition as SSQM$^{\natural}$-concavity.}
They show that the optimal solution is non-increasing in the parameters when the objective function is ordinally concave,
and illustrate when ordinal concavity is preserved. \cite{chenli2020} analyze neither choice rules nor matching problems.

Discrete convex analysis is relatively new in economic theory. \cite{murota:dca:2016} provides an excellent review with some applications in economics. In addition to the papers mentioned above, see, for example, \cite{paesleme2017}, \cite{hatfield2019full}, \cite{kojimaconstraints2020,kojima2020job}, \cite{candogan2016competitive},  and
\cite{kushnir2021}. We substantially contribute to this literature by establishing results for economics in general and market design in particular (and also by introducing new notions of concavity, such as the notions of pseudo M$^\natural$-concavity$^+$ and semistrict pseudo M$^\natural$-concavity).

We introduce our model in the next section. We study the distribution-conscious choice rule in Section \ref{sec:diversity} and its generalization,
which maximizes merit subject to attaining a given level of the distributional objective, and the trace algorithm in Section \ref{sec:frontier}.
In Section \ref{sec:math}, we provide two new characterizations of matroids and the proof sketch of Theorem \ref{thm:diversitychoice}.
In Section \ref{sec:conclusion}, we conclude the paper. We present comparisons of
discrete concavity notions in Appendix \ref{app:compare} and the proofs
of our main results in Appendix \ref{app:proofs}. We introduce another notion of concavity
called semi-strict pseudo M$^{\natural}$-concavity and present the remaining proofs
 and examples in Appendix \ref{app:auxiliary}.

\section{Model}\label{sec:model}

\subsection{Agents, Distributions, and Types}
Let $\calc$ denote a finite set of academic \df{schools} (or colleges/divisions) in a university and $\cals$
a finite set of \df{students} applying to the university. Each school represents a
major or program that students can apply to. For example, when students are admitted
only as ``undecided'' without specifying a major or program, the set $\calc$ is a singleton.

There exist a finite set $\calt$ of \df{student types} and a \df{type function}
$\tau : \cals \rightarrow \calt$
that specifies a type $\tau(s)\in \calt$ for each student $s\in \cals$.
A type specifies distribution-related student traits that the university cares about. For example,
it can specify gender, race, ethnicity, disability status, veteran
status, nationality, and socioeconomic status.


Each application is represented by a \df{contract} specifying a school, a student, and
the terms of admissions that may include financial aid information.
The set of all contracts is finite and denoted by $\mathcal{X}$.
The university has a \df{merit ranking} $\succ$ of contracts, which is a
strict preference relation (linear order) over $\mathcal{X}$.\footnote{Applications that
are strictly less preferred than having an empty seat for the university are
dropped from $\mathcal{X}$. Thus, without loss of generality, we assume that the university
strictly prefers each application in $\mathcal{X}$ to having an empty seat.} The corresponding
weak preference is denoted by $\succeq$, that is, for each
$x,y\in \mathcal{X}$,
$x \mathrel{\succeq} y$ if $x=y$ or $x \mathrel{\succ} y$.

Let contracts in $X=\{x_1,\ldots,x_{|X|}\} \subseteq \mathcal{X}$ and
$X'=\{x'_1,\ldots,x'_{|X'|}\} \subseteq \mathcal{X}$
be enumerated such that,
\begin{align*}
\mbox{for each} \; i,j\in \{1,\ldots,|X|\}, \qquad  i < j \; &\implies \; x_i \succ x_j, \; \mbox{ and}\\
\mbox{for each} \; i,j\in \{1,\ldots,|X'|\}, \qquad  i < j \; &\implies \; x'_i \succ x'_j.
\end{align*}
Then, $X$ \textbf{merit dominates} $X'$ if $|X|\geq |X'|$ and,
for each $i\in \{1,\ldots,|X'|\}$, $x_i \mathrel{\succeq} x'_i$. 

A \df{distribution} $\xi \in \mathbb Z_+^{|\mathcal{C}|\times |\mathcal{T}|}$ is
a vector such that the entry for school $c\in \mathcal{C}$ and type $t\in \mathcal{T}$
is denoted by $\xi_c^t$.\footnote{$\mathbb Z_+$ is the set of non-negative integers including zero.} The entry $\xi_c^t$ is interpreted as the number of
students of type $t\in \calt$ assigned to school $c\in \calc$ at
$\xi$. For a set of contracts $X\subseteq \mathcal{X}$, $\xi(X) \in \mathbb Z_+^{|\mathcal{C}|\times |\mathcal{T}|}$ denotes
the distribution induced from $X$ so that $\xi_c^t(X)$ denotes the number of
contracts between students of type $t\in \mathcal{T}$ and school
$c\in \mathcal{C}$ in $X$. For each distribution
$\xi \in \mathbb Z_+^{|\mathcal{C}|\times |\mathcal{T}|}$, $\norm{\xi}$ denotes the sum of coordinates of $\xi$. There may be
feasibility constraints on distributions, such as capacity constraints for schools.
The set of \textbf{feasible distributions} is denoted by $\Xi^0 \subseteq \mathbb Z_+^{|\mathcal{C}|\times |\mathcal{T}|}$. We assume that the zero vector is in $\Xi^0$.
For each school $c\in \mathcal{C}$
and type $t\in \mathcal{T}$, let $\chi_{c,t}$ denote the distribution where there is one type-$t$ student at school $c$ and there are no other students.


There exists a \textbf{distributional objective} $f: \Xi^0 \rightarrow \mathbb{R}_+$.\footnote{$\mathbb{R}_+$
is the set of non-negative real numbers including zero.} The function $f$ measures how good a distribution of students is in terms of an objective. Therefore, if $f(\xi)\geq f(\xi')$, then it means that distribution $\xi$ is at least as good as distribution $\xi'$ in terms of the distributional objective.

Two remarks on the distributional objective are in order. First, our analysis only depends on the ordinal content of $f$ and not on the cardinal values
it takes. Therefore, a distributional objective $f$ and any strictly increasing transformation of $f$ are equivalent for our purposes.\footnote{A function
$g: \mathbb{R} \rightarrow \mathbb{R}$ is strictly increasing if, for each $x,y\in \mathbb{R}$ such that $x>y$, we have $g(x)>g(y)$. We say that a function
$h: \Xi^0 \rightarrow \mathbb{R}_+$ is a strictly increasing transformation of $f$ if, for each $\xi \in \Xi^0$, $h(\xi)=g(f(\xi))$  where $g$ is a strictly increasing function.} Second, we allow for indifferences, i.e., $f$ can take the same value at different distributions. This is a natural assumption in practice because school authorities often introduce coarse distributional goals rather than strict ordering over distributions. We offer concrete examples in Section \ref{sec:choice}.\footnote{In the literature on matching theory, a standard approach for addressing the indifference issue is to break ties and create a linear order. In the context of balancing distributional objective and merit, tie-breaking is not appropriate because indifferences in the distributional objective leave room for merit to be taken into account. Consider two sets of contracts $X$ and $X'$ such that (i) they are equally desirable in terms of distributional objective (i.e., $f(\xi(X))=f(\xi(X'))$) but
(ii) $X$ merit dominates $X'$. In this case, we should choose $X$ rather than $X'$.
Tie-breaking might lead to an undesirable choice. Furthermore, several real-life applications with reserves and quotas
have ties naturally. Finally, even when the distributional objective takes distinct values, the choice rule that maximizes merit subject to attaining a given level of distributional objective does not differentiate between the value of the distributional objective above the threshold level.}


\subsection{Choice Rules}
We introduce the concept of a choice rule, which specifies the subset accepted from each possible set of applications.

\begin{definition}
A \textbf{choice rule} is a function $C: 2^{\mathcal{X}} \rightarrow 2^{\mathcal{X}}$ such that,
for each $X \subseteq \mathcal{X}$,
\[ C(X)\subseteq X \quad \mbox{ and } \quad  \xi(C(X))\in \Xi^0.\]
\end{definition}

A choice rule must be such that the distribution of a chosen set is feasible.\footnote{It is possible that $X \subseteq \mathcal{X}$ contains two
contracts associated with the same student. We do not impose the assumption that $C(X)$ chooses at most
one contract per student because it is not necessary for practical purposes. In Theorem \ref{thm:pi}, we show that our new choice rule $C$ satisfies
path independence, under which the deferred-acceptance algorithm results in a stable matching where each student signs at most one contract.
It is also worth mentioning that students in the U.S. can submit at most one application to most universities.}
Next, we consider a highly desirable property of choice rules.

\begin{definition}
A choice rule $C$ satisfies \textbf{path independence} if, for
each $X, X' \subseteq \calx$,
\[C(X \cup X')=C(C(X) \cup X').\footnote{\cite{plott1973path} introduces
path independence as an axiom of rationality in a model of
social choice. See \cite{chayen17} for an application of path independence in a
matching context.}\]
\end{definition}

Path independence guarantees that applications can be viewed in batches in
any order without changing the final outcome, thereby implying that the university
is applying consistent admissions policies regardless of the sequence or composition
of the batches that are considered during the admissions process. Therefore, it
is a desirable property in college admissions (and other applications). Path
independence is equivalent to the conjunction of the \emph{substitutes condition} and
a mild consistency axiom routinely used in matching theory.\footnote{See the proof of Theorem \ref{thm:pi} in Appendix \ref{app:proofs} 
for the definitions of these two notions.}

\begin{definition}
A choice rule $C$ satisfies the \textbf{law of aggregate demand} if, for each $X,X' \subseteq \calx$,
\[X \supseteq X' \; \implies \; |C(X)|\geq |C(X')|.\footnote{\cite{hatmil05} introduce the law of aggregate demand in a matching market with contracts. \cite{alkan03} calls this property \emph{size monotonicity} in a matching context without contracts.}\]
\end{definition}

The law of aggregate demand states that when a university gets more applications, the number of chosen applications cannot decrease.

In the context of assigning students to schools in a centralized clearinghouse,
path independence guarantees that the most commonly used method, the deferred-acceptance algorithm, works well, e.g., it
produces the \emph{student-optimal stable matching}; and if the law of aggregate demand is also satisfied, it is
\emph{strategy-proof} \citep{hatmil05}.\footnote{In this context, only students are strategic agents.
Therefore, a direct revelation mechanism is strategy-proof if, for each student, reporting their
true preference ranking over schools is a weakly dominant strategy.}

\section{A Lexicographic Approach to Distributional Objective and Merit}\label{sec:diversity}
In this section, we introduce a choice rule that lexicographically maximizes a distributional objective first and merit second,
and establish some further desirable properties of this choice rule. In Section \ref{sec:frontier}, we generalize
this admissions policy so that the university maximizes merit subject to attaining a given level of the 
distributional objective.

\subsection{Distribution-Conscious Choice Rule}\label{sec:choice}
In the following choice rule, we first find all maximizers of the distributional objective.
Then we choose contracts one by one according to their merit ranking as long as the chosen set of contracts can be 
completed to a subset of contracts whose distribution maximizes the distributional objective.

\paragraph{\textbf{Distribution-Conscious Choice Rule} $\mathbf{C^d}$}

\begin{description}
  \item[Input] Let $X$ be a set of contracts.
  \item[Step 1] Find the set of distributions in $\{\xi : 0 \leq \xi \leq \xi(X)\}$
  that maximize the distributional objective $f$ and denote it by $\Xi^*(X)$.
  Set $X_0=\emptyset$ and $k=0$.
  \item[Step 2] If there exist $x \in X \setminus X_k$ and $\xi \in \Xi^*(X)$ such that $\xi(X_k \cup \{x\}) \leq \xi$, then choose such a contract
      $x_{k+1}$ of highest merit, let $X_{k+1}=X_k \cup \{x_{k+1}\}$, and go
      to Step 3. Otherwise, go to Step 4.
  \item[Step 3] Add 1 to $k$ and go to Step 2.
  \item[Step 4] Return $X_k$ and stop.
\end{description}

The algorithm ends at a finite index $k$ since the number of contracts is finite.

By construction, the distribution-conscious choice rule always produces an outcome that maximizes
the distributional objective.
However, Step 2 of the distribution-conscious choice rule is myopic in choosing contracts, so
it need not produce an outcome that maximizes merit
among the groups maximizing the distributional objective.
To address this problem, we make the following assumption on the distributional objective.

Let $\chi_{\emptyset}$ denote the distribution with zero entries.

\begin{definition}\label{def:ordinal}
The distributional objective $f: \Xi^0 \rightarrow \mathbb{R}_+$ is \textbf{ordinally concave} if, for each $\xi,\tilde{\xi}\in \Xi^0$
and $(c,t) \in \calc \times \calt$ with  $\xi_c^t>\tilde{\xi}_c^t\:$, there exists $(c',t') \in (\calc \times \calt) \cup \{\emptyset\}$
(with $\xi_{c'}^{t'}<\tilde{\xi}_{c'}^{t'}$ whenever $(c',t')\neq \emptyset$) such that
\begin{enumerate}[(i)]
\item $f(\xi-\chi_{c,t}+\chi_{c',t'})> f(\xi)$, or
\item $f(\tilde{\xi}+\chi_{c,t}-\chi_{c',t'}) > f(\tilde{\xi})$, or
\item $f(\tilde{\xi}+\chi_{c,t}-\chi_{c',t'})=f(\tilde{\xi})$ and $f(\xi-\chi_{c,t}+\chi_{c',t'})=f(\xi)$.
\end{enumerate}
\end{definition}
Each condition in the definition above not only imposes the stated inequality or equations, but also that the arguments of $f$ are
in the domain $\Xi^0$.\footnote{As we discuss in the Introduction ordinal concavity has been studied in operations research. However, to our knowledge,
it is new to the economics literature. In Appendix \ref{app:compare}, we show that ordinal concavity is weaker than M-concavity and M$^{\natural}$-concavity.}

To give the intuition for ordinal concavity, let us consider a special
case when there are only one school and one type. Hence, a distribution
specifies how many students are assigned to the university. For
simplicity, take $\Xi^0=\mathbb Z_+$. Consider $\xi, \tilde{\xi}
\in \mathbb Z_+$ such that $\xi>\tilde{\xi}$.
Since the distributions have only one coordinate, ordinal concavity implies that either
\begin{enumerate}[(i)]
  \item $f(\xi-1)>f(\xi)$, or
  \item $f(\tilde{\xi}+1)>f(\tilde{\xi})$, or
  \item $f(\xi-1)=f(\xi)$ and $f(\tilde{\xi}+1)=f(\tilde{\xi})$.
\end{enumerate}

In words, when we move $\xi$ and $\tilde{\xi}$ towards each other
by one, either the value of $f$ increases on at least one side or the value of $f$
stays the same on both sides. 
For example, if $f$ is a concave or strictly increasing (or decreasing)
function on the real line, then its restriction on integers is ordinally concave.
It is also satisfied when $f$ represents a single-peaked preference
relation.

When there are more schools and types so that distributions are multidimensional,
moving closer to each other may mean either moving in one direction as in the one-dimensional case above,
or it may mean the existence of another dimension
so that from one distribution, we remove one in one direction and add one in
the other direction and we do the opposite operations on the other distribution.

Our first result shows that, when $f$ is \oconcave{}, the distribution-conscious choice rule
lexicographically maximizes the distributional objective first and merit second in a computationally efficient way. 

\begin{theorem}\label{thm:diversitychoice}
Suppose that the distributional objective $f$ is \oconcave{}.\footnote{By inspection of the proof,
one can verify that the conclusions of parts (i) and (ii) of the result hold under
a weaker condition than ordinal concavity. More specifically, these conclusions hold
if one of the conditions in the definition of ordinal concavity holds when $f(\xi)=f(\tilde \xi)$.} Then, for each set of contracts $X\subseteq \calx$,
\begin{enumerate}[(i)]
\item $C^d(X)$ maximizes the distributional objective $f$ among subsets of $X$,
\item $C^d(X)$ merit dominates each subset of $X$ that maximizes the
distributional objective $f$, and
\item
$C^d(X)$ can be calculated in $O(|\mathcal{C}| \times |\mathcal{T}| \times |X|^2)$, assuming
$f$ can be evaluated in a constant time.
\end{enumerate}
\end{theorem}
We prove the result in Appendix \ref{app:proofs} and provide a proof sketch in Section \ref{sec:proofsketch}.
Next, we provide examples of ordinally concave distributional objectives. The first is a simple illustrative example that we use throughout
the paper. Subsequent examples are more practical and motivated by ``reserves'' in various forms.

\begin{example}\label{ex:ladfail}
Suppose that there are three students of different types and one school, say $c$.
There is only one contract between each student and the university.
Denote these contracts by $x$, $y$, and $z$. 
The university has a capacity of two, so $\Xi^0=\{\xi:\norm{\xi} \leq 2\}$
is the set of feasible distributions.

Let the distributional objective $f$ be defined as follows:
\[f(\xi(\emptyset))=0, f(\xi(\{x\}))=1, f(\xi(\{y\}))=1, f(\xi(\{z\}))=n,\]
\[f(\xi(\{x,y\}))=1, f(\xi(\{x,z\}))=5, \text{ and } f(\xi(\{y,z\}))=5\]
where $n \geq 5$.\footnote{We consider different values of $n$
in the subsequent sections to illustrate different results.}
To see that $f$ is ordinally concave, we need to consider different cases
depending on the value of $\xi$ in the definition.
Here, we  only consider the first of several cases for illustration,
namely the case with $\xi=\xi(\{x,y\})$, whereas
in Appendix \ref{app:auxiliary}, we provide a full proof.

Let $\xi=\xi(\{x,y\})$. Let $t\in \calt$ be the type of the
student associated with contract $x$ and $t'\in \calt$ be the type of the student
associated with contract $z$. Consider $\tilde{\xi}\in \Xi^0$ with $\tilde{\xi}^t_c < \xi^t_c$.  
If $\tilde{\xi}_{c}^{t'}=0$, then $\tilde{\xi}=\xi(\emptyset)$
or $\tilde{\xi}=\xi(\{y\})$. For $\tilde \xi=\xi(\emptyset)$, we have
\begin{align*}
f(\tilde \xi + \chi_{c,t})=f(\xi(\{x\}))=1>0=f(\xi(\emptyset))=f(\tilde \xi).
\end{align*} 
Therefore, condition (ii) in the definition of ordinal concavity is satisfied. For $\tilde \xi=\xi(\{y\})$, we have 
\begin{align*}
&f(\xi-\chi_{c,t})=f(\xi(\{y\}))=1=f(\xi(\{x,y\}))=f(\xi), \\
&f(\tilde \xi+\chi_{c,t})=f(\xi(\{x,y\}))=1=f(\xi(\{y\}))=f(\tilde \xi). 
\end{align*}
Therefore, condition (iii) in the definition of ordinal concavity is satisfied.
If $\tilde{\xi}_{c}^{t'}=1$, then $\tilde{\xi}=\xi(\{z\})$ or $\tilde{\xi}=\xi(\{y,z\})$. 
For both values of $\tilde \xi$, 
\begin{align*}
f(\xi-\chi_{c,t}+\chi_{c,t'})=f(\xi(\{y,z\}))=5>1=f(\xi(\{x,y\}))=f(\xi), 
\end{align*}
which means that
condition (i) in the definition of ordinal concavity is satisfied.
\end{example}

In the second example, we consider settings in which a number of seats
are reserved for each student type at each school.

\begin{example}[\emph{Saturated Diversity}\label{ex:saturated}]
For each school $c\in \calc$ and type $t\in \calt$, let $r^t_c\in \mathbb Z_+$ be
the number of reserved seats for type-$t$ students at school $c$.
Suppose that $\Xi^0=\{\xi\in \mathbb{Z}^{|\mathcal{C}|\times |\mathcal{T}|}_+ \mid \sum_{(c,t) \in \mathcal{C}\times \mathcal{T}} \xi_c^t \leq q\}$ for some $q\in \mathbb{Z}_+$; namely, $\Xi^0$ is the set of distributions satisfying a capacity constraint.
Then, for each $\xi\in \Xi^0$,
\[f^s(\xi)=\sum_{(c,t) \in \mathcal{C} \times \mathcal{T}} \min\{ \xi^t_c, r^t_c\},
\]
is an ordinally concave function.
\end{example}

As noted in the Introduction, reserves have been studied in the literature and employed in school choice programs in the real world, e.g., in
Chile \citep{correa19} or India \citep{sonmez2022affirmative}.\footnote{See also \cite{aygtur17,aytur20} for affirmative action in India.}

The next example generalizes reserves so that the marginal value of
each type of student at every school is non-increasing.

\begin{example}[\emph{Marginally Decreasing Diversity}]\label{ex:marginal}
For each school $c\in \calc$ and type $t\in \calt$, let $g_{c,t}$ be
a univariate concave function.
Suppose that
$\Xi^0$ is defined as in Example \ref{ex:saturated}. Then, for each $\xi\in \Xi^0$,
\[f^m(\xi)=\sum_{(c,t)\in \mathcal{C} \times \mathcal{T}} g_{c,t}(\xi^t_c)
\]
is an ordinally concave function.
\end{example}

Marginally decreasing diversity allows for the marginal value of an additional student of a given type to depend on the number of other admitted students of the same type. The case of saturated diversity is a special case in which the marginal value is positive and constant up to reserves and then drops to zero. The additional flexibility of marginally decreasing diversity allows college admissions offices to make a more nuanced tradeoff between students of different types than saturated diversity.

We further generalize the example so that the distributional objective also depends
on the number of minority students (or international students) at the university level.

\begin{example}[\emph{University Diversity}]\label{ex:total}
Let $\mathcal{M} \subseteq \mathcal{T}$ be a set of minority types.
For each school $c\in \calc$ and type $t\in \calt$,
let $g_{c,t}$ be a univariate concave function. Likewise, let $h$ be a univariate concave function. Suppose that
$\Xi^0$ is defined as in Example \ref{ex:saturated}. Then, for each $\xi\in \Xi^0$,
\[f^u(\xi)=h\left(\sum_{(c,t)\in \mathcal{C} \times \mathcal{M}} \xi^t_c\right) +
\sum_{(c,t) \in \mathcal{C} \times \mathcal{T}} g_{c,t}(\xi^t_c)
\]
is an ordinally concave function.
\end{example}

In Appendix \ref{app:compare}, we show that the distributional objectives
defined in Examples \ref{ex:saturated}-\ref{ex:total} satisfy ordinal
concavity (see Proposition \ref{prop:comparison} and the subsequent paragraph). 
Of course, ordinal concavity excludes some cases of interest. Examples include minimum guarantee requirements with overlapping types as well as the requirement to have a fixed proportion of students of different types.

\subsection{Path Independence and the Law of Aggregate Demand}
In this section, we investigate further desirable properties of the
distribution-conscious choice rule. We first establish the following result.

\begin{theorem}\label{thm:pi}
Suppose that the distributional objective $f$ is \oconcave{}. Then the
distribution-conscious choice rule $C^d$ satisfies path independence.
\end{theorem}

Even though the distribution-conscious choice rule satisfies path independence
when the distributional objective $f$ is ordinally concave, it need not satisfy
the law of aggregate demand. We show this claim simply by providing an
example. Let $C^d$ be the distribution-conscious choice rule corresponding to the
distributional objective in Example \ref{ex:ladfail} when $n>5$ for a merit ranking
of contracts. Then $C^d(\{x,y,z\})=\{z\}$ and $C^d(\{x,y\})=\{x,y\}$
show that $C^d$ does not satisfy the law of aggregate demand because
$|C^d(\{x,y,z\})| < |C^d(\{x,y\})|$.

To get the law of aggregate demand, we introduce an additional concavity assumption.
\begin{definition}
The distributional objective $f: \Xi^0 \rightarrow \mathbb{R}_+$ is \textbf{size-restricted concave} if, for each $\xi, \tilde \xi\in \Xi^0$ with $||\xi||>||\tilde \xi||$, there exists $(c,t)\in \mathcal{C}\times \mathcal{T}$ with $\xi_c^t>\tilde \xi_c^t$ such that
\begin{enumerate}
\item[(i)] $f(\xi-\chi_{c,t})>f(\xi)$, or
\item[(ii)] $f(\tilde \xi+\chi_{c,t})>f(\tilde \xi)$, or
\item[(iii)] $f(\xi-\chi_{c,t})=f(\xi)$ and $f(\tilde \xi+\chi_{c,t})=f(\tilde \xi)$.
\end{enumerate}
\end{definition}
This condition was originally introduced by \cite{hakoyeyo24} for functions defined over $\{0,1\}^{|\mathcal{C}| \times |\mathcal{T}|}$.\footnote{In \cite{hakoyeyo24},
we consider a combinatorial choice problem without student types and prove that ordinal concavity and size-restricted concavity are
necessary for the induced choice rule to satisfy path-independence and the law of aggregate demand; see Theorem \ref{thm:pi} therein for a formal statement.} We generalize it to functions over $\mathbb{Z}^{|\mathcal{C}|\times |\mathcal{T}|}_+$.
Like ordinal concavity, this condition starts with two given distributions and imposes that the function value either goes up when one of the distributions is made closer to
the other one or the function values stay the same when both distributions are made closer to each other. Different from ordinal concavity, size-restricted concavity
requires that the first distribution has a larger sum of coordinates than the second distribution and, furthermore, when distributions are made closer only a single coordinate changes.



%

In Example \ref{ex:ladfail}, when $n>5$, size-restricted concavity fails. To see this, let $\xi=\xi(\{x,y\})$ and $\tilde \xi=\xi(\{z\})$  and note that $||\xi||>||\tilde \xi||$. Then, $\xi_c^t>\tilde \xi_c^t$ holds if $\chi_{c,t}=\xi(\{x\})$ or $\chi_{c,t}=\xi(\{y\})$. If $\chi_{c,t}=\xi(\{x\})$,
\begin{align*}
&f(\xi-\chi_{c,t})=f(\xi(\{y\}))=1=f(\xi(\{x,y\}))=f(\xi), \text{ and }  \\
&f(\tilde \xi+\chi_{c,t})=f(\xi(\{x,z\})=5<n=f(\xi(\{z\}))=f(\tilde \xi),
\end{align*}
showing that neither conditions (i)-(iii) of size-restricted concavity holds. The same conclusion follows when $\chi_{c,t}=\xi(\{y\})$. One can verify that size-restricted concavity holds when $n=5$.
Other examples in Section \ref{sec:choice} also satisfy size-restricted concavity; see the discussion after Proposition \ref{prop:comparison} in Appendix \ref{app:compare}.

%
%


Assuming size-restricted concavity of the distributional objective, in addition to ordinal concavity,
delivers the law of aggregate demand for the distribution-conscious choice rule.\footnote{Theorem  \ref{prop:structure} also holds under an alternative assumption that $f$ is ordinally concave and {\it monotone}; see the discussion after the proof of Theorem \ref{prop:structure} in Appendix \ref{app:proofs}.}


\begin{theorem}\label{prop:structure}
Suppose that the distributional objective $f$ is \oconcave{} and size-restricted concave. Then
the distribution-conscious choice rule $C^d$ satisfies the law of aggregate demand.
\end{theorem}

\section{Maximizing Merit Subject to a Level of Distributional Objective}\label{sec:frontier}
A university administration may want to maximize merit of an incoming freshman
class subject to attaining a given level of the distributional objective instead of lexicographically
maximizing these two objectives. In this section, we introduce a family of choice rules,
parameterized by the level of the distributional objective, that achieves this goal. Using this family, we provide
an algorithm that produces the Pareto frontier of the distributional objective and merit.

\subsection{Maximizing Merit Subject to a Level of Distributional Objective} \label{sec:choice-target}
Let $\lambda\in \mathbb{R}_+$ be a target level of the distributional objective.
If the level is achievable for a given set of applications,
then the goal is to choose a subset that maximizes merit subject to attaining that level.
Otherwise, if the  level is not achievable, then the goal is to maximize the distributional objective first and merit second
as in the distribution-conscious choice rule in Section \ref{sec:choice}.

Key to our analysis is to formulate a new distributional objective so that the distribution-conscious choice rule developed in the previous section for the new objective maximizes
merit subject to achieving the distributional-objective level. Specifically,
consider the following modification of the original distributional objective $f$, denoted as $f_\lambda$: for each $\xi\in \Xi^0$,
\[f_{\lambda}(\xi)=\min\{f(\xi),\lambda\}.\]
Therefore, $f_{\lambda}: \Xi^0 \rightarrow \mathbb{R}_+$ is the distributional objective
that flattens the top part of the distributional objective $f$ by $\lambda$.
For each $X'\subseteq X$, $f(\xi(X'))\geq \min\{f(\xi(C^d(X))),\lambda\}$ is equivalent to $\xi(X')\in \underset {\xi\in \Xi^0} {\arg\max} \: f_\lambda(\xi)$, so our goal is to choose $X'\subseteq X$ that maximizes merit subject to $\xi(X')$ 
being an optimum of $f_\lambda$. 
This is exactly what the distribution-conscious choice rule does when it is defined using the distributional objective $f_{\lambda}$, which we denote by $C^d_{\lambda}$.
For example, if $\lambda\geq f(\xi(C^d(X))$, then $C^d_{\lambda}(X)=C^d(X)$.
If $\lambda=0$, then $C^d_{\lambda}$ maximizes the merit ranking subject to attaining a feasible distribution in $\Xi^0$.

If $f_\lambda$ is ordinally concave, then the desirable properties of $C^d$ established in Theorems \ref{thm:diversitychoice} and \ref{thm:pi}
hold for $C_\lambda^d$ as well. Unfortunately, ordinal concavity of $f$ does not necessarily imply ordinal concavity of $f_\lambda$; see Example \ref{ex:oc-truncation-fail} in Appendix \ref{app:auxiliary}.
%
To guarantee that $f_\lambda$ is ordinally concave for each $\lambda$, we explore other concavity conditions.

\begin{definition}[\cite{hakoye2022}]\label{def:pseudo}
The distributional objective $f: \Xi^0 \rightarrow \mathbb{R}_+$ is \textbf{pseudo M$^\natural$-concave} if, for each
$\xi, \tilde{\xi} \in \Xi^0$ and $(c, t) \in \mathcal{C} \times \mathcal{T}$ with $\xi_c^t>\tilde{\xi}_{c}^t\:$,
there exists $\left(c^{\prime}, t^{\prime}\right) \in (\mathcal{C} \times \mathcal{T})\cup\{\emptyset\}$ (with $\xi_{c^{\prime}}^{t^{\prime}}<\tilde{\xi}_{c^{\prime}}^{t^{\prime}}$ whenever $(c',t')\neq \emptyset$) such that
$$
\min\{f(\xi),f(\tilde{\xi})\}\leq\min\{f(\xi-\chi_{c,t}+\chi_{c',t'}), f(\tilde{\xi}+\chi_{c,t}-\chi_{c',t'})\}.
$$
\end{definition}
Pseudo M$^\natural$-concavity is similar in spirit to ordinal concavity in the sense that both conditions
require the value of $f$ to increase when $\xi$ and $\tilde \xi$ move toward each other (recall the interpretation of Definition \ref{def:ordinal}). 

\begin{remark}\label{rem: pseudo}
One can check that
the first two statements in Theorem \ref{thm:diversitychoice} also hold under pseudo M$^\natural$-concavity. 
\end{remark}

Pseudo M$^\natural$-concavity of $f$ is logically independent of ordinal concavity of $f$,\footnote{The distributional objective in Example \ref{ex:oc-truncation-fail} satisfies ordinal concavity but violates pseudo M$^\natural$-concavity. The distributional objective in Example \ref{ex:truncation-sufficient-fail} satisfies pseudo M$^\natural$-concavity but violates ordinal concavity.
These examples are presented in Appendix \ref{app:auxiliary}.
} but it is related to ordinal concavity of $f_\lambda$ for each $\lambda\geq 0$.

\begin{proposition}\label{prop:truncation-necessary}
If $f_\lambda$ is ordinally concave for each $\lambda\geq 0$, then $f$ is pseudo M$^\natural$-concave.
\end{proposition}

Unfortunately, the converse of Proposition \ref{prop:truncation-necessary} does not hold; see Example \ref{ex:truncation-sufficient-fail} in Appendix \ref{app:auxiliary}.
%
%
%
To guarantee the equivalence to ordinal concavity of $f_\lambda$ for each $\lambda\geq 0$, we strengthen pseudo M$^\natural$-concavity as follows.

\begin{definition}\label{def:semistrict-Mc}
The distributional objective $f: \Xi^0 \rightarrow \mathbb{R}_+$ is \textbf{pseudo M$^\natural$-concave$^+$} if, for each $\xi, \tilde{\xi} \in \Xi^0$ and $(c, t) \in \mathcal{C} \times \mathcal{T}$ with $\xi_c^t>\tilde{\xi}_{c}^t\:$, there exists $\left(c^{\prime}, t^{\prime}\right) \in (\mathcal{C} \times \mathcal{T})\cup\{\emptyset\}$ (with $\xi_{c^{\prime}}^{t^{\prime}}<\tilde{\xi}_{c^{\prime}}^{t^{\prime}}$ whenever $(c',t')\neq \emptyset$) such that
\begin{align*}
\min\{f(\xi),f(\tilde{\xi})\}\leq\min\{f(\xi-\chi_{c,t}+\chi_{c',t'}), f(\tilde{\xi}+\chi_{c,t}-\chi_{c',t'})\}.
\end{align*}
Moreover,
\begin{enumerate}[(A)]
\item \label{semistrict-a}
If $f(\xi)>f(\xi-\chi_{c,t}+\chi_{c',t'})$ and $f(\tilde{\xi})=f(\tilde{\xi}+\chi_{c,t}-\chi_{c',t'})$ hold, then there exists $\left(c'', t''\right) \in (\mathcal{C} \times \mathcal{T})\cup\{\emptyset\}$ (with $\xi_{c''}^{t''}<\tilde{\xi}_{c''}^{t''}$ whenever $(c'',t'')\neq \emptyset$) such that
\begin{align*}
f(\tilde{\xi})<f(\tilde{\xi}+\chi_{c,t}-\chi_{c'',t''}).
\end{align*}
\item \label{semistrict-b}
If $f(\tilde{\xi})>f(\tilde{\xi}+\chi_{c,t}-\chi_{c',t'})$ and $f(\xi)=f(\xi-\chi_{c,t}+\chi_{c',t'})$ hold, then there exists $\left(c'', t''\right) \in (\mathcal{C} \times \mathcal{T})\cup\{\emptyset\}$ (with $\xi_{c''}^{t''}<\tilde{\xi}_{c''}^{t''}$ whenever $(c'',t'')\neq \emptyset$) such that
\begin{align*}
f(\xi)<f(\xi-\chi_{c,t}+\chi_{c'',t''}).
\end{align*}
\end{enumerate}
\end{definition}
By the displayed weak inequality in the definition, the if-clause of (\ref{semistrict-a})
is true only if $f(\xi)>f(\tilde \xi)$ and that of (\ref{semistrict-b}) is true only if $f(\tilde \xi)>f(\xi)$.
Hence, the if-clauses concern the case in which the higher function value decreases and the lower function value remains the same when $\xi$ and $\tilde \xi$ move towards each other.
In such a case, pseudo M$^\natural$-concavity$^+$ requires that there is another coordinate $(c'',t'')$ for which
the lower function value {\it strictly} increases as indicated by the displayed strict inequality.

One might find the definition of pseudo-M$^\natural$-concavity$^+$ complicated. In Appendix \ref{app:auxiliary}, we introduce a new condition that implies
pseudo M$^\natural$-concavity$^+$ and is more easily interpretable due to its analogy to the notion of {\it quasi-concavity}, an important assumption on utility functions in the analysis of markets with continuous commodities. In the subsequent analysis, we focus on pseudo M$^\natural$-concavity$^+$ because it allows us to establish an equivalence result and accommodate a canonical
distributional objective given in Section \ref{sec:diversity}, as formalized below.

\begin{proposition} \label{prop:ordinal-equivalence}
Function $f_\lambda$ is ordinally concave for each $\lambda\geq 0$ if, and only if, $f$ is pseudo M$^\natural$-concave$^+$.
\end{proposition}
Now we present two distributional objectives that are pseudo  M$^\natural$-concave$^+$.

\begin{claim}\label{claim:ex1-pMC}
The distributional objective $f$ in Example \ref{ex:ladfail} is pseudo M$^\natural$-concave$^+$.
\end{claim}

\begin{claim} \label{claim:ex2-pMC}
The saturated diversity $f^s$ in Example \ref{ex:saturated} is pseudo M$^\natural$-concave$^+$ if $\Xi^0=\{\xi\in \mathbb{Z}^{|\mathcal{C}|\times |\mathcal{T}|}_+ \mid \sum_{(c,t) \in \mathcal{C}\times \mathcal{T}} \xi_c^t \leq q\}$ for some $q\in \mathbb{Z}_+$.
\end{claim}
Claim \ref{claim:ex2-pMC} implies that the analysis of this section is applicable to the choice rule of a single school with saturated diversity and a capacity constraint.
In Appendix \ref{app:auxiliary}, we provide proofs of Claims \ref{claim:ex1-pMC} and \ref{claim:ex2-pMC} as well as a counterexample to Claim \ref{claim:ex2-pMC}
when $\Xi^0$ is not given as in the statement. We note that the distributional objectives in Examples \ref{ex:marginal} and \ref{ex:total} violate pseudo M$^\natural$-concavity$^+$.

We obtain the following corollary by combining Proposition \ref{prop:ordinal-equivalence} and Theorem \ref{thm:diversitychoice}.
\begin{corollary}\label{corollary:diversitychoice}
Suppose that the distributional objective $f$ is pseudo M$^\natural$-concave$^+$. Then, for each $\lambda\geq 0$ and  set of contracts $X\subseteq \calx$,
\begin{enumerate}[(i)]
\item $C^d_\lambda(X)$ maximizes the distributional objective $f_\lambda$ among subsets of $X$. In particular, if $\lambda\leq f(\xi(C^d(X)))$, then $C^d_\lambda(X)$ attains a level of distributional objective of at least $\lambda$.
\item $C^d_\lambda(X)$ merit dominates each subset $X'$ of $X$ with $f(\xi(X'))\geq \lambda$, and
\item $C^d_\lambda(X)$ can be calculated in $O(|\mathcal{C}| \times |\mathcal{T}| \times |X|^2)$ time, assuming $f$ can be evaluated in a constant time.
\end{enumerate}
\end{corollary}
Hence, if $f$ is pseudo M$^\natural$-concave$^+$, then $C^d_{\lambda}$ maximizes merit subject to attaining a distributional-objective level of at least $\lambda$, and its outcome can be constructed in quadratic time in the number of contracts. Under the weaker notion of
pseudo M$^\natural$-concavity, the first two parts of Corollary \ref{corollary:diversitychoice} continue to hold because pseudo
M$^\natural$-concavity of $f$ implies that, for each $\lambda$, $f_{\lambda}$ satisfies pseudo M$^\natural$-concavity.
Recall that the first two statements in Theorem \ref{thm:diversitychoice} continue to
hold under pseudo M$^\natural$-concavity (Remark \ref{rem: pseudo}). 

\subsection{Pareto Frontier of the Distributional Objective and Merit}
A university administration may not have a particular target level of the distributional objective in
mind but may want to know the Pareto frontier of the distributional objective and merit (PFDM) and choose the incoming 
class from this Pareto frontier. Therefore, identifying PFDM is important, especially for institutions that do not have 
a target level. In this section, we provide an algorithm to find the PFDM
by using the choice rule developed in Section \ref{sec:choice-target}.

For a given set of applications $X$, we define the PFDM of $X$, denoted by $\mathcal{P}(X)$, as follows:
\[\mathcal{P}(X)=\{Y\subseteq X : \not\exists
    Z\subseteq X  \mbox { s.t. } Z\neq Y, Z \mbox{ merit dominates } Y, \mbox { and }
    f(\xi(Z))\geq f(\xi(Y))\}.\]

Throughout this section, we assume that the distributional objective
$f$ takes integer values.\footnote{We make this assumption for expositional simplicity of the computational part.
Alternatively, we can assume that $f$ takes real values as in the preceding sections and that we know in advance the minimum difference between values of $f$. Our algorithm defined below works by changing the cutoff $\lambda$ over the set of possible function values instead of over integers.}
We introduce a new algorithm that traces the PFDM. The algorithm takes a set of contracts $X\subseteq \mathcal{X}$ as input and produces a collection of subsets of $X$.

\paragraph{\textbf{Trace Algorithm}} 

\begin{description}
  \item[Input] Let $X$ be a set of contracts.
  \item[Step 1] Set $k=0$, $\lambda_0=0$, and $\mathcal{X}_0=\emptyset$.
  \item[Step 2] Let $\mathcal{X}_{k+1}=\mathcal{X}_k \cup \{C^d_{\lambda_k}(X)\}$. If
   $C^d_{\lambda_k}(X) = C^d(X)$, go to Step 4. Otherwise, set $\lambda_{k+1}=f(C^d_{\lambda_k}(X))+1$ and go to Step 3.
  \item[Step 3] Add 1 to $k$ and go to Step 2.
  \item[Step 4] Return $\mathcal{X}_{k+1}$ and stop.
\end{description}

Since the number of contracts is finite, the distributional objective $f$ can take only a finite number
of values. Therefore, the algorithm ends at some finite $k$ because
$C^d(X)$ maximizes the distributional objective among subsets of $X$, and it merit
dominates any subset with a distributional objective of $\xi(C^d(X))$ (Theorem \ref{thm:diversitychoice}).

The main result of this section is the following.

\begin{theorem}\label{thm:trace}
Suppose that $f$ is pseudo M$^\natural$-concave$^+$.
Then, for each $X\subseteq \mathcal{X}$, the trace algorithm outcome is the Pareto frontier of the 
distributional objective and merit $\mathcal{P}(X)$. 
The time complexity of the algorithm is $O(\alpha \times |\mathcal{C}| \times |\mathcal{T}| \times |X|^2)$, assuming $f$
can be evaluated in a constant time and $\alpha$ is the maximum value of $f$.
\end{theorem}

Hence, the trace algorithm finds all subsets of the set of applications
that generate the PFDM. The computational part states that the trace algorithm is {\it pseudo polynomial} in the sense that the time complexity is polynomial in the largest integer present in the input data describing the matching problem.
We observe that the first part of the result that the trace algorithm finds the PFDM also holds under pseudo M$^\natural$-concavity.
We illustrate the trace algorithm in Example \ref{ex:frontier} in Appendix \ref{app:auxiliary}.

\section{Key Mathematical Results and Proof Sketch of Theorem \ref{thm:diversitychoice}}\label{sec:math}
In this section, we present key mathematical results used in our proofs which may be of independent interest.
Then, we sketch the proof of Theorem \ref{thm:diversitychoice}.

\subsection{Matroids and the Greedy Rule}\label{sec:matroid}
Here, we first follow \cite{oxley} to introduce some basic definitions. Then we provide two novel characterizations of matroids.

A \emph{matroid} is a pair $(\calx,\calf)$ where $\calx$ is a finite set of contracts
and $\calf$ is a collection of subsets of $\calx$ that satisfies the following
three properties.
\begin{description}
  \item[I1] $\emptyset \in \calf$.
  \item[I2] If $X \in \calf$ and $X' \subseteq X$, then $X' \in \calf$.
  \item[I3] If $X, X' \in \calf$ and $|X|<|X'|$, then there is $x \in X' \setminus X$
  such that $X\cup \{x\} \in \calf$.
\end{description}

Set $\calx$ is called the \emph{ground set} of the matroid. Every set
in $\calf$ is called an \emph{independent set}. An independent set
is called a \emph{base} if no proper superset of it is independent.
\emph{I3} implies that all bases of a matroid have the same cardinality.
In addition, the set of bases $\mathcal{B}$ is characterized by the
following two properties.
\begin{description}
  \item[B1] $\mathcal{B}$ is non-empty.
  \item[B2] If $X, X' \in \mathcal{B}$ and $x \in X \setminus X'$, then there is $x' \in X' \setminus X$ such that
  $(X \setminus \{x\}) \cup \{x'\} \in \mathcal{B}$.
\end{description}
More precisely, if $(\calx,\calf)$ is a matroid, then the set of its bases satisfies
\emph{B1} and \emph{B2}; moreover, if a collection of subsets $\mathcal{B}$
satisfies \emph{B1} and \emph{B2}, then there exists a matroid of which
$\mathcal{B}$ is the set of bases.
The stronger version of \emph{B2} where the implication is
$(X \setminus \{x\}) \cup \{x'\} \in \mathcal{B}$
and $(X'\setminus \{x'\}) \cup \{x\} \in \mathcal{B}$ also holds \citep{brualdi1969}. \label{strongerB2}
We next consider a weaker version of \emph{B2} that we call \emph{B2'}.
\begin{description}
  \item[B2'] If $X, X' \in \mathcal{B}$ and $x \in X \setminus X'$, then there are $x' \in X' \setminus X$ and $Y\in \mathcal{B}$ such that
  $(X \setminus \{x\}) \cup \{x'\} \subseteq Y$.
\end{description}
That is, we weaken the condition \emph{B2} by requiring $(X \setminus \{x\}) \cup \{x'\}$ is only a \textit{subset} of
an element of $\mathcal B$.

In the next lemma, we provide a new characterization for the set of bases of a matroid.

\begin{lemma}\label{lem:matchar}
Let $\mathcal{B}$ be a collection of subsets of $\mathcal{X}$. Then
$\mathcal{B}$ is the collection of bases of a matroid on $\mathcal{X}$
if, and only if, B1 and B2' hold.
\end{lemma}

As already mentioned, it is well known that \emph{B1} and \emph{B2} provide
a characterization for the set of bases. In our proof, we show that
\emph{B1} and \emph{B2'} imply \emph{B2}. Therefore, \emph{B1} and \emph{B2'}
provide another characterization of the set of bases, which is easier to
check than \emph{B1} and \emph{B2} since \emph{B2'} is weaker than \emph{B2}.
We use this characterization in our proofs, and we note that this is
a novel characterization that may be of independent interest and prove
useful elsewhere.

The following is a well-known algorithm, referred to as the greedy algorithm in the combinatorial-optimization literature.
To define it, we assume that
there exists a weight function on the set of contracts that assigns a distinct real
number to every contract. By changing the set of available contracts,
we get a well-defined choice rule. Therefore, we refer to it as the \emph{greedy rule}.

\paragraph{\textbf{Greedy Rule}}
\begin{description}
  \item[Input] Let $X\subseteq \mathcal{X}$ and $\calf$ be a collection of subsets of $\mathcal{X}$.
  \item[Step 1] Set $X_0=\emptyset$ and $k=0$.
  \item[Step 2] If there exist $x \in X \setminus X_k$ and $Y\in \calf$ such that $X_k \cup \{x\} \subseteq Y$,
        then choose such a contract $x_{k+1}$ with the highest non-negative weight,
        let $X_{k+1}=X_k \cup \{x_{k+1}\}$, and go to Step 3.\footnote{A more common definition of the greedy rule requires  $X_k \cup \{x\} \in \calf$ instead of the existence of $Y \in \calf$ with $X_k \cup \{x\} \subseteq Y$. Clearly, that definition is equivalent to the present definition if $(X, \calf)$ satisfies \emph{I2}, and hence in particular, if it is a matroid.} Otherwise, go to Step 4.
  \item[Step 3] Add 1 to $k$ and go to Step 2.
  \item[Step 4] Return $X_{k+1}$ and stop.
\end{description}

When $(\calx,\calf)$ is a matroid, the greedy rule produces an independent set
that maximizes the total weight among all independent sets that can be
chosen. We next provide a new
characterization of matroids using properties of the greedy rule.

\begin{proposition}\label{prop:matroid}
Let $\mathcal{F}$ be a nonempty collection of subsets of $\mathcal{X}$. The following statements are equivalent.
\begin{enumerate}
\item $(\mathcal{X},\mathcal{F})$ is a matroid.
\item For all weight functions on $\mathcal{X}$, the greedy rule satisfies path independence.
\item For all weight functions on $\mathcal{X}$, the greedy rule satisfies path independence and the law of aggregate demand.
\end{enumerate}
\end{proposition}

If $(\mathcal{X},\mathcal{F})$ is a matroid, then the greedy rule satisfies
path independence \citep{fleiner2001} and the law of aggregate
demand \citep{yokoi2019}. Therefore, (1) implies (3). Furthermore, (3) implies
(2) trivially. In our proof, we show that if the greedy rule satisfies
path independence for all weight functions on $\mathcal{X}$, then
$(\mathcal{X},\mathcal{F})$ is a matroid using our matroid characterization
above (Lemma \ref{lem:matchar}), completing the proof.

\subsection{Convexity for Discrete Sets}\label{sec:convexity}
We use two notions of convexity for discrete sets. See \cite{Murota:SIAM:2003} for intuition and details. The first one is M-convexity.

\begin{definition}
A set of distributions $\Xi$ is \textbf{M-convex} if, for any $\xi,\tilde{\xi} \in \Xi$ and $(c,t) \in \calc \times \calt$ with $\xi_c^t>\tilde{\xi}_c^t\:$,
there exists $(c',t') \in \calc \times \calt$ with $\xi_{c'}^{t'}<\tilde{\xi}_{c'}^{t'}$ such that
\[\xi-\chi_{c,t}+\chi_{c',t'}\in \Xi \; \mbox{ and } \; \tilde{\xi}+\chi_{c,t}-\chi_{c',t'} \in \Xi.\]
\end{definition}

The second convexity notion is a weakening of M-convexity.

\begin{definition}
A set of distributions $\Xi$ is M$^{\natural}$\textbf{-convex} if, for any $\xi,\tilde{\xi} \in \Xi$ and $(c,t) \in \calc \times \calt$
with $\xi_c^t>\tilde{\xi}_c^t\:$, then there exists $(c',t') \in (\calc \times \calt) \cup \{\emptyset\}$
(with $\xi_{c'}^{t'}<\tilde{\xi}_{c'}^{t'}$ whenever $(c',t')\neq \emptyset$) such that
\[\xi-\chi_{c,t}+\chi_{c',t'}\in \Xi \; \mbox{ and } \; \tilde{\xi}+\chi_{c,t}-\chi_{c',t'} \in \Xi.\]
\end{definition}

Given a set of distributions $\Xi$ and a distribution $\xi\in \Xi$, we say
that $\xi$ is \textbf{maximal} in $\Xi$ if there exists no $\xi' \in \Xi\setminus \{\xi\}$ such that $\xi' \geq \xi$.
Therefore, the set of maximal distributions in $\Xi$ is given by
$\{\xi\in \Xi|\nexists \xi'\in \Xi \mbox{ such that } \xi'\geq \xi \mbox{ and } \xi'\neq \xi\}$.

The following lemma shows that a similar relation to the one between independent sets and bases also holds between M$^{\natural}$-convex sets
and M-convex sets.\footnote{Theorem 2.3 in \cite{fujishige2005submodular} proves that the set of maximal elements in an integral g-polymatroid is an integral base polyhedron.
An integral g-polymatroid is a convex hull of an M$^\natural$-convex set, and an integral base polyhedron is a convex hull of an M-convex set. One can prove Lemma \ref{lem:mconvex} by using this result. In Appendix \ref{app:auxiliary}, we provide an independent proof.}

\begin{lemma}\label{lem:mconvex}
The set of maximal distributions in an M$^{\natural}$-convex set is M-convex.
\end{lemma}

\subsection{Sketch of the Proof of Theorem \ref{thm:diversitychoice}} \label{sec:proofsketch}
In Theorem \ref{thm:diversitychoice}, ordinal concavity matters in two ways: to choose a set that
maximizes merit among those that maximize the distributional objective and to make this selection computationally tractable.

The first statement in Theorem \ref{thm:diversitychoice} that the distribution-conscious
choice rule outcome maximizes the distributional objective among all subsets of the set of
applications follows by construction. Therefore, we discuss the proofs for
the second and third statements.
We provide a high-level explanation of our proofs
and also illustrate each step of the construction in the distribution-conscious choice rule using
Example \ref{ex:ladfail}. 

Fix a set of contracts $X\subseteq \mathcal{X}$. The proof that
$C^d(X)$ maximizes merit among all subsets
of $X$ that maximize the distributional objectives has three main steps
and uses discrete convexity notions as well as matroid theory.

\noindent
\emph{Step 1:} The set of maximal distributions in $\Xi^*(X)$ is an M-convex set.

First, we study the structure of  $\Xi^*(X)$ that we
find in the distribution-conscious choice rule construction. We show
that if the distributional objective $f$ is ordinally
concave, then $\Xi^{*}(X)$ satisfies M$^{\natural}$-convexity. Since
the distribution-conscious choice rule produces an outcome that is maximal
in $\Xi^{*}(X)$, we focus on maximal distributions in $\Xi^{*}(X)$.
By Lemma \ref{lem:mconvex}, the set of maximal distributions in an
M$^{\natural}$-convex set is M-convex; therefore, the
set of maximal distributions in $\Xi^{*}(X)$ is M-convex (Lemma \ref{lem:pareto}).

Consider Example \ref{ex:ladfail}. Let $n=5$ and $X=\{x,y,z\}$. For the first step,
we maximize $f$ on
$\Xi^0=\{\xi:\norm{\xi} \leq 2\}$ and get $\Xi^*(X)=\{\xi(\{z\}),\xi(\{x,z\}),
\xi(\{y,z\})\}$, which is an M$^{\natural}$-convex set. The set of
maximal distributions in $\Xi^*(X)$ is equal to $\{\xi(\{x,z\}),\xi(\{y,z\})\}$,
which is an M-convex set.


\noindent
\emph{Step 2:} Let $\mathcal{F}(X) \equiv \{X' \subseteq X | \xi(X')
\leq \xi \mbox{ for some } \xi \in \Xi^*(X)\}$. $(X,\mathcal{F}(X))$ is a matroid.

Next, we consider subsets of $X$ that have a distribution less than or
equal to a distribution in $\Xi^*(X)$, and, hence, these sets have
a distribution less than or equal to a maximal distribution in $\Xi^*(X)$.
$\mathcal{F}(X)$ is the collection of such sets.
Depending on the merit ranking, the distribution-conscious choice rule can produce
any maximal set in $\mathcal{F}(X)$ because in Step 2 of the distribution-conscious choice rule
construction contracts are chosen so
that the outcome has a maximal distribution in $\Xi^*(X)$.
Therefore, the structure of maximal
sets in $\mathcal{F}(X)$ plays a crucial role. We show M-convexity of
the set of maximal distributions in $\Xi^*(X)$ implies that the maximal
sets in $\mathcal{F}(X)$ satisfy the base axioms \emph{B1} and \emph{B2}',
which we use in Lemma \ref{lem:matchar} to characterize
the set of bases of a matroid, so $(X, \mathcal{F}(X))$ is a matroid
(Lemma \ref{lem:matroid}).

In Example \ref{ex:ladfail}, when $n=5$ and $X=\{x,y,z\}$, the set
of maximal distributions in
$\Xi^*(X)$ is equal to $\{\xi(\{x,z\}),\xi(\{y,z\})\}$. Therefore, the
collection of maximal sets in $\mathcal{F}(X)$ is equal to $\{\{x,z\},\{y,z\}\}$,
which satisfy the base axioms. Hence, $(X,\mathcal{F}(X))=
(X, \{\emptyset, \{x\}, \{y\}, \{z\},\{x,z\},\{y,z\}\})$ is a matroid.

\noindent
\emph{Step 3:} The greedy rule on $(X,\mathcal{F}(X))$ produces $C^d(X)$.

Finally, we show that the greedy rule on matroid $(X,\mathcal{F}(X))$ produces
$C^d(X)$ (Lemma \ref{lem:equal}). Thus, $C^d(X)$ is a base of the matroid $(X,\mathcal{F}(X))$.
\cite{gale1968} shows that the greedy rule outcome merit dominates any independent set.
Therefore, $C^d(X)$ merit dominates any set in $\mathcal{F}(X)$, which includes
subsets of $X$ that maximize the distributional objective.

In Example \ref{ex:ladfail}, when $n=5$ and $X=\{x,y,z\}$, the greedy rule on
$(X,\mathcal{F}(X))$ may produce $\{x,z\}$ and $\{y,z\}$ depending on the relative
merit ranking of $x$ and $y$. Therefore, if $x\succ y$, then the distribution-conscious
choice rule produces $\{x,z\}$, and, if $y\succ x$, then the distribution-conscious choice
rule produces $\{y,z\}$.


The proof of the third statement in Theorem \ref{thm:diversitychoice} works in
two main steps. In the first step, we generalize a technique used in discrete convex
analysis
to our setting to find a distribution that maximizes the distributional objective.
Step 1 of the distribution-conscious choice rule involves the problem of finding a distribution in
$\Xi^*(X)$, i.e., a maximizer of $f(\xi)$ subject to
$0\leq \xi\leq \xi(X)$. Clearly, checking all distributions is computationally hard
because the size of the domain depends exponentially on the number of colleges and
types (recall $\Xi^0\subseteq \mathbb{Z}^{|\mathcal{C}|\times |\mathcal{T}|}_+$).
We instead
generalize the so-called {\it domain-reduction algorithm} to our setting.

We illustrate the algorithm in Example \ref{ex:ladfail}. Let $n=5$ and $X=\{x,y,z\}$.
Since $|\mathcal{C}|\times |\mathcal{T}|=3$, we identify $\mathbb{Z}^{|\mathcal{C}|\times |\mathcal{T}|}_+$ with $\mathbb{Z}^3_+$
and assume that $\xi(\{x\})=(1,0,0)$, $\xi(\{y\})=(0,1,0)$, and $\xi(\{z\})=(0,0,1)$.
The algorithm starts from $\xi=(0,0,0)$ and iteratively updates $\xi$ until it reaches a maximizer of $f$.
In every iteration, we identify a direction $d \in \{(1,0,0), (0,1,0), (0,0,1)\}$ in which $f(\xi+d)$ is maximized.
By the definition of $f$,
\begin{align*}
f((0,0,0)+d)=\begin{cases} 1 & \text{ if } d=(1,0,0) \text{ or } (0,1,0), \\
                                   5 & \text{ if } d=(0,0,1).
                \end{cases}
\end{align*}
The maximum function value is attained when $(0,0,0)$ moves toward the direction $d=(0,0,1)$, so we update $\xi=(0,0,0)$ to $\xi+d=(0,0,1)$. Importantly, $d=(0,0,1)$ being a solution to the maximization problem implies that there exists a maximizer $\xi^*$ of $f$ with $\xi^*\geq (0,0,1)$ due to the {\it maximizer-cut theorem} that we establish for ordinally concave 
functions (Theorem \ref{thm:maximizer-cut}). 
In words, we can ``cut off''  distributions that have zero as their third coordinate
 and reduce the set of distributions we search
for from $\{\xi:\xi\geq (0,0,0)\}$ to $\{\xi:\xi\geq (0,0,1)\}$.\footnote{We verify the maximizer-cut theorem in the current example. The maximizers of $f$ are
\begin{align*}
(0,0,1)(=\xi(\{z\})), \: (1,0,1)(=\xi(\{x,z\})), \: (0,1,1)(=\xi(\{y,z\})),
\end{align*}
showing that there exists a maximizer with the third coordinate being one (each maximizer satisfies the condition).}

\begin{remark}\label{rem:maximizer-cut}
The domain-reduction algorithm and the maximizer-cut theorem are originally established for M-convex functions; see Section 10.1.3 and Theorem 6.28 in \cite{Murota:SIAM:2003}. 
Our proof generalizes Murota's proof. 
\end{remark}

The algorithm terminates when $\xi$ does not increase in any direction, which is interpreted as $\xi$ locally maximizing the distributional objective.
We prove that local maximization implies global maximization, i.e., the final distribution $\xi$ is
a global maximizer and, hence, included in $\Xi^*(X)$ (Lemma \ref{lem:outcome-maximizer}).\footnote{This implication is reminiscent of the same property under the standard concavity for univariate continuous functions. In the formal proof, we show that the final distribution $\xi$ is a {\it maximal} distribution in $\Xi^*(X)$ (Lemma \ref{lem:outcome-maximal}).}
At each iteration, the number of directions that we search for is $|\mathcal{C}| \times |\mathcal{T}|$.
Furthermore, since the domain for maximization is restricted to $\{\xi:0\leq \xi\leq \xi(X)\}$ and shrinks in every iteration, the number of iterations is at most $\norm{\xi(X)}$, which is bounded by $|X|$, a linear function of the number of applications. Hence, the domain-reduction algorithm finds a maximizer in $O(|\mathcal{C}| \times |\mathcal{T}| \times |X|)$.

The domain-reduction algorithm finds {\it one} maximizer, but Step 2 of the distribution-conscious choice rule searches for {\it all} maximizers.
It turns out that the process of checking all maximizers can be simplified to checking only local distributions around a maximizer
 (Lemma \ref{lem:modified}), which is more computationally tractable.
Building upon this finding, we develop a modified version of the distribution-conscious choice rule
and show that the new choice rule produces the same outcome as the original one (Lemma \ref{lem:choice-equivalence}) and can be
calculated in $O(|\mathcal{C}| \times |\mathcal{T}| \times |X|^2)$.

\section{Conclusion}\label{sec:conclusion}
When institutions hire workers or admit students, they often need to balance a distributional objective and merit, which may sometimes conflict with each other. In this context, we have identified
a family of institutional choice rules with appealing properties. First, the choice rules maximize merit
subject to attaining a level of the distributional objective in a computationally efficient manner. 
Second, they satisfy path-independence, 
which guarantees that the chosen set of applicants does not depend on the order of selection. Finally, 
a subclass of those choice rules satisfy the law of aggregate demand. The latter two properties guarantee that 
there exists a stable and strategy-proof matching mechanism when there are multiple institutions. We have also introduced the trace algorithm to find the Pareto frontier of the distributional objective and merit. We anticipate that our results will be useful in markets where there are dual objectives, such as diversity and meritocracy.

We assume that the desirability of a group of agents with respect to the distributional objective is measured by a function
satisfying \emph{ordinal concavity}, a notion of discrete concavity that is new to economics to our knowledge. Since ordinal concavity allows the \emph{greedy algorithm} to be
effectively used in discrete optimization problems faced in economics, operations research, and computer science, 
ordinal concavity and its desirable properties may prove useful in other applications in the future.


Lastly, our analysis has highlighted an intimate connection between the theories of
discrete convexity and matroids. For instance, we have provided two novel characterizations
of matroids, which may be of independent interest. Moreover, we introduced and analyzed
different concavity notions such as pseudo M$^\natural$-concavity$^+$.\footnote{See
also semistrict pseudo M$^\natural$-concavity in Appendix \ref{app:auxiliary}.}
We envision that those concavity notions may prove useful in other studies.

\bibliographystyle{aer}
\bibliography{matching}

\newpage

\appendix



\section{Concavity Notions for Discrete Functions}\label{app:compare}
There are two notions of concavity for discrete
functions that are commonly used in discrete mathematics.
The first one is the following.

\begin{definition}\label{def:natural}
A function $f$ is \textbf{M-concave} if, for each $\xi,\tilde{\xi}\in \Xi^0$ and $(c,t) \in \calc \times \calt$ with  $\xi_c^t>\tilde{\xi}_c^t$,
there exists $(c',t') \in \calc \times \calt$ with  $\xi_{c'}^{t'}<\tilde{\xi}_{c'}^{t'}$ such that
\[f(\xi-\chi_{c,t}+\chi_{c',t'})+f(\tilde{\xi}+\chi_{c,t}-\chi_{c',t'})
\geq f(\xi)+f(\tilde \xi).\]
\end{definition}

A weaker version of M-concavity is also used.

\begin{definition}\label{def:natural}
A function $f$ is \textbf{M$^{\natural}$-concave} if, for each $\xi,\tilde{\xi}\in \Xi^0$ and $(c,t) \in \calc \times \calt$ with  $\xi_c^t>\tilde{\xi}_c^t$,
there exists $(c',t') \in (\calc \times \calt) \cup \{\emptyset\}$
(with $\xi_{c'}^{t'}<\tilde{\xi}_{c'}^{t'}$ whenever $(c',t')\neq \emptyset$) such that
\[f(\xi-\chi_{c,t}+\chi_{c',t'})+f(\tilde{\xi}+\chi_{c,t}-\chi_{c',t'})
\geq f(\xi)+f(\tilde \xi).\]

\end{definition}

Even though our ordinal concavity is an ordinal concept,
M-concavity and M$^{\natural}$-concavity both depend on the cardinal
values that the distributional objective takes. Furthermore, both M$^{\natural}$-concavity
and M-concavity imply ordinal concavity.

\begin{proposition}\label{prop:comparison}
If a function is M$^{\natural}$-concave, then it is ordinally concave. There exists
an ordinally concave function that is not M$^{\natural}$-concave.
\end{proposition}

We prove this proposition in Appendix \ref{app:auxiliary}. 
The distributional objective defined in Examples \ref{ex:saturated}-\ref{ex:total}
satisfy M$^{\natural}$-concavity
(see page 140 of \cite{Murota:SIAM:2003}).\footnote{These distributional objectives satisfy M$^\natural$-concavity if $\Xi^0$ is an M$^\natural$-convex set (recall the definition in Section \ref{sec:convexity}), which is true if $\Xi^0$ is given by the set of distributions satisfying a capacity constraint as in the examples.}
Therefore, by Proposition \ref{prop:comparison}, they also satisfy
ordinal concavity.

M$^\natural$-concavity of $f$ implies the following condition (see \cite{murotashioura2018}): for each $\xi, \tilde \xi\in \Xi^0$ with $||\xi||>||\tilde \xi||$, there exists $(c,t)\in \mathcal{C}\times \mathcal{T}$ with $\xi_c^t>\tilde \xi_c^t$ such that
\begin{align*}
f(\xi-\chi_{c,t})+f(\tilde \xi+\chi_{c,t})\geq f(\xi)+f(\tilde \xi).
\end{align*}
One can easily verify that this condition is stronger than size-restricted concavity. Hence, the distributional objective defined in Examples \ref{ex:saturated}-\ref{ex:total} satisfy size-restricted concavity.

\section{Main Proofs}\label{app:proofs}
In this section, we include the proofs of our main result.

For each contract $x\in \mathcal{X}$, the
school associated with the contract is denoted by $\gamma(x) \in \mathcal{C}$ and the
student associated with the contract is denoted by $\sigma(x) \in \mathcal{S}$.

\subsection*{Proof of Lemma \ref{lem:matchar}}
The collection of bases of a matroid satisfies \emph{B1} and \emph{B2}. Furthermore,
\emph{B2} implies \emph{B2'}. Therefore, the collection of bases of a matroid
satisfies \emph{B1} and \emph{B2'}. To finish
the proof, we need to show that \emph{B1} and \emph{B2'} imply \emph{B2}.

Suppose, for contradiction, that \emph{B2} does not hold. Then, there exist
$X_1, X_2 \in \mathcal{B}$ and $x_1 \in X_1 \setminus X_2$ such that
for each $x\in X_2 \setminus X_1$ we have
$(X_1\setminus \{x_1\}) \cup \{x\} \not\in \mathcal{B}$.
\emph{B2'} implies that there exist $x_2 \in X_2 \setminus X_1$
and $Y \in \mathcal{B}$ such that
$(X_1\setminus \{x_1\}) \cup \{x_2\} \subseteq Y$. Note that we also have
$(X_1\setminus \{x_1\}) \cup \{x_2\} \not\in \mathcal{B}$ since
$x_2\in X_2 \setminus X_1$ and, therefore, we can take $x=x_2$ in
$(X_1\setminus \{x_1\}) \cup \{x\} \not\in \mathcal{B}$. Furthermore,
since \emph{B2'} implies that there cannot be two sets in $\mathcal{B}$ such
that one is a proper subset of the other, $X_1$ is not a
subset of $Y$. Therefore, $x_1\notin Y$ because otherwise $X_1$ would be
a proper subset of $Y$.

Let $Z=Y \setminus (X_1\setminus \{x_1\})$. Then $Z=Y \setminus X_1$ since $Y$ does
not include $x_1$. Furthermore, $x_2\in Y$ and $x_2\notin X_1$ imply that $x_2\in Z$.

Now let $X_1^{*} = Y$ and $X_2^{*} = X_1$. We have
\begin{enumerate}[(i)]
\item $X_1^{*}, X_2^{*} \in \mathcal B$,
\item $X_1^{*} \setminus X_2^{*} = Y \setminus X_1 = Z$, and
\item $X_2^{*} \setminus X_1^{*} = X_1 \setminus Y = \{x_1\}$.
\end{enumerate}
By \emph{B2'}, since $x_2\in X_1^{*} \setminus X_2^{*}=Z$, there exists
$y \in X_2^{*} \setminus X_1^{*}=\{x_1\}$ such that
$(X_1^{*} \setminus \{x_2\}) \cup \{y\} \subseteq Y'$ for
some $Y'\in \mathcal{B}$. However, $y=x_1$ implies
$(X_1^{*} \setminus \{x_2\}) \cup \{y\}
= (Y \setminus \{x_2\}) \cup \{x_1\} \supseteq X_1$. Since $Y'\supseteq X_1$ and $Y',X_1\in \mathcal{B}$, \emph{B2'} implies that $Y'=X_1$. Hence,
\[X_1=Y' \supseteq (Y \setminus \{x_2\}) \cup \{x_1\},\]
which implies that $Y=(X_1\setminus \{x_1\}) \cup \{x_2\}$ because, by
construction, $Y \supseteq (X_1\setminus \{x_1\}) \cup \{x_2\}$ and $x_1\notin Y$.
This is a contradiction since $(X_1\setminus \{x_1\}) \cup \{x_2\} \notin \mathcal{B}$ and $Y\in \mathcal{B}$. Hence, \emph{B1} and \emph{B2'} imply \emph{B2}.
Therefore, \emph{B1} and \emph{B2'} provide a characterization of the collection
of bases of a matroid.
\qed


\subsection*{Proof of Theorem \ref{thm:diversitychoice}}
We first prove parts (i) and (ii) using the following lemmas.
Recall that $\Xi^*(X)$ denotes the set of distributions in $\{\xi\in \Xi^0 : 0 \leq \xi \leq \xi(X)\}$ that maximize $f$.

\begin{lemma}\label{lem:pareto}
Suppose that the distributional objective $f$ is \oconcave{}.
For each set of contracts $X\subseteq \mathcal{X}$, the set of maximal distributions in $\Xi^*(X)$ is M-convex.
\end{lemma}

\begin{proof}[Proof of Lemma \ref{lem:pareto}]\renewcommand{\qedsymbol}{$\blacksquare$}
Let $\xi,\tilde \xi \in \Xi^*(X)$ be two distinct distributions, $c \in \calc$ a school,
and $t \in \calt$ a type such that $\xi_c^t>\tilde{\xi}_c^t$.
By ordinal concavity, 
either (i)
\begin{center}
$f(\xi-\chi_{c,t})=f(\xi)$ and $f(\tilde{\xi}+\chi_{c,t})=f(\tilde{\xi})$
\end{center}
or (ii) there exist school $c'\in \mathcal{C}$ and type $t'\in \mathcal{T}$ with $\xi_{c'}^{t'}<\tilde{\xi}_{c'}^{t'}$ such that
  \begin{center}
$f(\xi-\chi_{c,t}+\chi_{c',t'})=f(\xi)$ and $f(\tilde{\xi}+\chi_{c,t}-\chi_{c',t'}) = f(\tilde{\xi})$.
  \end{center}
If (i) holds, then $\xi-\chi_{c,t} \in \Xi^*(X)$ and $\tilde{\xi}+\chi_{c,t} \in \Xi^*(X)$.
Otherwise, if (ii) holds, then $\xi-\chi_{c,t}+\chi_{c',t'} \in \Xi^*(X)$ and $\tilde{\xi}+\chi_{c,t}-\chi_{c',t'} \in \Xi^*(X)$. Therefore, $\Xi^*(X)$ is an M$^{\natural}$-convex set.

We finish the proof by using Lemma \ref{lem:mconvex}:
M$^{\natural}$-convexity of $\Xi^*(X)$ implies that the set of maximal distributions
in $\Xi^*(X)$ is M-convex.
\end{proof}

Recall the definition of $\mathcal{F}(X) \equiv \{Y \subseteq X | \xi(Y) \leq \xi \mbox{ for some } \xi \in \Xi^*(X)\}$.

\begin{lemma}\label{lem:matroid}
Suppose that the distributional objective $f$ is \oconcave{}.
For each set of contracts $X\subseteq \mathcal{X}$, $(X,\mathcal{F}(X))$ is a matroid.
\end{lemma}

\begin{proof}[Proof of Lemma \ref{lem:matroid}]\renewcommand{\qedsymbol}{$\blacksquare$}
We show that the maximal sets in $\mathcal{F}(X)$ satisfy \emph{B1} and \emph{B2'},
which together with Lemma \ref{lem:matchar} implies that they are the bases of a matroid. Since $\mathcal{F}(X)$ satisfies \emph{I2}, $\mathcal{F}(X)$ is the collection of subsets of the bases, which implies that $(X,\mathcal{F}(X))$ is a matroid (see Theorem 1.2.3 of \cite{oxley}).
Since $X$ is a finite set, $\Xi^*(X)$ is nonempty. Therefore, \emph{B1} is satisfied.

We now show \emph{B2'}. Let $X_1$ and $X_2$ be two distinct maximal sets in $\mathcal{F}(X)$. Then, by construction, $\xi(X_1)$ and $\xi(X_2)$ are
maximal distributions in $\Xi^*(X)$. 
We consider two cases in the rest of the proof.

In the first case, for each school $c\in \calc$ and type $t\in \calt$, $\xi_c^t(X_1)=\xi_c^t(X_2)$.
Since $X_1 \neq X_2$, $|X_1 \setminus X_2|>0$. Then,
for each $x_1 \in X_1 \setminus X_2$,
there exists $x_2 \in X_2 \setminus X_1$ such that
$\gamma(x_1)=\gamma(x_2)$ and $\tau(\sigma(x_1))=\tau(\sigma(x_2))$. Therefore, $\xi((X_1\setminus \{x_1\}) \cup \{x_2\})=\xi(X_1)$ and so $f(\xi((X_1\setminus \{x_1\})\cup \{x_2\})=f(\xi(X_1))$,
which implies that $(X_1\setminus \{x_1\})\cup \{x_2\}\in \mathcal{F}(X)$. Therefore, \emph{B2'} is satisfied.

In the second case, there exist school $c\in \calc$ and type $t\in \calt$ such that
$\xi_c^t(X_1)>\xi_c^t(X_2)$. Since $\xi(X_1),\xi(X_2) \in \Xi^*(X)$ and
the set of maximal distributions in $\Xi^*(X)$ is an M-convex set (Lemma \ref{lem:mconvex}),
there exist school $c'\in \calc$ and type $t'\in \calt$  with $\xi_{c'}^{t'}(X_1)<\xi_{c'}^{t'}(X_2)$ such that
$\xi(X_1)-\chi_{c,t}+\chi_{c',t'} \in \Xi^*(X)$ and
$\xi(X_2)+\chi_{c,t}-\chi_{c',t'} \in \Xi^*(X)$. Since $\xi_c^t(X_1)>\xi_c^t(X_2)$ and $\xi_{c'}^{t'}(X_1)<\xi_{c'}^{t'}(X_2)$, there exist
$x_1 \in X_1 \setminus X_2$ and $x_2 \in X_2 \setminus X_1$ such that
$\gamma(x_1)=c$, $\tau(\sigma(x_1))=t$, $\gamma(x_2)=c'$,
and $\tau(\sigma(x_2))=t'$. Therefore,
\[\xi((X_1\setminus \{x_1\})\cup \{x_2\})=\xi(X_1)-\chi_{c,t}+\chi_{c',t'}\in \Xi^*(X),\]
which implies that $(X_1\setminus \{x_1\})\cup \{x_2\} \in \mathcal{F}(X)$.
Therefore, \emph{B2'} is satisfied.

In both cases, we have shown \emph{B1} and \emph{B2'} and $(X,\mathcal{F}(X))$ is a matroid.
\end{proof}

\begin{lemma}\label{lem:equal}
Suppose that the distributional objective $f$ is \oconcave. Then, for each set of contracts
$X\subseteq \mathcal{X}$, the greedy rule on matroid $(X,\mathcal{F}(X))$ produces
$C^d(X)$ when the set of available contracts is $X$.\footnote{To define the greedy rule, we set a weight function in such a way that a contract with a higher merit has a higher weight.}
\end{lemma}

\begin{proof}[Proof of Lemma \ref{lem:equal}]\renewcommand{\qedsymbol}{$\blacksquare$}
We show by induction
that $C^d$ and the greedy rule choose the same set of contracts for each index $k$
used in the definitions of both choice rules and terminate at the same index. Let $X_k$ be defined as in the construction of $C^d(X)$ and $X'_k$ be analogously
defined for the greedy rule. For $k=0$, we have $X_k=\emptyset=X'_k$.
By mathematical induction hypothesis, suppose that $X_j=X'_j$ for
each $j=0,\ldots,k$. We now show the hypothesis for $j=k+1$.

By the induction hypothesis, $\{x\in X\setminus X_k| \exists \xi\in \Xi^*(X) \mbox{ s.t. } \xi(X_k \cup \{x\}) \leq \xi\}$ used in the construction of $C^d$ is the same as
$\{x \in X\setminus X'_k|\exists Y \subseteq \mathcal{F}(X)
\mbox{ s.t. } X'_k \cup \{x\} \subseteq Y\}$ used in the greedy rule description.  Therefore, either both algorithms terminate at index $k$ and produce $X_k=X'_k$
or the same contract $x$ is chosen so that $X_{k+1}=X'_{k+1}$. This
finishes the proof of the mathematical induction hypothesis.

Therefore, the greedy rule on matroid $(X,\mathcal{F}(X))$ produces $C^d(X)$.
\end{proof}

Now, we finish the proofs of parts (i) and (ii).
By Lemma \ref{lem:equal}, $C^d(X)$ is a base of the matroid $(X,\mathcal{F}(X))$.
Therefore, by construction of $\mathcal{F}(X)$, $\xi(C^d(X))\in \Xi^*(X)$,
which means that $C^d(X)$ maximizes the distributional objective $f$ among subsets of $X$.
Furthermore, by \citep{gale1968}, $C^d(X)$ merit dominates each set in $\mathcal{F}(X)$, which includes all subsets of $X$ that maximizes the distributional objective.

We continue with the proof of part (iii). We prove the result in a number of steps.

\paragraph{Step 1: } We prove the so-called {\it maximizer-cut theorem} for ordinally concave functions.
\footnote{As noted in Remark \ref{rem:maximizer-cut}, our proof builds on that of \cite{Murota:SIAM:2003}.}
\begin{lemma}\label{lem:maximizer-cut-1}
Let $f$ be ordinally concave, $\xi\in \Xi^0$, $(c,t)\in (\mathcal{C}\times \mathcal{T})\cup\{\emptyset\}$, and $(c',t')\in (\mathcal{C}\times \mathcal{T})\cup\{\emptyset\}$ be such that
\begin{align*}
f(\xi-\chi_{c',t'}+\chi_{c,t})=\max_{(\tilde c', \tilde t')\in (\mathcal{C}\times \mathcal{T})\cup \{\emptyset\}}f(\xi-\chi_{\tilde c' \tilde t'}+\chi_{c,t}).
\end{align*}
Then, there exists $\xi^*\in \underset {\xi\in \Xi^0} {\arg\max} \: f(\xi)$ with $(\xi^*)_{c'}^{t'}\leq \xi_{c'}^{t'}-1+(\chi_{c,t})_{c'}^{t'}$.
\end{lemma}
\begin{proof}[Proof of Lemma \ref{lem:maximizer-cut-1}]\renewcommand{\qedsymbol}{$\blacksquare$}
Let $\xi'=\xi-\chi_{c',t'}+\chi_{c,t}$. Suppose, for contradiction, that there does not exist $\xi^*\in \underset {\xi\in \Xi^0} {\arg\max} \: f(\xi)$ with $(\xi^*)_{c'}^{t'}\leq (\xi')_{c'}^{t'}$. Let $\xi^*$ be an element of $\underset {\xi\in \Xi^0} {\arg\max} \: f(\xi)$ that minimizes the $(c',t')$ coordinate. By assumption, we have $(\xi^*)_{c'}^{t'}>(\xi')_{c'}^{t'}$. By ordinal concavity, there exists $(c'', t'') \in (\mathcal{C}\times \mathcal{T})\cup\{\emptyset\}$ (with $(\xi')_{c''}^{t''}>(\xi^*)_{c''}^{t''}$ if $(c'',t'')\neq \emptyset$) such that
\begin{enumerate}
\item $f(\xi^*-\chi_{c',t'}+\chi_{c'',t''})>f(\xi^*)$ or
\item $f(\xi'+\chi_{c',t'}-\chi_{c'',t''})>f(\xi')$ or
\item $f(\xi^*-\chi_{c',t'}+\chi_{c'',t''})=f(\xi^*)$ and $f(\xi'+\chi_{c',t'}-\chi_{c'',t''})=f(\xi')$.
\end{enumerate}
If condition (3) holds, then $\xi^*-\chi_{c',t'}+\chi_{c'',t''}\in \underset {\xi\in \Xi^0} {\arg\max} \: f(\xi)$ and $(\xi^*-\chi_{c',t'}+\chi_{c'',t''})_{c'}^{t'}<(\xi^*)_{c'}^{t'}$, a contradiction to the choice of $\xi^*$. Condition (1) is impossible because $\xi^*\in \underset {\xi\in \Xi^0} {\arg\max} \: f(\xi)$. If condition (2) holds,
\begin{align*}
f(\xi-\chi_{c'',t''}+\chi_{c,t})=f(\xi'+\chi_{c',t'}-\chi_{c'',t''})>f(\xi')=f(\xi-\chi_{c',t'}+\chi_{c,t}),
\end{align*}
a contradiction to the choice of $(c',t')$.
\end{proof}

\begin{lemma} \label{lem:maximizer-cut-2}
Let $f$ be ordinally concave, $\xi \in \Xi^0$ with $\xi\notin \underset {\xi\in \Xi^0} {\arg\max} \: f(\xi)$, and $(c,t), (c',t')\in (\mathcal{C}\times \mathcal{T})\cup\{\emptyset\}$ be such that
\begin{align*}
f(\xi-\chi_{c',t'}+\chi_{c,t})=\max_{(\tilde c', \tilde t')\in (\mathcal{C}\times \mathcal{T})\cup\{\emptyset\}}\max_{(\tilde c, \tilde t)\in (\mathcal{C}\times \mathcal{T})\cup\{\emptyset\}}f(\xi-\chi_{\tilde c', \tilde t'}+\chi_{\tilde c, \tilde t}).
\end{align*}
Then, $(c,t)\neq \emptyset$ or $(c',t')\neq \emptyset$ holds.
\end{lemma}
\begin{proof}[Proof of Lemma \ref{lem:maximizer-cut-2}]\renewcommand{\qedsymbol}{$\blacksquare$}
Suppose, for contradiction, that $(c,t)=(c',t')=\emptyset$, i.e.,
$$
f(\xi)=\max_{(\tilde c', \tilde t')\in (\mathcal{C}\times \mathcal{T})\cup\{\emptyset\}}\max_{(\tilde c, \tilde t)\in (\mathcal{C}\times \mathcal{T})\cup\{\emptyset\}}f(\xi-\chi_{\tilde c', \tilde t'}+\chi_{\tilde c, \tilde t}).
$$
Let $\xi^*$ be an element of $\underset {\xi\in \Xi^0} {\arg\max} \: f(\xi)$ that minimizes $\sum_{(\tilde c, \tilde t)}|(\xi^*)_{\tilde c}^{\tilde t}-\xi_{\tilde c}^{\tilde t}|$. Since $\xi\notin \underset {\xi\in \Xi^0} {\arg\max} \: f(\xi)$, there exists $(c'',t'')\in \mathcal{C}\times \mathcal{T}$ with $(\xi^*)_{c''}^{t''}\neq \xi_{c''}^{t''}$. Suppose that  $(\xi^*)_{c''}^{t''}>\xi_{c''}^{t''}$ (the other case $(\xi^*)_{c''}^{t''}<\xi_{c''}^{t''}$ can be handled analogously). By ordinal concavity, there exists $(c''',t''')\in (\mathcal{C}\times \mathcal{T})\cup\{\emptyset\}$ (with $\xi_{c'''}^{t'''}>(\xi^*)_{c'''}^{t'''}$ if $(c''',t''')\neq \emptyset$) such that
\begin{enumerate}
\item $f(\xi^*-\chi_{c'',t''}+\chi_{c''',t'''})>f(\xi^*)$ or
\item $f(\xi+\chi_{c'',t''}-\chi_{c''',t'''})>f(\xi)$ or
\item $f(\xi^*-\chi_{c'',t''}+\chi_{c''',t'''})=f(\xi^*)$ and $f(\xi+\chi_{c'',t''}-\chi_{c''',t'''})=f(\xi)$.
\end{enumerate}
If condition (3) holds, then $\xi^*-\chi_{c'',t''}+\chi_{c''',t'''}\in \underset {\xi\in \Xi^0} {\arg\max} \: f(\xi)$ and
$$
\sum_{(\tilde c, \tilde t)}|(\xi^*-\chi_{c'',t''}+\chi_{c''',t'''})_{\tilde c}^{\tilde t}-\xi_{\tilde c}^{\tilde t}|<\sum_{(\tilde c, \tilde t)}|(\xi^*)_{\tilde c}^{\tilde t}-\xi_{\tilde c}^{\tilde t}|,
$$
which is a contradiction to the choice of
$\xi^*$. Condition (1) is impossible because $\xi^*\in \underset {\xi\in \Xi^0} {\arg\max} \: f(\xi)$. If condition (2) holds, we obtain a contradiction to the assumption made in the beginning of the proof.
\end{proof}

\begin{theorem}[Maximizer-cut theorem] \label{thm:maximizer-cut}
Let $f$ be ordinally concave, $\xi\in \Xi^0$ with $\xi \not\in \underset {\xi\in \Xi^0} {\arg\max} \: f(\xi)$, and $(c,t), (c',t') \in (\mathcal C \times \mathcal T) \cup \{\emptyset\}$ be such that
$$
f(\xi-\chi_{c',t'}+\chi_{c,t}) = \max_{(\tilde c,\tilde t), (\tilde c',\tilde t') \in (\mathcal C \times \mathcal T) \cup \{\emptyset\}} f(\xi-\chi_{\tilde c',\tilde t'}+\chi_{\tilde c,\tilde t}).
$$
Then, $(c,t) \neq \emptyset$ or $(c',t') \neq \emptyset$ holds and the following statements hold:
\begin{enumerate}[(i)]
\item If $(c,t) \neq \emptyset$ and $(c',t') = \emptyset$, then there exists $\xi^*\in \underset {\xi\in \Xi^0} {\arg\max} \: f(\xi)$ with $(\xi^*)^t_c \ge  \xi^t_c+1$,
\item If $(c,t) = \emptyset$ and $(c',t') \neq \emptyset$, then there exists $\xi^* \in \underset {\xi\in \Xi^0} {\arg\max} \: f(\xi)$ with $(\xi^*)^{t'}_{c'} \le \xi^{t'}_{c'}-1$,
\item If $(c,t) \neq \emptyset$ and $(c',t') \neq \emptyset$, then there exists $\xi^* \in \underset {\xi\in \Xi^0} {\arg\max} \: f(\xi)$ with $(\xi^*)^t_c \ge \xi^t_c+1$ and $(\xi^*)^{t'}_{c'} \le \xi^{t'}_{c'}-1$.
\end{enumerate}
\end{theorem}

\begin{proof}[Proof of Theorem \ref{thm:maximizer-cut}]
Note that $(c,t)\neq \emptyset$ or $(c',t')\neq \emptyset$ follows from Lemma \ref{lem:maximizer-cut-2}.

\noindent
\emph{Proof of (i):}
Let $\xi'=\xi+\chi_{c,t}$. Suppose, for contradiction, that there does not exist $\xi^*\in \underset {\xi\in \Xi^0} {\arg\max} \: f(\xi)$ with $(\xi^*)_{c}^{t}\geq (\xi')_{c}^{t}$.  Let $\xi^*$ be an element of $\underset {\xi\in \Xi^0} {\arg\max} \: f(\xi)$ that maximizes the $(c,t)$ coordinate. By assumption, we have $(\xi^*)_{c}^{t}<(\xi')_{c}^{t}$. By ordinal concavity, there exists $(c'',t'') \in (\mathcal{C}\times \mathcal{T})\cup\{\emptyset\}$ (with $(\xi^*)_{c''}^{t''}>(\xi')_{c''}^{t''}$ if $(c'',t'')\neq \emptyset$) such that
\begin{enumerate}
\item $f(\xi'-\chi_{c,t}+\chi_{c'',t''})>f(\xi')$ or
\item $f(\xi^*+\chi_{c,t}-\chi_{c'',t''})>f(\xi^*)$ or
\item $f(\xi'-\chi_{c,t}+\chi_{c'',t''})=f(\xi')$ and $f(\xi^*+\chi_{c,t}-\chi_{c'',t''})=f(\xi^*)$.
\end{enumerate}
If condition (3) holds, then $\xi^*+\chi_{c,t}-\chi_{c'',t''}\in \underset {\xi\in \Xi^0} {\arg\max} \: f(\xi)$ and $(\xi^*+\chi_{c,t}-\chi_{c'',t''})_{c}^{t}>(\xi^*)_{c}^{t}$, a contradiction to the choice of $\xi^*$. Condition (2) is impossible because $\xi^*\in \underset {\xi\in \Xi^0} {\arg\max} \: f(\xi)$. If condition (1) holds,
$$
f(\xi+\chi_{c'',t''})=f(\xi'-\chi_{c,t}+\chi_{c'',t''})>f(\xi')=f(\xi+\chi_{c,t}),
$$
which is a contradiction to the choices of $(c,t)$ and $(c',t')$.

\noindent
\emph{Proof of (ii):} The proof is similar to that for (i).

\noindent
\emph{Proof of (iii):}  Let $\xi'=\xi-\chi_{c',t'}+\chi_{c,t}$. By Lemma \ref{lem:maximizer-cut-1}, there exists $\xi^*\in \underset {\xi\in \Xi^0} {\arg\max} \: f(\xi)$ such that $(\xi^*)_{c'}^{t'}\leq (\xi')_{c'}^{t'}$; we assume $\xi^*$ maximizes $(\xi^*)_{c}^{t}$ among all such vectors. Suppose, for contradiction, that $(\xi^*)_{c}^{t}\geq (\xi')_{c}^{t}$ is not satisfied, i.e., $(\xi^*)_{c}^{t}<(\xi')_{c}^{t}$. By ordinal concavity, there exists $(c'',t'')\in (\mathcal{C}\times \mathcal{T})\cup\{\emptyset\}$ (with $(\xi^*)_{c''}^{t''}>(\xi')_{c''}^{t''}$ if $(c'',t'')\neq \emptyset$) such that
\begin{enumerate}
\item $f(\xi'-\chi_{c,t}+\chi_{c'',t''})>f(\xi')$ or
\item $f(\xi^*+\chi_{c,t}-\chi_{c'',t''})>f(\xi^*)$ or
\item $f(\xi'-\chi_{c,t}+\chi_{c'',t''})=f(\xi')$ and $f(\xi^*+\chi_{c,t}-\chi_{c'',t''})=f(\xi^*)$.
\end{enumerate}
Suppose that condition (3) holds, which implies $\xi^*+\chi_{c,t}-\chi_{c'',t''}\in \underset {\xi\in \Xi^0} {\arg\max} \: f(\xi)$. By Lemma \ref{lem:maximizer-cut-2}, we have $(c,t)\neq (c',t')$ and hence $(\xi^*+\chi_{c,t}-\chi_{c'',t''})_{c'}^{t'}\leq (\xi^*)_{c'}^{t'}$. Together with $(\xi^*+\chi_{c,t}-\chi_{c'',t''})_c^t>(\xi^*)_c^t$, we obtain a  contradiction to the choice of $\xi^*$. Condition (2) is impossible because $\xi^*\in \underset {\xi\in \Xi^0} {\arg\max} \: f(\xi)$. If condition (1) holds,
\begin{align*}
f(\xi-\chi_{c',t'}+\chi_{c'',t''})=f(\xi'-\chi_{c,t}+\chi_{c'',t''})>f(\xi')=f(\xi-\chi_{c',t'}+\chi_{c,t}),
\end{align*}
which is a contradiction to the choices of $(c,t)$ and $(c',t')$.
\end{proof}
Two remarks on Theorem \ref{thm:maximizer-cut} are in order.
\begin{itemize}
\item Although we assume that $\Xi^0\subseteq \mathbb{Z}^{|\mathcal{C}|\times |\mathcal{T}|}_+$, $\mathbf{0}\in \Xi^0$, and $f(\xi)\geq 0$ for each
$\xi\in \Xi^0$, neither of these assumptions is used in the proof. Hence,
the maximizer-cut theorem holds for ordinally concave functions more generally.
\item Among the three statements (i)-(iii), we use only the first one in the proof below.
\end{itemize}

\paragraph{Step 2: } We develop a variation of the {\it domain-reduction algorithm} that produces a maximizer of the distributional objective that is maximal in the set of maximizers.%
\footnote{As noted in Remark \ref{rem:maximizer-cut}, the construction of this algorithm builds on that of \cite{Murota:SIAM:2003}.}
Fix an ordinally concave $f$.
\paragraph{\textbf{Domain-reduction algorithm}}
\begin{description}
\item[Input] Let $X$ be a set of contracts.
\item[Step 1] Set $\xi_0=\mathbf{0}$ and $k=0$.
\item[Step 2] Check if
$$
f(\xi_k) \leq \max\{f(\xi_k+\chi_{\tilde c, \tilde t})\mid (\tilde c, \tilde t) \in \mathcal C \times \mathcal T, \xi_k+\chi_{\tilde c, \tilde t}\leq \xi(X)\}.
$$
If this is the case, then choose a maximizer $(c_{k+1}, t_{k+1})$ of the right-hand side, let $\xi_{k+1}=\xi_k+\chi_{c_{k+1}, t_{k+1}}$, and go to Step 3. Otherwise, go to Step 4.
\item[Step 3] Add 1 to $k$ and go to Step 2.
\item[Step 4] Return $\xi_{k}$ and stop.
\end{description}
Let $k^*$ denote the value of $k$ at the end of the algorithm.
For each $k\in \{0, \dots, k^*\}$, let $\Xi^0_k=\{\xi \in \Xi^0\mid \xi_k\leq \xi\leq \xi(X)\}$ and $f_k: \Xi^0_k\rightarrow \mathbb{R}_+$ be defined as $f_k(\xi)=f(\xi)$ for all $\xi\in \Xi^0_k$. One can verify that ordinal concavity of $f$ is inhereted to $f_k$ for each $k$.
We prove that the algorithm produces a maximal distribution in $\underset {\xi\in \Xi^0_0} {\arg\max} \: f_0(\xi)=\Xi^*(X)$ (recall the notation in the definition of the distribution-conscious choice rule) by establishing three lemmas.
\begin{lemma}\label{lem:maximum-equal}
For each $k\in \{0, \dots, k^*\},$ $\underset {\xi\in \Xi^0_k} {\max} \: f_k(\xi)=\underset {\xi\in \Xi^0_0} {\max} \: f_0(\xi)$.
\end{lemma}
\begin{proof}[Proof of Lemma \ref{lem:maximum-equal}]\renewcommand{\qedsymbol}{$\blacksquare$}
The proof is by mathematical induction. The claim trivially holds for $k=0$. Suppose that it holds for $k-1$. We show the claim for $k$.

\noindent
\emph{Case 1:}
Suppose that $\xi_{k-1}$ is a maximizer of $f_{k-1}$. Then, $f_{k-1}(\xi_{k-1}+\chi_{c_{k}, t_{k}})\leq f_{k-1}(\xi_{k-1})$. Together with $f(\xi_{k-1}+\chi_{c_{k}, t_{k}})\geq f(\xi_{k-1})$ (which follows from the choice of $(c_{k}, t_{k})$) and $f(\xi)=f_{k-1}(\xi)$ for each $\xi\in \Xi^0_{k-1}$, we obtain $f_{k-1}(\xi_{k-1}+\chi_{c_{k}, t_{k}})=f_{k-1}(\xi_{k-1})$. Substituting $f_{k-1}(\xi_{k-1}+\chi_{c_{k}, t_{k}})=f_k(\xi_k)$, we get $f_k(\xi_k)=f_{k-1}(\xi_{k-1})$. Together with $\Xi^0_{k-1}\supseteq \Xi^0_k$ and the assumption of Case 1, $\xi_k$ is a maximizer of $f_k$ and $\underset {\xi\in \Xi^0_k} {\max} \: f_k(\xi)=\underset {\xi\in \Xi^0_{k-1}} {\max} \: f_{k-1}(\xi)$. This equality and mathematical induction hypothesis give us the desired claim.

\noindent
\emph{Case 2:}
Suppose that $\xi_{k-1}$ is not a maximizer of $f_{k-1}$.
\begin{align*}
f_{k-1}(\xi_{k-1}+\chi_{c_{k}, t_{k}})&=f(\xi_{k-1}+\chi_{c_{k}, t_{k}}) \\
&= \max_{(\tilde c,\tilde t) \in (\mathcal C \times \mathcal T)\cup\{\emptyset\}}  f(\xi_{k-1}+\chi_{\tilde c, \tilde t}) \\
&=\max_{(\tilde c,\tilde t) \in (\mathcal C \times \mathcal T)\cup\{\emptyset\}}  f_{k-1}(\xi_{k-1}+\chi_{\tilde c, \tilde t}) \\
&= \max_{(\tilde c,\tilde t), (\tilde c',\tilde t') \in (\mathcal C \times \mathcal T) \cup \{\emptyset\}} f_{k-1}(\xi_{k-1}-\chi_{\tilde c',\tilde t'}+\chi_{\tilde c,\tilde t}),
\end{align*}
where the second equality follows from the choice of $(c_{k}, t_{k})$ and the last equality follows from the fact that every distribution in $\Xi^0_{k-1}$ is greater than or equal to $\xi_{k-1}$.
By Theorem \ref{thm:maximizer-cut}, there exists a maximizer $\xi^*$ of $f_{k-1}$ such that $\xi^* \geq \xi_{k-1}+\chi_{c_{k}, t_{k}}=\xi_k$, which implies $\xi^*\in \Xi^0_k$. Together with $\Xi^0_{k-1}\supseteq \Xi^0_k$, we obtain $\underset {\xi\in \Xi^0_k} {\max} \: f_k(\xi)=\underset {\xi\in \Xi^0_{k-1}} {\max} \: f_{k-1}(\xi)$. This equality and mathematical induction hypothesis give us the desired claim.
\end{proof}

\begin{lemma}\label{lem:outcome-maximizer}
$\xi_{k^*} \in \underset {\xi\in \Xi^0_0} {\arg\max} \: f_0(\xi)$.
\end{lemma}
\begin{proof}[Proof of Lemma \ref{lem:outcome-maximizer}]\renewcommand{\qedsymbol}{$\blacksquare$}
Suppose, for contradiction, that $\xi_{k^*} \notin \underset {\xi\in \Xi^0_0} {\arg\max} \: f_0(\xi)$.
By Lemma \ref{lem:maximum-equal} and $\Xi^0_{k^*}\subseteq \Xi^0_0$, there exists $\xi^* \in \Xi^0_{k^*}$ such that $\xi^*\in \underset {\xi\in \Xi^0_0} {\arg\max} \: f_0(\xi)$ and $f_{k^*}(\xi_{k^*})<f_{k^*}(\xi^*)$, which implies $f(\xi_{k^*})<f(\xi^*)$.
Assume that  $(\xi^*)^t_c >(\xi_{k^*})^t_c$ (such $c$ and $t$ exist because $\xi^*>\xi_{k^*}$ by the definition of $\Xi^0_{k^*}$).
By ordinal concavity of $f$, there exists $(c',t') \in (\mathcal C \times \mathcal T) \cup \{\emptyset\}$, with $(\xi^*)^{t'}_{c'} <(\xi_{k^*})^{t'}_{c'}$ if $(c',t') \neq \emptyset$, such that one of the inequalities required of  ordinal concavity holds.
However, because $\xi^* > \xi_{k^*}$, it follows that $(c',t')=\emptyset$, so we have
\begin{enumerate}
\item $f(\xi_{k^*}+\chi_{c,t})>f(\xi_{k^*})$ or
\item $f(\xi^*-\chi_{c,t})>f(\xi^*)$ or
\item $f(\xi_{k^*}+\chi_{c,t})=f(\xi_{k^*})$ and $f(\xi^*-\chi_{c,t})=(\xi^*)$.
\end{enumerate}
Condition (2) is impossible because $\xi^* \in \underset {\xi\in \Xi^0_0} {\arg\max} \: f_0(\xi)$. Therefore, condition (1) or (3) holds. In either case, because $\xi_{k^*}+\chi_{c,t}\leq \xi^*\leq \xi(X)$, we have
$$
f(\xi_{k^*}) \leq \max\{f(\xi_{k^*}+\chi_{\tilde c, \tilde t})\mid (\tilde c, \tilde t) \in \mathcal C \times \mathcal T, \xi_{k^*}+\chi_{\tilde c, \tilde t}\leq \xi(X)\}.
$$
We obtain a contradiction to the fact that the algorithm terminates when $k=k^*$.
\end{proof}

\begin{lemma}\label{lem:outcome-maximal}
$\xi_{k^*}$ is a maximal distribution in $\underset {\xi\in \Xi^0_0} {\arg\max} \: f_0(\xi)$.
\end{lemma}
\begin{proof}[Proof of Lemma \ref{lem:outcome-maximal}]\renewcommand{\qedsymbol}{$\blacksquare$}
Suppose, for contradiction, that the statement does not hold. By Lemma \ref{lem:outcome-maximizer}, $\xi_{k^*}\in \underset {\xi\in \Xi^0_0} {\arg\max} \: f_0(\xi)$.
Since it is not a maximal distribution, there exists $\xi^*$ such that $\xi^* \in \underset {\xi\in \Xi^0_0} {\arg\max} \: f_0(\xi)$ and $\xi^*>\xi_{k^*}$.
Assume that  $(\xi^*)^t_c >(\xi_{k^*})^t_c$ (such $c$ and $t$ exist because $\xi^*>\xi_{k^*}$).
By ordinal concavity of $f$, there exists $(c',t') \in (\mathcal C \times \mathcal T) \cup \{\emptyset\}$, with $(\xi^*)^{t'}_{c'} <(\xi_{k^*})^{t'}_{c'}$ if $(c',t') \neq \emptyset$, such that one of the inequalities required of  ordinal concavity holds.
However, because $\xi^* > \xi_{k^*}$, it follows that $(c',t')=\emptyset$, so we have
\begin{enumerate}
\item $f(\xi_{k^*}+\chi_{c,t})>f(\xi_{k^*})$ or
\item $f(\xi^*-\chi_{c,t})>f(\xi^*)$ or
\item $f(\xi_{k^*}+\chi_{c,t})=f(\xi_{k^*})$ and $f(\xi^*-\chi_{c,t})=f(\xi^*)$.
\end{enumerate}
If condition (1) or (2) holds, then together with $\xi_{k^*}+\chi_{c,t}\leq \xi^*\leq \xi(X)$ and $\xi^*-\chi_{c,t}\leq \xi^*\leq \xi(X)$, we obtain a contradiction to $\xi_{k^*}, \xi^* \in \underset {\xi\in \Xi^0_0} {\arg\max} \: f_0(\xi)$. Therefore, condition (3) holds, implying that
$$
f(\xi_{k^*}) \leq \max\{f(\xi_{k^*}+\chi_{\tilde c, \tilde t})\mid (\tilde c, \tilde t) \in \mathcal C \times \mathcal T, \xi_{k^*}+\chi_{\tilde c, \tilde t}\leq \xi(X)\}.
$$
We obtain a contradiction to the fact that the algorithm terminates when $k=k^*$.
\end{proof}

\paragraph{Step 3: } We develop a modified version of the distribution-conscious choice rule that produces the same outcome as the original one and is more tractable from a computational viewpoint.
Fix an ordinally concave $f$.

\paragraph{\textbf{Modified Distribution-Conscious Choice Rule}}
\begin{description}
  \item[Input] Let $X$ be a set of contracts. Let $\xi$ be a maximal distribution in $\Xi^*(X)$.
  \item[Step 1]  Set $X_0=\emptyset$, $\xi_0=\xi$, and $k=0$.
  \item[Step 2] Check whether there exists $x\in X\backslash X_k$ that satisfies one of the following conditions:
  \begin{itemize}
  \item[(i)]  $\xi(X_{k}\cup\{x\})\leq \xi_{k}$, or
  \item[(ii)] there exists $(c',t')\in \mathcal{C}\times \mathcal{T}$ such that $\xi_{k}+\chi_{c,t}-\chi_{c',t'}\in \Xi^*(X)$ and $\xi(X_{k}\cup \{x\})\leq \xi_{k}+\chi_{c,t}-\chi_{c',t'}$, where $\chi_{c,t}=\xi(\{x\})$.
  \end{itemize}
  If there exists such a contract, then choose the one $x_{k+1}$ with the highest merit and let
\begin{align*}
X_{k+1}&=X_{k}\cup \{x_{k+1}\}, \;  \\
\xi_{k+1}&=\begin{cases} \xi_{k} \text{ (if (i) holds), } \\
                              \xi_{k}+\chi_{c,t}-\chi_{c',t'} \text{ (if (ii) holds), }
               \end{cases}
\end{align*}
and go to Step 3. Otherwise, go to Step 4.
  \item[Step 3] Add 1 to $k$ and go to Step 2.
  \item[Step 4] Return $X_{k}$ and stop.
\end{description}
The process of modifying $\xi_k$ is motivated by the following lemma.

\begin{lemma} \label{lem:modified}
Let $X'\subseteq X$, $x\in X\backslash X'$, and $(c,t)\in \mathcal{C}\times \mathcal{T}$ be such that $\xi(\{x\})=\chi_{c,t}$. Suppose that
there exists a maximal distribution $\xi$ in $\Xi^*(X)$ with $\xi(X')\leq \xi$.
Then, the following implication holds: if there exists a maximal distribution $\xi^*$ in $\Xi^*(X)$ such that $\xi(X'\cup \{x\})\leq \xi^*$, then either (i) $\xi(X'\cup \{x\})\leq \xi$, or (ii) there exists $(c',t')\in \mathcal{C}\times \mathcal{T}$ such that $\xi+\chi_{c,t}-\chi_{c',t'}\in \Xi^*(X)$ and $\xi(X'\cup \{x\})\leq \xi+\chi_{c,t}-\chi_{c',t'} $.
\end{lemma}
\begin{proof}[Proof of Lemma \ref{lem:modified}]\renewcommand{\qedsymbol}{$\blacksquare$}
We consider two cases.

\noindent
\emph{Case 1:}
Suppose that $\xi^{t}_{c}(X')<\xi^{t}_{c}$. Then,
\begin{align*}
&\xi^{t}_{c}(X'\cup \{x\})=\xi^{t}_{c}(X')+1\leq \xi^{t}_{c}, \text{ and } \\
&\xi^{\tilde t}_{\tilde c}(X'\cup \{x\})=\xi^{\tilde t}_{\tilde c}(X')\leq \xi^{\tilde t}_{\tilde c} \text{ for all } (\tilde c, \tilde t)\in \mathcal{C}\times \mathcal{T} \text{ with } (\tilde c, \tilde t)\neq (c,t).
\end{align*}
Thus, (i) holds.

\noindent
\emph{Case 2:}
Suppose that $\xi^{t}_{c}(X')=\xi^{t}_{c}$. By the sufficient condition of the implication, there exists a maximal maximizer $\xi^*$ in $\Xi^{*}(X)$ with $\xi(X'\cup \{x\})\leq \xi^*$. Then,
\begin{align*}
\xi^{t}_{c}+1=\xi^{t}_{c}(X')+1=\xi^{t}_{c}(X'\cup \{x\})\leq (\xi^*)^{t}_{c},
\end{align*}
which implies $\xi^{t}_{c}<(\xi^*)^{t}_{c}$.
By Lemma \ref{lem:pareto} (M-convexity of the set of maximal distributions in $\Xi^*(X)$), there exists $(c',t') \in (\mathcal{C}\times \mathcal{T})$ with $\xi^{t'}_{c'}>(\xi^*)^{t'}_{c'}$ such that $\xi+\chi_{c,t}-\chi_{c',t'}$ is a maximal distribution in $\Xi^*(X)$.
It holds that
\begin{align*}
&(\xi+\chi_{c,t}-\chi_{c',t'})^{t'}_{c'}\geq (\xi^*)^{t'}_{c'} \geq \xi^{t'}_{c'}(X'\cup \{x\}), \\
&(\xi+\chi_{c,t}-\chi_{c',t'})^{t}_{c}=\xi^{t}_{c}+1=\xi^{t}_{c}(X')+1=\xi^{t}_{c}(X'\cup \{x\}), \\
&(\xi+\chi_{c,t}-\chi_{c',t'})^{\tilde t}_{\tilde c}=\xi^{\tilde t}_{\tilde c} \geq \xi^{\tilde t}_{\tilde c}(X')=\xi^{\tilde t}_{\tilde c}(X'\cup \{x\}) \\
&\hspace{30mm}\text{ for all } (\tilde c, \tilde t)\in \mathcal{C}\times \mathcal{T} \text{ with } (\tilde c, \tilde t) \neq (c,t) \text{ and } (\tilde c, \tilde t)\neq (c',t').
\end{align*}
Thus, (ii) holds.
\end{proof}

\begin{lemma} \label{lem:choice-equivalence}
The modified distribution-conscious choice rule and the (original) distribution-conscious choice rule produce the same outcome.
\end{lemma}
\begin{proof}[Proof of Lemma \ref{lem:choice-equivalence}]\renewcommand{\qedsymbol}{$\blacksquare$}
Let $X_k$ be defined as in the construction of the distribution-conscious choice rule and let $X'_k$ and $\xi_k$ be defined as in the construction of the modified distribution-conscious choice rule.
We show by induction that $X_k=X'_k$ and $\xi_k$ is a maximal distribution in $\Xi^*(X)$ for each index $k$ used in the definitions of both rules and terminate at the same index.
For $k=0$, we have $X_k=\emptyset=X'_k$ and, by the definition of the modified distribution-conscious choice rule, $\xi_0$ is a maximal distribution in $\Xi^*(X)$.
By mathematical induction hypothesis, suppose that $X_k=X'_k$ and $\xi_k$ is a maximal distribution. We now show the hypothesis for $k+1$.

\emph{Case 1:} Suppose that the distribution-conscious choice rule does not terminate when the index is $k$.
By the induction hypothesis and the definition of the modified distribution-conscious choice rule,
$\xi(X_k)=\xi(X'_k)\leq \xi_k$. By the induction hypothesis, $\xi_k$ is a maximal distribution in $\Xi^*(X)$. Let $x_{k+1}$ be such that $X_{k+1}=X_k\cup\{x_{k+1}\}$. By the definition of the distribution-conscious choice rule, $\xi(X_k\cup\{x_{k+1}\})\leq \xi^*$ for some $\xi^*\in \Xi^*(X)$; let us choose $\xi^*$ so that it is maximal. By Lemma \ref{lem:modified}, either (i) $\xi(X_k \cup \{x_{k+1}\})\leq \xi_k$, or (ii) there exists $(c',t')\in \mathcal{C}\times \mathcal{T}$ such that $\xi_k+\chi_{c,t}-\chi_{c',t'}\in \Xi^*(X)$ and $\xi(X_k\cup \{x_{k+1}\})\leq \xi_k+\chi_{c,t}-\chi_{c',t'}$, where $\chi_{c,t}=\xi(\{x_{k+1}\})$. It follows that $x_{k+1}$ satisfies one of the two conditions stated in Step 2 of the modified distribution-conscious choice rule. Suppose, for contradiction, that $X'_{k+1}\neq X'_k\cup \{x_{k+1}\}$. Then, by the deifnition of the modified distribution-conscious choice rule, there exists $x'\in X\backslash X'_k$ such that $x'$ has a higher merit than $x_{k+1}$ and $\xi(X'_k\cup\{x'\})\leq \xi^{**}$ for some $\xi^{**}\in \Xi^*(X)$. By the induction hypothesis, we have $X'_k=X_k$, which implies $x'\in X\backslash X_k$ and $\xi(X_k\cup\{x'\})\leq \xi^{**}$. Since $x'$ has a higher merit than $x_{k+1}$, we obtain a contradiction to the fact that $x_{k+1}$ is chosen when the index of the distribution-conscious choice rule is $k+1$. Therefore, $X'_{k+1}=X'_k\cup \{x_{k+1}\}=X_k\cup\{x_{k+1}\}=X_{k+1}$, where the second equality follows from the induction hypothesis. It remains to show that $\xi_{k+1}$ is a maximal distribution in $\Xi^*(X)$. By Lemma \ref{lem:pareto} (M-convexity of the maximal distributions in $\Xi^*(X)$) and Proposition 4.1 of \cite{Murota:SIAM:2003}, every maximal distribution in $\Xi^*(X)$ has the same sum of coordinates. Since $\xi_k$ is a maximal distribution (which follows from the induction hypothesis) and $\xi_k$ and $\xi_{k+1}$ have the same sum of coordinates (which follows from the definition of the modified distribution-conscious choice rule), $\xi_{k+1}$ is a maximal distribution.

\emph{Case 2:} Suppose that the distribution-conscious choice rule terminates when the index is $k$. Then, there does not exist $x\in X\backslash X_k$ and $\xi\in \Xi^*(X)$ such that $\xi(X_k\cup\{x\})\leq \xi$. Then, for each $x\in X\backslash X_k=X\backslash X'_k$ (where the equality follows from the induction hypothesis), neither (i) nor (ii) in Step 2 of the modified distribution-conscious choice rule holds true. Theorefore, the modified distribution-conscious choice rule termines when the index is $k$.

\end{proof}

\paragraph{Step 4: }
We derive the time complexity of the distribution-conscious choice rule. The first step for calculating the choice rule is to find one maximal distribution in $\Xi^*(X)$. By Lemma \ref{lem:outcome-maximal}, we can use the domain-reduction algorithm. We assume that $f$ can be evaluated in a constant time in what follows.
Step 2 of the algorithm takes $O(|\mathcal{C}\times \mathcal{T}|)$ time.
Let $\xi_{k^*}$ denote the outcome of the algorithm. Since the algorithm starts from $\mathbf{0}$ and adds $1$ to some coordinate toward $\xi_{k^*}$ at every round, the number of iterations is $||\xi_{k^*}||$, which is bounded by $||\xi(X)||$ because $\xi_{k^*}\leq \xi(X)$. Since
\begin{align*}
||\xi(X)||=\sum_{(c,t)}\xi_c^t(X)=\sum_{(c,t)}\sum_{x\in X}\xi_c^t(\{x\})=\sum_{x\in X}\sum_{(c,t)}\xi_c^t(\{x\})=|X|,
\end{align*}
the number of iterations is bounded by $O(|X|)$. Thus, finding an outcome of the algorithm takes $O(|\mathcal{C}|\times |\mathcal{T}|\times |X|)$ time.

Given a maximal distribution in $\Xi^*(X)$, we can run the distribution-conscious choice rule. By Lemma \ref{lem:choice-equivalence}, it suffices to examine the computational time of the modified rule. Step 2 of the rule takes $O(|C|\times |T|\times |X|)$ time.
The number of iterations is equal to $|X|$. Hence, the modified distribution-conscious choice rule finds an outcome in $O(|\mathcal{C}|\times |\mathcal{T}|\times |X|^2)$ time.
Together with the time complexity of executing the domain-reduction algorith, we conclude that finding an outcome of the distribution-conscious choice rule takes $O(|\mathcal{C}|\times |\mathcal{T}|\times |X|^2)$ time.
\qed


\subsection*{Proof of Theorem \ref{thm:pi}}

We need the following properties of choice rules in our proofs.
A choice rule $C$ satisfies the \textbf{irrelevance of rejected contracts} condition, if, for each $X \subseteq \calx$
and $x \in \calx \setminus X$, $x\notin C(X\cup \{x\}) \; \implies \;  C(X\cup \{x\})=C(X)$ \citep{aygson12a}.
A choice rule $C$ satisfies the \textbf{substitutes} condition, if, for each $X \subseteq \calx$
and $x \in \calx \setminus X$, $C(X) \supseteq  C(X\cup \{x\}) \cap X$ \citep{kelso82,roth84}. 

\begin{lemma}[\cite{aizmal81}]\label{lem:pi}
A choice rule $C$ is path independent if, and only if, it satisfies the irrelevance of rejected contracts condition
and the substitutes condition.
\end{lemma}

By this lemma path independence is equivalent to the conjunction of the
irrelevance of rejected contracts condition (IRC) and the substitutes condition,
so we show these two properties to prove path independence.

\noindent
\emph{Proof of IRC:}
Let $X \subseteq \calx$ and $x \in \calx \setminus X$ such that $x\notin C^d(X\cup \{x\})$. We need to show $C^d(X\cup \{x\})=C^d(X)$.

Let $c=\gamma(x)$, $t=\tau(\sigma(x))$, $\xi_1= \xi(C^d(X))$, and $\xi_2=\xi(C^d(X\cup\{x\}))$.

Since  $x \notin C^d(X\cup \{x\})$, $\xi_2 \leq \xi(X)$. Together with Theorem \ref{thm:diversitychoice} (i), we get $f(\xi_1)=f(\xi_2)$.
Furthermore, $C^d(X\cup\{x\})$ is in $\mathcal{F}(X)$ and
$\mathcal{F}(X\cup \{x\})$. Likewise, $C^d(X)$ is in $\mathcal{F}(X\cup\{x\})$ because
$(\mathcal{X},\mathcal{F}(\mathcal{X}))$ is a matroid (Lemma \ref{lem:matroid}).
Therefore, $C^d(X),C^d(X\cup\{x\}) \in \mathcal{F}(X) \cap \mathcal{F}(X\cup \{x\})$.
By Theorem \ref{thm:diversitychoice} (ii), $C^d(X)$ merit
dominates $C^d(X\cup\{x\})$ and $C^d(X\cup\{x\})$ merit dominates $C^d(X)$.
Therefore, $C^d(X)=C^d(X\cup\{x\})$, which follows from the \emph{antisymmetry} of merit domination that if two sets merit dominate each other they have
to be the same. The antisymmetry of merit domination is straightforward
because if two sets merit dominate each other,
then they have the same number of contracts and, furthermore, because different
contracts have distinct merit rankings, they need to have the
same set of contracts.

To finish the proof, we show that $C^d$ satisfies the substitutes condition.

\noindent
\emph{Proof of Substitutability:}
Let $X \subseteq \calx$ and $x \in \calx \setminus X$. We need
to show $C^d(X)\supseteq C^d(X \cup \{x\}) \cap X$.

Let $c=\gamma(x)$, $t=\tau(\sigma(x))$, $\xi_1= \xi(C^d(X))$, and $\xi_2=\xi(C^d(X\cup\{x\}))$.

If $x\notin C^d(X \cup \{x\})$, then by the irrelevance of rejected contracts condition
we have $C^d(X)=C^d(X\cup\{x\})$. Therefore, $C^d(X)\supseteq C^d(X \cup \{x\}) \cap X= C^d(X)$.

For the rest of the proof suppose that $x\in C^d(X \cup \{x\})$.
We consider several cases depending on the value of $\xi_2$.

\emph{Case 1:} Consider the case $\xi_2 \leq \xi(X)$. Then $f(\xi_1)=f(\xi_2)$.
By construction of $C^d$, $\xi_1$ is maximal in $\Xi^*(X)$.
Likewise, $\xi_2$ is maximal in
$\Xi^*(X\cup\{x\})$. Since $\xi_2 \leq \xi(X)$, we get that $\xi_2$ is also
maximal in $\Xi^*(X)$. By Lemma \ref{lem:pareto}, $\xi_1$ and $\xi_2$ belong
to an M-convex set, so $\norm{\xi_2} = \norm{\xi_1}$.\footnote{Members of an M-convex set
have the same sum of coordinates, see Proposition 4.1 in \citet{Murota:SIAM:2003}.} Therefore,
\[\abs{C^d(X\cup\{x\})\setminus C^d(X)}=\abs{C^d(X)\setminus C^d(X\cup\{x\})}.\]
Since $x\in C^d(X\cup\{x\})\setminus C^d(X)$, we have $|C^d(X\cup\{x\})\setminus C^d(X)|\geq 1$. We show that
$|C^d(X\cup\{x\})\setminus C^d(X)|=1$.

Suppose, for contradiction, that $\abs{C^d(X\cup\{x\})\setminus C^d(X)}\geq 2$. Then,
there exists $x_1 \in X\setminus \{x\}$ such that $x_1 \in C^d(X\cup\{x\})\setminus C^d(X)$. Since
$f(\xi_1)=f(\xi_2)$, $C^d(X\cup\{x\})$ and $C^d(X)$ are bases in $\mathcal{F}(X\cup \{x\})$.
By the stronger version of \emph{B2}, which is stated on page \pageref{strongerB2}, there exists $x_2 \in C^d(X)\setminus C^d(X\cup\{x\})$ such
that $(C^d(X\cup\{x\}) \setminus \{x_1\}) \cup \{x_2\}$ and $(C^d(X) \setminus \{x_2\}) \cup \{x_1\}$
are also bases in $\mathcal{F}(X\cup \{x\})$.
Theorem \ref{thm:diversitychoice} implies that $C^d(X\cup\{x\})$ merit dominates $(C^d(X\cup\{x\})\setminus \{x_1\}) \cup \{x_2\}$, so
$x_1 \mathrel{\succ} x_2$. Furthermore, since
$(C^d(X) \setminus \{x_2\}) \cup \{x_1\}$ is a base in
$\mathcal{F}(X\cup \{x\})$ it must also be a base in $\mathcal{F}(X)$. By Theorem \ref{thm:diversitychoice},
$C^d(X)$ merit dominates $(C^d(X) \setminus \{x_2\}) \cup \{x_1\}$, therefore,
$x_2 \mathrel{\succ} x_1$, which is a
contradiction to $x_1 \mathrel{\succ} x_2$. Therefore, $|C^d(X\cup\{x\})\setminus C^d(X)|=1$ and
$C^d(X \cup\{x\})= (C^d(X) \cup \{x\})\setminus \{y\}$ for some $y\in C^d(X)$. As a result,
$C^d(X)\supseteq C^d(X \cup \{x\}) \cap X = C^d(X) \setminus \{y\}$ for some $y\in C^d(X)$.
This finishes the proof of Case 1.

\emph{Case 2:} Consider the case $\xi_2 \not \leq \xi(X)$. Since $C^d(X\cup \{x\})\subseteq X\cup \{x\}$,
it must be that $(\xi_2)^t_c> \xi^t_c(X)$, so $C^d(X\cup \{x\})$ includes $x$ and all
contracts in $X$ with type-$t$ students and school $c$. Furthermore, $(\xi_2)^t_c=\xi^t_c(X)+1$ and
$(\xi_1)^t_c\leq \xi^t_c(X)$.

\begin{claim}\label{claim:types}
For each school $c'\in \mathcal{C}$ and type $t'\in \mathcal{T}$ such that $(c',t') \neq (c,t)$, we have $(\xi_1)^{t'}_{c'}\geq (\xi_2)^{t'}_{c'}$.
\end{claim}

\begin{proof}[Proof of Claim \ref{claim:types}]\renewcommand{\qedsymbol}{$\blacksquare$}
Suppose, for contradiction, that there exist school $c'\in \calc$ and type $t'\in \calt$ with $(c',t') \neq (c,t)$
such that $(\xi_1)^{t'}_{c'} < (\xi_2)^{t'}_{c'}$. Then, by ordinal concavity,
either (i)
\begin{enumerate}
\item $f(\xi_2-\chi_{c',t'})>f(\xi_2)$ or
\item $f(\xi_1+\chi_{c',t'})>f(\xi_1)$ or
\item $f(\xi_2-\chi_{c',t'})=f(\xi_2)$ and $f(\xi_1+\chi_{c',t'})=f(\xi_1)$
\end{enumerate}
or (ii) there exist school $\hat c\in \calc$ and type $\hat t\in \calt$  with
$(\xi_2)_{\hat c}^{\hat t}<(\xi_1)_{\hat c}^{\hat t}$ such that
\begin{enumerate}
\item $f(\xi_2-\chi_{c',t'}+\chi_{\hat c, \hat t})>f(\xi_2)$ or
\item $f(\xi_1+\chi_{c',t'}-\chi_{\hat c, \hat t})>f(\xi_1)$ or
\item $f(\xi_2-\chi_{c',t'}+\chi_{\hat c, \hat t})=f(\xi_2)$ and $f(\xi_1+\chi_{c',t'}-\chi_{\hat c, \hat t}) = f(\xi_1)$.
\end{enumerate}
Condition (i) cannot hold because under (i)(1) $\xi_2-\chi_{c',t'}\leq \Xi(X\cup \{x\})$ and
$f(\xi_2-\chi_{c',t'})>f(\xi_2)$ give us a contradiction to the result that the outcome of $C^d$
maximizes the distributional objective among feasible subsets of $X\cup \{x\}$ (Theorem \ref{thm:diversitychoice}),
because a contract in $(X\cup \{x\}) \setminus C^d(X\cup \{x\})$ with type-$t'$ student and
school $c'$ can be added to $C^d(X\cup \{x\})$ and increase the value of $f$.
Under (i)(2) $\xi_1+\chi_{c',t'} \leq \xi(X)$ and $f(\xi_1+\chi_{c',t'}) > f(\xi_1)$ give us a contradiction to the result that the outcome of $C^d$ maximizes the distributional objective among subsets of $X$ (Theorem \ref{thm:diversitychoice}), because a
contract in $X\setminus C^d(X)$ with type-$t'$ student and
school $c'$ can be added to $C^d(X)$ and increase the value of $f$. Under (i)(3),
$\xi_1+\chi_{c',t'} \leq \xi(X)$ and $f(\xi_1+\chi_{c',t'})=f(\xi_1)$ give us a contradiction to the result that the outcome of $C^d$ merit dominates any
feasible subset of $X$ that maximizes the distributional objective
(Theorem \ref{thm:diversitychoice}), because a contract in $X\setminus C^d(X)$ with type-$t'$ student and
school $c'$ can be added to $C^d(X)$ without changing the value of $f$.

Likewise condition (ii) cannot hold because under (ii)(1) $\xi_2-\chi_{c',t'}+\chi_{\hat c, \hat t} \leq \xi(X\cup \{x\})$ and $f(\xi_2-\chi_{c',t'}+\chi_{\hat c, \hat t}) > f(\xi_2)$ give us a contradiction to the result that the outcome of $C^d$ maximizes the distributional objective among feasible subsets of $X\cup \{x\}$
(Theorem \ref{thm:diversitychoice}), because a contract in $(X\cup\{x\}) \setminus C^d(X\cup\{x\})$ with
type-$\hat t$ student and school $\hat c$ can be added to $C^d(X\cup\{x\})$ and a contract from
$C^d(X\cup\{x\})$ with type-$t'$ student and school $c'$ can be removed from $C^d(X\cup\{x\})$ to increase the value of $f$.
Under (ii)(2) $\xi_1+\chi_{c',t'}-\chi_{\hat c, \hat t} \leq \xi(X)$ and $f(\xi_1+\chi_{c',t'}-\chi_{\hat c, \hat t}) > f(\xi_1)$ give us a contradiction to the result that the outcome of $C^d$ maximizes the distributional objective among feasible subsets of $X$
(Theorem \ref{thm:diversitychoice}), because a contract in $X\setminus C^d(X)$ with type-$t'$ student and
school $c'$ can be added to $C^d(X)$ and a contract from $C^d(X)$ with type-$\hat t$ student and
school $\hat c$ can be removed from $C^d(X)$ to increase the value of $f$.
Under (ii)(3), $f(\xi_2-\chi_{c',t'}+\chi_{\hat c, \hat t}) = f(\xi_2)$ and $\xi_2-\chi_{c',t'}+\chi_{\hat c, \hat t} \leq \xi(X\cup\{x\})$ imply that the lowest merit ranked type-$t'$ student with a contract at school $c'$ in $C^d(X\cup \{x\})\setminus C^d(X)$ has a higher merit ranking than the lowest merit ranked type-$\hat t$ student with a contract at school $\hat c$ in
$C^d(X) \setminus C^d(X\cup \{x\})$. Similarly, $\xi_1+\chi_{c',t'}-\chi_{\hat c, \hat t}\leq \xi(X)$ and $f(\xi_1+\chi_{c',t'}-\chi_{\hat c, \hat t}) = f(\xi_1)$
imply that the lowest merit ranked type-$\hat t$ student with a contract
at school $\hat{c}$ in $C^d(X)\setminus C^d(X\cup \{x\})$ has a higher
merit ranking than the lowest merit type-$t'$ student with a contract at school $c'$ in $C^d(X\cup \{x\}) \setminus C^d(X)$, which is a contradiction since the merit
ranking is strict and $(\hat c, \hat t)\neq (c', t')$.
\end{proof}

\begin{claim}\label{claim:comparetc}
$(\xi_1)^t_c=(\xi_2)^t_c-1$.
\end{claim}

\begin{proof}[Proof of Claim \ref{claim:comparetc}]\renewcommand{\qedsymbol}{$\blacksquare$}
Suppose, for contradiction, that $(\xi_1)^t_c\neq (\xi_2)^t_c-1$. Since $(\xi_1)^t_c\leq \xi^t_c(X)$ and
$(\xi_2)^t_c = \xi^t_c(X)+1$, we get $(\xi_1)^t_c < \xi^t_c(X)=(\xi_2)^t_c-1$.

By ordinal concavity,
either (i)
\begin{enumerate}
\item $f(\xi_2-\chi_{c,t})>f(\xi_2)$, or
\item $f(\xi_1+\chi_{c,t})>f(\xi_1)$, or
\item $f(\xi_2-\chi_{c,t})=f(\xi_2)$ and $f(\xi_1+\chi_{c,t}) = f(\xi_1)$
\end{enumerate}
or (ii) there exist school $c'\in \calc$ and type $t'\in \calt$ with $(\xi_2)_{c'}^{t'}<(\xi_1)_{c'}^{t'}$ such that
\begin{enumerate}
\item $f(\xi_2-\chi_{c,t}+\chi_{c',t'})>f(\xi_2)$
\item $f(\xi_1+\chi_{c,t}-\chi_{c',t'})>f(\xi_1)$ or
\item $f(\xi_2-\chi_{c,t}+\chi_{c',t'})=f(\xi_2)$ and $f(\xi_1+\chi_{c,t}-\chi_{c',t'}) = f(\xi_1)$.
\end{enumerate}
Condition (i) cannot hold because under (i)(1) $\xi_2-\chi_{c,t} \leq \xi(X\cup \{x\})$
and $f(\xi_2-\chi_{c,t})>f(\xi_2)$ give us a contradiction to the result that the outcome of $C^d$
maximizes the distributional objective among feasible subsets of $X\cup \{x\}$ (Theorem \ref{thm:diversitychoice}),
because a contract in $(X\cup \{x\}) \setminus C^d(X\cup \{x\})$ with type-$t$ student and
school $c$ can be added to $C^d(X\cup \{x\})$ and increase the value of $f$.
Similarly, under (i)(2) $\xi_1+\chi_{c,t} \leq \xi(X)$ and $f(\xi_1+\chi_{c,t})>f(\xi_1)$
give us a contradiction to the result that the outcome of $C^d$ maximizes the distributional objective among feasible subsets of $X$
(Theorem \ref{thm:diversitychoice}), because a contract in $X\setminus C^d(X)$ with type-$t$ student and
school $c$ can be added to $C^d(X)$ to increase the value of $f$. Under (i)(3) $\xi_1+\chi_{c,t} \leq \xi(X)$ and
$f(\xi_1+\chi_{c,t}) = f(\xi_1)$ give us a contradiction to the result that the outcome of $C^d$ merit dominates each feasible
subset of $X$ that maximizes the distributional objective (Theorem \ref{thm:diversitychoice}), because a contract in
$X\setminus C^d(X)$ with type-$t$ student and school $c$ can be added to $C^d(X)$ without changing the value of $f$.
Therefore, condition (ii) must hold.

Under condition (ii)(1) $\xi_2-\chi_{c,t}+\chi_{c',t'} \leq \xi(X\cup \{x\})$
and $f(\xi_2-\chi_{c,t}+\chi_{c',t'})>f(\xi_2)$ give a contradiction to the result that the
outcome of $C^d$ maximizes the distributional objective among feasible subsets of $X\cup \{x\}$
(Theorem \ref{thm:diversitychoice}), because a contract in $(X\cup\{x\}) \setminus C^d(X\cup\{x\})$ with
type-$t$ student and school $c$ can be added to $C^d(X\cup\{x\})$ and a contract from
$C^d(X\cup\{x\})$ with type-$t'$ student and school $c'$ can be removed from $C^d(X\cup\{x\})$ to increase the value of $f$.
Likewise, under (ii)(2) $\xi_1+\chi_{c,t}-\chi_{c',t'}\leq \xi(X)$ and $f(\xi_1+\chi_{c,t}-\chi_{c',t'})>f(\xi_1)$
give us a contradiction to the result that the outcome of $C^d$ maximizes the distributional objective among feasible subsets of $X$
(Theorem \ref{thm:diversitychoice}), because a contract in $X\setminus C^d(X)$ with type-$t$ student and
school $c$ can be added to $C^d(X)$ and a contract in $C^d(X)$ with type-$t'$ student and
school $c'$ can be removed to increase the value of $f$. Under (ii)(3)
$f(\xi_2-\chi_{c,t}+\chi_{c', t'}) = f(\xi_2)$ and $\xi_2-\chi_{c,t}+\chi_{c', t'} \leq \xi(X\cup\{x\})$ imply that the lowest merit ranked type-$t$ student with a contract
at school $c$ in $C^d(X\cup \{x\})\setminus C^d(X)$ has a higher merit ranking than the lowest merit ranked type-$t'$ student with a contract at school $c'$ in
$C^d(X) \setminus C^d(X\cup \{x\})$. Similarly, $f(\xi_1+\chi_{c,t}-\chi_{c',t'}) = f(\xi_1)$
and $\xi_1+\chi_{c,t}-\chi_{c',t'}\leq \xi(X)$ imply that the lowest merit ranked type-$t'$ student with a contract at school $c'$ in $C^d(X)\setminus C^d(X\cup \{x\})$ has a higher merit ranking than
the lowest merit ranked type-$t'$ student with a contract at school $c'$ in $C^d(X\cup \{x\}) \setminus C^d(X)$, which is a contradiction since the merit ranking is strict and $(c,t)\neq (c', t')$.

Both conditions cannot hold. Therefore, $(\xi_1)^t_c=(\xi_2)^t_c-1$.
\end{proof}

To finish the proof of Case 2, we combine the results that we have established so far:
$(\xi_2)^c_t=\xi^t_c(X\cup \{x\})=\xi^t_c(X)+1$,
$(\xi_1)^c_t=\xi^t_c(X)$, and, for each type $t'\in \calt$ and school $c'\in \calc$
with $(t',c')\neq (t,c)$, $(\xi_1)^{c'}_{t'} \geq (\xi_2)^{c'}_{t'}$.
For a fixed type $t'\in \calt$ and school $c'\in \calc$ and the number of contracts of type-$t'$ students with school $c'$, choice rule $C^d$ chooses contracts with the highest merit ranking.
Therefore, $C^d(X) \supseteq C^d(X \cup \{x\}) \cap X$, which finishes the proof of Case 2. Therefore, $C^d$ satisfies the substitutes condition.
\qed


\subsection*{Proof of Theorem \ref{prop:structure}}
\begin{subequations}
Suppose, for contradiction, that $C^d$ violates the law of aggregate demand, i.e., there exist $X, X'\subseteq \mathcal{X}$ such that
\begin{align}
&X\subseteq X', \text{ and } \label{sr-1} \\
&|C^d(X)|>|C^d(X')|. \label{sr-2}
\end{align}
By (\ref{sr-2}),
\begin{align*}
||\xi(C^d(X))||>||\xi(C^d(X'))||.
\end{align*}
By size-restricted concavity, there exists $(c,t)\in \mathcal{C}\times \mathcal{T}$ such that
\begin{align}
\xi_c^t(C^d(X))>\xi_c^t(C^d(X')),
\label{sr-3}
\end{align}
and one of the following conditions holds:
\begin{enumerate}
\item[(i)] $f(\xi(C^d(X))-\chi_{c,t})>f(\xi(C^d(X)))$, or
\item[(ii)] $f(\xi(C^d(X'))+\chi_{c,t})>f(\xi(C^d(X')))$, or
\item[(iii)] $f(\xi(C^d(X))-\chi_{c,t})=f(\xi(C^d(X)))$ and $f(\xi(C^d(X'))+\chi_{c,t})=f(\xi(C^d(X')))$.
\end{enumerate}
If (i) holds, then we obtain a contradiction to $\xi(C^d(X))$ maximizing $f$ among all distributions $\xi$ with $\xi\leq \xi(X)$ (Theorem 1 (i)).
Hence, (ii) or (iii) holds. In either case, we get
\begin{align}
f(\xi(C^d(X'))+\chi_{c,t})\geq f(\xi(C^d(X'))).
\label{sr-4}
\end{align}
By (\ref{sr-3}), there exists $x\in C^d(X)\backslash C^d(X')$ such that
\begin{align}
&\xi(\{x\})=\chi_{c,t}. \label{sr-6}
\end{align}
Since $x\in C^d(X)$ and $C^d(X)\subseteq X \subseteq X'$ (where the latter set-inclusion follows from \ref{sr-1}), we get
\begin{align}
x\in X'.
\label{sr-7}
\end{align}
By (\ref{sr-7}) and $\xi(C^d(X'))\leq \xi(X')$ (which follows from the definition of choice rules), we get
\begin{align}
\xi(C^d(X')\cup\{x\})\leq \xi(X').
\label{sr-9}
\end{align}
By (\ref{sr-4}) and (\ref{sr-6}),
\begin{align*}
f(\xi(C^d(X')\cup \{x\}))\geq f(\xi(C^d(X'))).
\end{align*}
Together with (\ref{sr-9}) and the fact that $\xi(C^d(X'))$ maximizes $f$ among all distributions $\xi$ with $\xi \leq \xi(X')$ (which is Theorem 1(i)),
it implies
\begin{align*}
\xi(C^d(X')\cup \{x\})\in \Xi^*(X').
\end{align*}
Since $x\notin C^d(X')$, we obtain a contradiction to the fact that $C^d(X')$ is the outcome of the distribution-conscious choice rule with input $X'$ (recall Step 2 of the rule). 
\qed
\end{subequations}
\smallskip

Note that size-restricted concavity is only used to derive (\ref{sr-4}). This observation implies that the proof remains valid under
the alternative assumption that $f$ is {\it monotone}.
\begin{definition}
The distributional objective $f: \Xi^0 \rightarrow \mathbb{R}_+$ is \textbf{monotone} if $f(\xi) \geq f(\tilde{\xi})$ for each $\xi,\tilde{\xi} \in \Xi^0$ with $\xi\geq \tilde{\xi}$.
\end{definition}
Therefore, if $f$ is ordinally concave and monotone, then the distribution-conscious choice rule satisfies path-independence and the law of aggregate demand.
\bigskip

\subsection*{Proof of Theorem \ref{thm:trace}}
The following result follows from Theorem \ref{thm:diversitychoice}.

\begin{lemma}\label{lem:lambdachoice}
Suppose that $\lambda\in \mathbb{R}_+$ is such that the distributional objective $f_{\lambda}$ is \oconcave{}. Then, for each set of contracts $X\subseteq \calx$,
\begin{enumerate}[(i)]
\item $\min\{f(\xi(C^d_{\lambda}(X))),\lambda\}=\min\{f(\xi(C^d(X))),\lambda\}$ and
\item $C^d_{\lambda}(X)$ merit dominates each $Y\subseteq X$ such that
\[ \min\{f(\xi(C^d_{\lambda}(Y))),\lambda\}=\min\{f(\xi(C^d(X))),\lambda\}.\]
\end{enumerate}
\end{lemma}

Now fix a set of contracts $X\subseteq \mathcal{X}$ and denote the outcome of
the trace algorithm as $C^{tr}(X)$. Using Lemma \ref{lem:lambdachoice}, first,
we show that $C^{tr}(X)\subseteq \mathcal{P}(X)$, and, then,
$\mathcal{P}(X) \subseteq C^{tr}(X)$ to finish the proof.

\begin{claim}\label{claim:pareto}
$C^{tr}(X)\subseteq \mathcal{P}(X)$.
\end{claim}

\begin{proof}[Proof of Claim \ref{claim:pareto}]\renewcommand{\qedsymbol}{$\blacksquare$}
Let $Y\in C^{tr}(X)$. Suppose, for contradiction, that $Y\notin \mathcal{P}(X)$, and, hence, there exists
$Z\subseteq X$ such that $Z\neq Y$, $Z$ merit dominates $Y$, and
$f(\xi(Z))\geq f(\xi(Y))$.

Suppose that $Z$ is chosen at index $k\in \mathbb{N}$ in the
construction of $C^{tr}(X)$. Therefore,
$Y=C^d_{\lambda_k}(X)$. Then, by Lemma \ref{lem:lambdachoice},
$Y=C^d_{\lambda_k}(X)$ merit dominates each subset of $X$ that attains the level of distributional objective of $f(\xi(C^d_{\lambda_k}(X)))=f(\xi(Y))$. Therefore, since
$f(\xi(Z)) \geq f(\xi(Y))$ and $Z\subseteq X$, we get $Y$ merit dominates $Z$.
As noted in the proof of Theorem \ref{thm:pi}, the merit domination is
antisymmetric, which is a contradiction because we have
$Y$ merit dominates $Z$, $Z$ merit dominates $Y$, and $Y\neq Z$.
Therefore, $Y\in \mathcal{P}(X)$. Since $Y$ is any set in $C^{tr}(X)$,
we conclude $C^{tr}(X) \subseteq \mathcal{P}(X)$.
\end{proof}

\begin{claim}\label{claim:trace}
$\mathcal{P}(X) \subseteq C^{tr}(X)$.
\end{claim}

\begin{proof}[Proof of Claim \ref{claim:trace}]\renewcommand{\qedsymbol}{$\blacksquare$}
Let $Y\in \mathcal{P}(X)$. Suppose, for contradiction, that $Y\notin C^{tr}(X)$.
Since $C^d(X)\in C^{tr}(X)$ and $C^{tr}(X)\subseteq \mathcal{P}(X)$, we get that
$C^d(X)\in \mathcal{P}(X)$. Since $Y\notin C^{tr}(X)$, we have $Y\neq C^d(X)$.
By Theorem \ref{thm:diversitychoice}, $f(\xi(C^d(X))\geq f(\xi(Y))$
and $C^d(X)$ merit dominates any subset of $X$ with level $f(\xi(C^d(X))$.
Therefore, since $Y \in \mathcal{P}(X)$, we cannot have $f(\xi(C^d(X)) = f(\xi(Y))$,
which implies $f(\xi(C^d(X))) > f(\xi(Y))$.

Since $\lambda_0=0$ and $f(\xi(C^d(X))) > f(\xi(Y))$, there exists an
index $k$ such that $f(\xi(C^d_{\lambda_k}(X))) > f(\xi(Y)) \geq \lambda_k$ where
$\lambda_k$ is defined as in the construction of $C^{tr}(X)$. By Lemma
\ref{lem:lambdachoice}, and because $\min\{f(\xi(C^d_{\lambda}(Y))),\lambda_k\}=\lambda_k=\min\{f(\xi(C^d(X))),\lambda_k\}$,
$C^d_{\lambda_k}(X)$ merit dominates $Y$. This is a contradiction because
$f(\xi(C^d_{\lambda_k}(X))) > f(\xi(Y))$, $C^d_{\lambda_k}(X)$ merit dominates $Y$,
and $Y\in \mathcal{P}(X)$. Hence, we get that $Y\in C^{tr}(X)$. Since $Y$
is an arbitrary set in $\mathcal{P}(X)$, we conclude that $\mathcal{P}(X) \subseteq C^{tr}(X)$.
\end{proof}
Claims \ref{claim:pareto} and \ref{claim:trace} imply that $\mathcal{P}(X) = C^{tr}(X)$.
\qed

\section{Proofs of Auxiliary Results, Definitions, Examples}\label{app:auxiliary}
In this appendix, we present a new definition of concavity, proofs of our auxiliary results, and omitted examples.

\begin{center}
\textbf{A New Concavity Notion}
\end{center}

\subsection*{Semi-strict Pseudo M$^\natural$-concavity}\label{app:semistrict}
We provide a new definition of concavity, which implies that for each
$\lambda \geq 0$, $f_{\lambda}$ is ordinally concave. Furthermore, this notion of concavity has a
clear interpretation.

\begin{definition}\label{def:strict-pseudo}
The distributional objective $f: \Xi^0 \rightarrow \mathbb{R}_+$ is \textbf{semistrictly pseudo M$^\natural$-concave} if, for each
$\xi, \tilde{\xi} \in \Xi^0$ and $(c, t) \in \mathcal{C} \times \mathcal{T}$ with $\xi_c^t>\tilde{\xi}_{c}^t$, then
there exists $\left(c^{\prime}, t^{\prime}\right) \in (\mathcal{C} \times \mathcal{T})\cup\{\emptyset\}$ (with $\xi_{c^{\prime}}^{t^{\prime}}<\tilde{\xi}_{c^{\prime}}^{t^{\prime}}$ whenever $(c',t')\neq \emptyset$) such that
$$
\min\{f(\xi),f(\tilde{\xi})\}\leq\min\{f(\xi-\chi_{c,t}+\chi_{c',t'}), f(\tilde{\xi}+\chi_{c,t}-\chi_{c',t'})\},
$$
with strict inequality holding whenever $f(\xi)\neq f(\tilde \xi)$ and $\xi-\chi_{c,t}+\chi_{c',t'}\neq \tilde \xi$.
\end{definition}
The difference from pseudo M$^\natural$-concavity is that the increase in the minimum value must be strict if the two function values are different and the two distributions do not coincide with each other as a result of moving toward each other.\footnote{Note that $\xi-\chi_{c,t}+\chi_{c',t'}\neq \tilde \xi$ is equivalent to $\{\xi, \tilde \xi\}\cap \{\xi-\chi_{c,t}+\chi_{c',t'}, \tilde \xi+\chi_{c,t}-\chi_{c',t'}\}=\emptyset$.} One can verify that semistrict pseudo M$^\natural$-concavity implies pseudo M$^\natural$-concavity$^+$.\footnote{The converse of this implication does not hold. Let $\mathcal{C}=\{c\}$ and $\mathcal{T}=\{t,t'\}$; we identify $\mathbb{Z}^{|\mathcal{C}|\times |\mathcal{T}|}_+$ with $\mathbb{Z}^2_+$. Let $f:\Xi^0\rightarrow \mathbb{R}_+$ be such that
$$
\Xi^0=\{(0,0),(0,1),(1,0),(1,1)\}\subseteq \mathbb{Z}^2_+, \: f(0,0)=f(1,0)=0, \:  f(0,1)=f(1,1)=1.
$$
This function satisfies pseudo M$^\natural$-concavity$^+$ but violates semistrict pseudo M$^\natural$-concavity. For $\xi=(1,1)$, $\tilde \xi=(0,0)$, and $(c,t)$ with $\chi_{c,t}=(1,0)$,
$$
f(\xi)=f(\xi-\chi_{c,t})=1, \: f(\tilde \xi)=f(\tilde \xi+\chi_{c,t})=0,
$$
showing that the minimum function value does not strictly increase although $f(\xi)\neq f(\tilde \xi)$ and $\xi-\chi_{c,t}\neq \tilde \xi$.
}

Semistrict pseudo M$^\natural$-concavity can be viewed as a variant of {\it quasi concavity},
which has been studied extensively in microeconomic theory.\footnote{In a model with a continuum of commodities, if a preference relation over the commodity space is convex, then any utility function representing the preference relation is quasi concave; see Section 3.C of \cite{masco95}.}
We say that a continuous function $f: \mathbb{R}^{|\mathcal{C}|\times |\mathcal{T}|}\rightarrow \mathbb{R}$ is {\it semistrictly quasi concave},\footnote{Precisely speaking, semistrict quasi concavity is defined for a possibly discontinuous function as follows: for each $\xi, \tilde \xi\in \mathbb{R}^{|\mathcal{C}|\times |\mathcal{T}|}$ and $\lambda \in (0,1)$,
$\min\{f(\xi), f(\tilde \xi)\}<f(\lambda \xi+(1-\lambda) \tilde \xi)$ whenever $f(\xi)\neq f(\tilde \xi)$. If $f$ is continuous, this condition is equivalent to the one in the main text.} if for each $\xi, \tilde \xi\in \mathbb{R}^{|\mathcal{C}|\times |\mathcal{T}|}$ and $\lambda \in (0,1)$,
\begin{align*}
\min\{f(\xi), f(\tilde \xi)\}\leq f(\lambda \xi+(1-\lambda) \tilde \xi),
\end{align*}
with strict inequality holding whenever $f(\xi)\neq f(\tilde \xi)$.
Both semistrict pseudo M$^\natural$-concavity and semistrict quasi concavity state that the minimum function value increases, with the increase being strict whenever the original function values are different.\footnote{There is a subtle difference between continuous and discrete domains. For each $\xi, \tilde \xi\in \mathbb{R}^{|\mathcal{C}|\times |\mathcal{T}|}$ with $\xi\neq \tilde \xi$, it always holds that $\lambda \xi+(1-\lambda) \tilde \xi\neq \tilde \xi$ if $\lambda \in (0,1)$. In a discrete domain, however, it is possible that $\xi-\chi_{c,t}+\chi_{c',t'}=\tilde \xi$. Hence, we add a condition that these two distributions are distinct in the definition of semistrict pseudo M$^\natural$-concavity.}
\bigskip

\begin{center}
\textbf{Proofs of the Propositions}
\end{center}

\bigskip

\begin{subequations}
\subsection*{Proof of Proposition \ref{prop:truncation-necessary}}
This proposition follows from Proposition \ref{prop:ordinal-equivalence} because pseudo M$^\natural$-concavity is weaker than pseudo M$^\natural$-concavity$^+$. \qed

\subsection*{Proof of Proposition \ref{prop:ordinal-equivalence}}
\medskip
\noindent
\emph{The ``only if'' direction:} Let $\xi, \tilde{\xi} \in \Xi^0$ and $(c, t) \in \mathcal{C} \times \mathcal{T}$ with $\xi_c^t>\tilde{\xi}_{c}^t$.
Our goal is to prove that there exists $\left(c^{\prime}, t^{\prime}\right) \in (\mathcal{C} \times \mathcal{T})\cup\{\emptyset\}$ (with $\xi_{c^{\prime}}^{t^{\prime}}<\tilde{\xi}_{c^{\prime}}^{t^{\prime}}$ whenever $(c',t')\neq \emptyset$) such that
\begin{align}
\min\{f(\xi),f(\tilde{\xi})\}\leq\min\{f(\xi-\chi_{c,t}+\chi_{c',t'}), f(\tilde{\xi}+\chi_{c,t}-\chi_{c',t'})\},
\label{TE-1}
\end{align}
and conditions (A) and (B) are satisfied.

Suppose that $f(\xi)=f(\tilde{\xi})$. Let $\lambda^*$ denote the equal value. By ordinal concavity of $f_{\lambda^*}$, there exists $\left(c^*, t^*\right) \in (\mathcal{C} \times \mathcal{T})\cup\{\emptyset\}$ (with $\xi_{c^*}^{t^*}<\tilde{\xi}_{c^*}^{t^*}$ whenever $(c^*,t^*)\neq \emptyset$) such that
\begin{itemize}
\item[(i$^*$)] $f_{\lambda^*}(\xi-\chi_{c, t}+\chi_{c^{*}, t^{*}})>f_{\lambda^*}(\xi)$, or
\item[(ii$^*$)] $f_{\lambda^*}(\tilde{\xi}+\chi_{c, t}-\chi_{c^{*}, t^{*}})>f_{\lambda^*}(\tilde{\xi})$, or
\item[(iii$^*$)] $f_{\lambda^*}(\tilde{\xi}+\chi_{c, t}-\chi_{c^{*}, t^{*}})=f_{\lambda^*}(\tilde{\xi})$ and $f_{\lambda^*}\left(\xi-\chi_{c, t}+\chi_{c^{*}, t^{*}}\right)=f_{\lambda^*}(\xi)$.
\end{itemize}
By the definition of $f_{\lambda^*}(\cdot)$, neither (i$^*$) nor (ii$^*$) holds. Thus, (iii$^*$) holds, which implies
\begin{align*}
f(\xi-\chi_{c,t}+\chi_{c^*,t^*})\geq \lambda^* \text{ and } f(\tilde{\xi}+\chi_{c,t}-\chi_{c^*,t^*})\geq \lambda^*.
\end{align*}
It follows that (\ref{TE-1}) holds. Note that neither the if-clause of (A) nor that of (B) holds.

In the remaining part, we assume $f(\xi)<f(\tilde{\xi})$ (the other case $f(\xi)>f(\tilde{\xi})$ can be handled analogously).
Under this assumption, for each $(c',t')\in (\mathcal{C}\times \mathcal{T})\cup \{\emptyset\}$ that satisfies (\ref{TE-1}), the if-clause of (A) never holds. Thus, it suffices to prove that (\ref{TE-1}) and condition (B) hold for some $(c',t')\in (\mathcal{C}\times \mathcal{T})\cup \{\emptyset\}$.

Let $\Phi \subseteq \{(c',t')\in (\mathcal{C}\times \mathcal{T}) \mid \xi_{c^{\prime}}^{t^{\prime}}<\tilde{\xi}_{c^{\prime}}^{t^{\prime}}\}\cup \{\emptyset\}$ be the set of coordinates that satisfy one of the following conditions:
\begin{itemize}
\item[(i)] $f(\xi-\chi_{c, t}+\chi_{c^{\prime}, t^{\prime}})>f(\xi)$, or
\item[(ii)] $f(\tilde{\xi}+\chi_{c, t}-\chi_{c^{\prime}, t^{\prime}})>f(\tilde{\xi})$, or
\item[(iii)] $f(\tilde{\xi}+\chi_{c, t}-\chi_{c^{\prime}, t^{\prime}})=f(\tilde{\xi})$ and $f\left(\xi-\chi_{c, t}+\chi_{c^{\prime}, t^{\prime}}\right)=f(\xi)$.
\end{itemize}
Note that $\Phi\neq \emptyset$ because $f_\lambda=f$ holds for a sufficiently large $\lambda$ and the function satisfies ordinal concavity.

\smallskip
\noindent
\emph{Case 1:}
Suppose there exists $(c',t')\in \Phi$ for which (iii) holds. Then, (\ref{TE-1}) immediately follows (the if-clause of (B) does not hold).

\smallskip
\noindent
\emph{Case 2:}  Suppose that there does not exist $(c',t')\in \Phi$ for which (iii) holds.

\smallskip
\noindent
\emph{Subcase 2-1:} Suppose that there does not exist $(c',t')\in \Phi$ for which (i) holds. In this case, every $(c',t')\in \Phi$ satisfies (ii). Let $\lambda'=f(\tilde{\xi})$. Since $f_{\lambda'}$ satisfies ordinal concavity, there exists $(c'',t'')\in (\mathcal{C}\times \mathcal{T})\cup\{\emptyset\}$ (with $\xi_{c''}^{t''}<\tilde{\xi}_{c''}^{t''}$ whenever $(c'',t'')\neq \emptyset$) such that
\begin{itemize}
\item[(iv)] $f_{\lambda'}(\xi-\chi_{c, t}+\chi_{c'', t''})>f_{\lambda'}(\xi)$, or
\item[(v)] $f_{\lambda'}(\tilde{\xi}+\chi_{c, t}-\chi_{c'', t''})>f_{\lambda'}(\tilde{\xi})$, or
\item[(vi)] $f_{\lambda'}(\tilde{\xi}+\chi_{c, t}-\chi_{c'', t''})=f_{\lambda'}(\tilde{\xi})$ and $f_{\lambda'}\left(\xi-\chi_{c, t}+\chi_{c'', t''}\right)=f_{\lambda'}(\xi)$.
\end{itemize}
By the definition of truncation, (v) never holds. If (iv) holds, then together with $\lambda'=f(\tilde{\xi})>f(\xi)$, we obtain a contradiction to the assumption of Subcase 2-1. Thus, (vi) holds, which establishes (\ref{TE-1}) (the if-clause of (B) does not hold).

\smallskip
\noindent
\emph{Subcase 2-2:} Suppose that there exists $(c',t')\in \Phi$ for which (i) holds. Let $\Phi' \subseteq \Phi$ be the set of coordinates for which (i) holds.

\smallskip
\noindent
\emph{Subcase 2-2-1:}
Suppose that there exists $(c',t')\in \Phi'$ such that
\begin{align}
f(\tilde{\xi}+\chi_{c,t}-\chi_{c',t'})\geq f(\xi) \; (=\min\{f(\xi), f(\tilde{\xi})\}).
\label{TE-3}
\end{align}
Then, (\ref{TE-1}) holds (the if-clause of (B) does not hold).

\smallskip
\noindent
\emph{Subcase 2-2-2:}
Suppose that there does not exist $(c',t')\in \Phi'$ that satisfies (\ref{TE-3}). Let $\lambda''=f(\xi)$. Since $f_{\lambda''}$ satisfies ordinal concavity, there exists $(c''',t''')\in (\mathcal{C}\times \mathcal{T})\cup\{\emptyset\}$ (with $\xi_{c'''}^{t'''}<\tilde{\xi}_{c'''}^{t'''}$ whenever $(c''',t''')\neq \emptyset$) such that
\begin{itemize}
\item[(vii)] $f_{\lambda''}(\xi-\chi_{c, t}+\chi_{c''', t'''})>f_{\lambda''}(\xi)$, or
\item[(viii)] $f_{\lambda''}(\tilde{\xi}+\chi_{c, t}-\chi_{c''', t'''})>f_{\lambda''}(\tilde{\xi})$, or
\item[(ix)] $f_{\lambda''}\left(\xi-\chi_{c, t}+\chi_{c''', t'''}\right)=f_{\lambda''}(\xi)$ and $f_{\lambda''}(\tilde{\xi}+\chi_{c, t}-\chi_{c''', t'''})=f_{\lambda''}(\tilde{\xi})$.
\end{itemize}
By the definition of $f_{\lambda''}(\cdot)$, neither (vii) nor (viii) holds. Thus, (ix) holds.

If $f(\xi)<f(\xi-\chi_{c,t}+\chi_{c''',t'''})$, then $(c''',t''')\in \Phi'$. By the assumption of Subcase 2-2-2, $f(\tilde{\xi}+\chi_{c,t}-\chi_{c''',t'''})<f(\xi)=\lambda''$. Then, $f_{\lambda''}(\tilde{\xi}+\chi_{c, t}-\chi_{c''', t'''})=f(\tilde{\xi}+\chi_{c,t}-\chi_{c''',t'''})<\lambda''= f_{\lambda''}(\tilde{\xi})$, where the last equality follows from $\lambda''=f(\xi)<f(\tilde{\xi})$.  We obtain a contradiction to $f_{\lambda''}(\tilde{\xi}+\chi_{c, t}-\chi_{c''', t'''})=f_{\lambda''}(\tilde{\xi})$ stated in (ix).

It follows that $f(\xi-\chi_{c,t}+\chi_{c''',t'''})\leq f(\xi)=\lambda''$. Together with $f_{\lambda''}\left(\xi-\chi_{c, t}+\chi_{c''', t'''}\right)=f_{\lambda''}(\xi)=\lambda''$ (the former equality follows from (ix)), we have
\begin{align}
f(\xi)=f(\xi-\chi_{c,t}+\chi_{c''',t'''}).
\label{TE-4}
\end{align}
By $f_{\lambda''}(\tilde{\xi}+\chi_{c, t}-\chi_{c''', t'''})=f_{\lambda''}(\tilde{\xi})\geq \lambda''$ (the equality follows from (ix)), we have $f(\tilde{\xi}+\chi_{c,t}-\chi_{c''',t'''})\geq \lambda''=f(\xi)$. This condition and (\ref{TE-4}) imply that (\ref{TE-1}) holds for $(c''',t''')$. Note that the if-clause of (B) holds if $f(\tilde{\xi}+\chi_{c,t}-\chi_{c''',t'''})<f(\tilde{\xi})$.
In this case, by the assumption of Subcase 2-2, there exists a coordinate in $\Phi'$, for which the desired strict inequality in (B) holds.

\medskip
\noindent
\emph{The ``if'' direction:} Let $\xi, \tilde{\xi} \in \Xi^0$ and $(c, t) \in \mathcal{C} \times \mathcal{T}$ with $\xi_c^t>\tilde{\xi}_{c}^t$. By pseudo M$^\natural$-concavity$^+$, there exists $\left(c^{\prime}, t^{\prime}\right) \in (\mathcal{C} \times \mathcal{T})\cup\{\emptyset\}$ (with $\xi_{c^{\prime}}^{t^{\prime}}<\tilde{\xi}_{c^{\prime}}^{t^{\prime}}$ whenever $(c',t')\neq \emptyset$) such that
\begin{align}
\min\{f(\xi),f(\tilde{\xi})\}\leq\min\{f(\xi-\chi_{c,t}+\chi_{c',t'}), f(\tilde{\xi}+\chi_{c,t}-\chi_{c',t'})\},
\label{TE-5}
\end{align}
and conditions (A) and (B) are satisfied. Let $\lambda\geq 0$.

\smallskip
\noindent
\emph{Case 1:} Suppose $f(\xi)=f(\tilde{\xi})$.

\smallskip
\noindent
\emph{Subcase 1-1:} Suppose $\lambda>f(\xi)=f(\tilde{\xi})$. Then, (\ref{TE-5}) implies that one of the following conditions holds:
\begin{itemize}
\item $f(\xi)<f(\xi+\chi_{c,t}-\chi_{c',t'}) \: \bigl(\Longleftrightarrow f_\lambda(\xi)<f_\lambda(\xi+\chi_{c,t}-\chi_{c',t'})\bigr)$, or
\item $f(\tilde{\xi})<f(\tilde{\xi}-\chi_{c,t}+\chi_{c',t'}) \: \bigl(\Longleftrightarrow f_\lambda(\tilde{\xi})<f_\lambda(\tilde{\xi}-\chi_{c,t}+\chi_{c',t'})\bigr)$, or
\item $f(\xi)=f(\xi+\chi_{c,t}-\chi_{c',t'}) \text{ and } f(\tilde{\xi})=f(\tilde{\xi}-\chi_{c,t}+\chi_{c',t'})$
\item[] $\bigl(\Longleftrightarrow f_\lambda(\xi)=f_\lambda(\xi+\chi_{c,t}-\chi_{c',t'}) \text{ and } f_\lambda(\tilde{\xi})=f_\lambda(\tilde{\xi}-\chi_{c,t}+\chi_{c',t'})\bigr)$.
\end{itemize}
Thus, ordinal concavity of $f_\lambda$ holds.

\smallskip
\noindent
\emph{Subcase 1-2:}
Suppose $\lambda\leq f(\xi)=f(\tilde{\xi})$. Then, (\ref{TE-5}) implies $f(\xi-\chi_{c,t}+\chi_{c',t'})\geq f(\xi)$ and $f(\tilde{\xi}+\chi_{c,t}-\chi_{c',t'})\geq f(\tilde{\xi})$, which in turn implies
\begin{align*}
f_\lambda(\xi)=f_\lambda(\xi+\chi_{c,t}-\chi_{c',t'}) \text{ and } f_\lambda(\tilde{\xi})=f_\lambda(\tilde{\xi}-\chi_{c,t}+\chi_{c',t'}).
\end{align*}
Thus, ordinal concavity of $f_\lambda$ holds.

\smallskip
\noindent
\emph{Case 2:}
Suppose $f(\xi)\neq f(\tilde{\xi})$. We assume $f(\xi)<f(\tilde{\xi})$ (the other case $f(\xi)>f(\tilde{\xi})$ can be handled analogously).

\smallskip
\noindent
\emph{Subcase 2-1:}
Suppose $\lambda>f(\tilde{\xi})$. Note that (\ref{TE-5}) implies $f(\xi)\leq f(\xi-\chi_{c,t}+\chi_{c',t'})$.

\smallskip
\noindent
\emph{Subcase 2-1-1:}
Suppose $f(\xi)<f(\xi-\chi_{c,t}+\chi_{c',t'})$. This inequality is equivalent to $f_\lambda(\xi)<f_\lambda(\xi-\chi_{c,t}+\chi_{c',t'})$, showing that ordinal concavity of $f_\lambda$ holds.

\smallskip
\noindent
\emph{Subcase 2-1-2:}
Suppose $f(\xi)=f(\xi-\chi_{c,t}+\chi_{c',t'})$. Equivalently,
\begin{align}
f_\lambda(\xi)=f_\lambda(\xi-\chi_{c,t}+\chi_{c',t'}).
\label{TE-5-5}
\end{align}
\begin{itemize}
\item If $f(\tilde{\xi})<f(\tilde{\xi}+\chi_{c,t}-\chi_{c',t'})$, then equivalently $f_\lambda(\tilde{\xi})<f_\lambda(\tilde{\xi}+\chi_{c,t}-\chi_{c',t'})$, showing that ordinal concavity of $f_\lambda$ holds.
\item If $f(\tilde{\xi})=f(\tilde{\xi}+\chi_{c,t}-\chi_{c',t'})$, then equivalently $f_\lambda(\tilde{\xi})= f_\lambda(\tilde{\xi}+\chi_{c,t}-\chi_{c',t'})$, which together with (\ref{TE-5-5}) implies that ordinal concavity of $f_\lambda$ holds.
\item If $f(\tilde{\xi})>f(\tilde{\xi}+\chi_{c,t}-\chi_{c',t'})$, then the if-clause of (B) holds. Thus, there exists $\left(c'', t''\right) \in (\mathcal{C} \times \mathcal{T})\cup\{\emptyset\}$ (with $\xi_{c''}^{t''}<\tilde{\xi}_{c''}^{t''}$ whenever $(c'',t'')\neq \emptyset$) such that
\begin{align*}
f(\xi)<f(\xi-\chi_{c,t}+\chi_{c'',t''}).
\end{align*}
This inequality is equivalent to $f_\lambda(\xi)<f_\lambda(\xi-\chi_{c,t}+\chi_{c'',t''})$, showing that ordinal concavity of $f_\lambda$ holds.
\end{itemize}

\smallskip
\noindent
\emph{Subcase 2-2:}
Suppose $f(\tilde{\xi})\geq \lambda>f(\xi)$. Note that (\ref{TE-5}) implies $f(\xi)\leq f(\xi-\chi_{c,t}+\chi_{c',t'})$.

\smallskip
\noindent
\emph{Subcase 2-2-1:}
Suppose $f(\xi)<f(\xi-\chi_{c,t}+\chi_{c',t'})$. This inequality is equivalent to $f_\lambda(\xi)<f_\lambda(\xi-\chi_{c,t}+\chi_{c',t'})$, showing that ordinal concavity of $f_\lambda$ holds.

\smallskip
\noindent
\emph{Subcase 2-2-2:}
Suppose $f(\xi)=f(\xi-\chi_{c,t}+\chi_{c',t'})$. Equivalently,
\begin{align}
f_\lambda(\xi)=f_\lambda(\xi-\chi_{c,t}+\chi_{c',t'}).
\label{TE-6}
\end{align}
\begin{itemize}
\item If $f(\tilde{\xi})\leq f(\tilde{\xi}+\chi_{c,t}-\chi_{c',t'})$, then $f_\lambda(\tilde{\xi})= f_\lambda(\tilde{\xi}+\chi_{c,t}-\chi_{c',t'})$, which together with (\ref{TE-6}) implies that ordinal concavity of $f_\lambda$ holds.
\item If $f(\tilde{\xi})>f(\tilde{\xi}+\chi_{c,t}-\chi_{c',t'})$, then the if-clause of (B) holds. Thus, there exists $\left(c'', t''\right) \in (\mathcal{C} \times \mathcal{T})\cup\{\emptyset\}$ (with $\xi_{c''}^{t''}<\tilde{\xi}_{c''}^{t''}$ whenever $(c'',t'')\neq \emptyset$) such that
\begin{align*}
f(\xi)<f(\xi-\chi_{c,t}+\chi_{c'',t''}).
\end{align*}
This inequality is equivalent to $f_\lambda(\xi)<f_\lambda(\xi-\chi_{c,t}+\chi_{c'',t''})$, showing that ordinal concavity of $f_\lambda$ holds.
\end{itemize}

\smallskip
\noindent
\emph{Subcase 2-3:}
Suppose $\lambda\leq f(\xi)$. By (\ref{TE-5}), we have $\lambda\leq f(\xi)\leq f(\xi-\chi_{c,t}+\chi_{c',t'})$ and $\lambda\leq f(\xi)\leq f(\tilde{\xi}-\chi_{c,t}+\chi_{c',t'})$, which implies
\begin{align*}
f_\lambda(\xi)=f_\lambda(\xi-\chi_{c,t}+\chi_{c',t'}) \text{ and } f_\lambda(\tilde{\xi})=f_\lambda(\tilde{\xi}+\chi_{c,t}-\chi_{c',t'}).
\end{align*}
Thus, ordinal concavity of $f_\lambda$ holds.
\qed
\end{subequations}

\bigskip

\subsection*{Proof of Proposition \ref{prop:matroid}}

If $(\mathcal{X},\mathcal{F})$ is a matroid, the greedy rule satisfies
path independence \citep{fleiner2001} and the law of aggregate demand \citep{yokoi2019}. Therefore, (1) implies (3). Furthermore, (3)
implies (2) trivially. To complete the proof, we show that
(2) implies (1).

Suppose that (2) is satisfied. Let $\mathcal B$ denote the
collection of maximal sets in $\mathcal F$.
By assumption, $\mathcal{F}$ is nonempty, which implies that $\mathcal{B}$
is nonempty. Hence, \emph{B1} holds. Before
showing \emph{B2'}, we establish that \emph{I2} is satisfied.

To show \emph{I2}, let $X \in \mathcal{F}$.
Consider a weight function that assigns all contracts in $\mathcal{X}$ a
distinct positive weight. Let $C$ be the greedy rule for such a weight function.
Then, by the greedy rule definition, $C(X)=X$ since $X\in \mathcal{F}$. For any
$X'\subseteq X$, path independence implies that $C(X')=X'$.
In addition, by the greedy rule definition, $C(X')\in \mathcal{F}$,
so we get $X'\in \mathcal{F}$. Therefore, \emph{I2} is satisfied.

Suppose, for contradiction, that \emph{B2'} is not satisfied. Therefore,
there exist $X_1, X_2 \in \mathcal B$ and $x_1 \in X_1 \setminus X_2$
such that for each $x_2 \in X_2 \setminus X_1$, $(X_1 \setminus \{x_1\}) \cup \{x_2\}$ is not included in a feasible set in $\mathcal{F}$.

Consider a weight function that assigns all contracts in $\mathcal{X}$ a
distinct and positive weight so that contracts in $X_1 \setminus \{x_1\}$
have higher weights than contracts in $X_2 \setminus X_1$, and contracts
in $X_2 \setminus X_1$ have higher weights than the weight of $x_1$.
Let $C'$ be the greedy rule for such a weight function.

When $X_1 \cup X_2$ is the set of available contracts for the greedy rule $C'$,
it chooses $X_1 \setminus \{x_1\}$ first because $X_1 \in \mathcal{F}$ and the weights of contracts in $X_1 \setminus \{x_1\}$ are greater
than the weights of other contracts in $X_1 \cup X_2$. Next the greedy rule chooses no $x_2\in X_2 \setminus X_1$ because, by construction, $(X_1 \setminus \{x_1\}) \cup \{x_2\}$ is not included in a feasible set in $\mathcal{F}$. Finally, the greedy rule chooses $x_1$ because $(X_1 \setminus \{x_1\}) \cup \{x_1\}=X_1 \in \calf$. Since $X_1\in \mathcal{B}$, no other contract
can be chosen. Therefore, we get
\[C'(X_1 \cup X_2) = X_1.\]

When $\{x_1\} \cup X_2$ is the set of available contracts for the greedy rule $C'$,
contracts in $X_2$ are chosen first because they have positive weights greater
than the weight of $x_1$ and $X_2 \in \mathcal{F}$. Furthermore, since
$X_2\in \mathcal{B}$, $x_1$ is not chosen and we get
\[C'(\{x_1\} \cup X_2)=X_2.\]

The two displayed equations provide a contradiction to path independence of $C'$: By
Lemma \ref{lem:pi}, path independence implies the substitutes condition. Now, by
the substitutes condition, $x_1\in X_1=C'(X_1\cup X_2)$ implies $x_1\in C'(\{x_1\}\cup X_2)=X_2$, which is a contradiction since $x_1\notin X_2$.

Therefore, we conclude that the maximal sets in $\mathcal{F}$ satisfy \emph{B1} and \emph{B2'},
which together with Lemma \ref{lem:matchar} implies that they are the bases of a matroid. Since $\mathcal{F}$ satisfies \emph{I2}, $\mathcal{F}$ is the collection of subsets of the bases, which implies that $(X,\mathcal{F})$ is a matroid (see Theorem 1.2.3 of \cite{oxley}).
%
%
\qed

\subsection*{Proof of Proposition \ref{prop:comparison}}
First we show that M$^{\natural}$-concavity implies ordinal concavity.

Let $\xi,\tilde{\xi}\in \Xi^0$ and $(c,t) \in \calc \times \calt$ be such that   $\xi_c^t>\tilde{\xi}_c^t$. Then, by M$^{\natural}$-concavity, one of
conditions (i) and (ii) in Definition \ref{def:natural} holds.

Suppose that condition (i) in Definition \ref{def:natural} holds. If
condition (i) or (ii) in Definition \ref{def:ordinal} holds, then ordinal concavity
is satisfied. If conditions (i) and (ii) in Definition \ref{def:ordinal}
do not hold, then we have
\[f(\xi-\chi_{c,t})\leq f(\xi) \mbox{ and } f(\tilde \xi+\chi_{c,t}) \leq f(\tilde \xi).\]
These two inequalities together with condition (i) in Definition \ref{def:natural}
imply that
\[f(\xi-\chi_{c,t})=f(\xi) \mbox{ and } f(\tilde \xi+\chi_{c,t}) = f(\tilde \xi),\]
which is condition (iii) in Definition of \ref{def:ordinal}, so ordinal concavity is satisfied.

Suppose that condition (i) in Definition \ref{def:natural} does not hold. By
M$^{\natural}$-concavity, condition (ii) in Definition \ref{def:natural} holds.
Therefore, there exists $(c',t') \in \calc \times \calt$ with  $\xi_{c'}^{t'}<\tilde{\xi}_{c'}^{t'}$ such that
\[f(\xi-\chi_{c,t}+\chi_{c',t'})+f(\tilde{\xi}+\chi_{c,t}-\chi_{c',t'})
\geq f(\xi)+f(\tilde \xi).\]
If condition (i) or (ii) in Definition \ref{def:ordinal} holds, then ordinal concavity
is satisfied. If conditions (i) and (ii) in Definition \ref{def:ordinal}
do not hold, then we have
\[f(\xi-\chi_{c,t}+\chi_{c',t'})\leq f(\xi) \mbox{ and }
f(\tilde \xi+\chi_{c,t}-\chi_{c',t'}) \leq f(\tilde \xi).\]
These two inequalities together with condition (ii) in Definition \ref{def:natural}
imply that
\[f(\xi-\chi_{c,t}+\chi_{c',t'})=f(\xi) \mbox{ and } f(\tilde \xi+\chi_{c,t}-\chi_{c',t'}) = f(\tilde \xi),\]
which is condition (iii) in Definition of \ref{def:ordinal}, so ordinal concavity is satisfied.

Now we provide a function that satisfies ordinal concavity but not M$^{\natural}$-concavity. Let the distributional objective $f$ be defined as $f(0)=0$, $f(1)=3$, and
$f(2)=10$. Since it is strictly increasing it is ordinally concave because
condition (ii) in Definition of \ref{def:ordinal} is satisfied. However,
M$^{\natural}$ concavity fails because for $\xi=2$, $\tilde \xi=0$, and $\chi=1$
we have $f(\xi-\chi)+f(\tilde{\xi}+\chi)=6 < 10=f(\xi)+f(\tilde \xi)$.
\qed

\bigskip

\begin{center}
\textbf{Proofs of the Lemmas and Claims}
\end{center}

\bigskip

\subsection*{Proof of Lemma \ref{lem:mconvex}}
In their Proposition 3.1, \cite{murotashioura2018} provide the following equivalent condition for M$^{\natural}$-convexity.


\begin{lemma}\label{lem:natalt}
A set of distributions $\Xi$ is M$^{\natural}$\textbf{-convex} if and only if, for each $\xi,\tilde{\xi} \in \Xi$,
\begin{enumerate}[(i)]
\item $\norm{\xi}>\norm{\tilde \xi}$ implies that there exists $(c,t) \in \calc \times \calt$ with $\xi_c^t>\tilde{\xi}_c^t$ such that $\xi-\chi_{c,t} \in \Xi$ and $\tilde{\xi}+\chi_{c,t} \in \Xi$, and
\item $\norm{\xi}=\norm{\tilde \xi}$ implies that for each $(c,t) \in \calc \times \calt$ with $\xi_c^t>\tilde{\xi}_c^t$, there exists $(c',t') \in \calc \times \calt$ with  $\xi_{c'}^{t'}<\tilde{\xi}_{c'}^{t'}$ such that
\[\xi-\chi_{c,t}+\chi_{c',t'}\in \Xi \; \mbox{ and } \; \tilde{\xi}+\chi_{c,t}-\chi_{c',t'} \in \Xi.\]
\end{enumerate}
\end{lemma}

Let $\Xi$ be a finite and non-empty M$^{\natural}$-convex set and
$\mathcal{M}$ the set of maximal distributions in $\Xi$. Then
there exists at least one distribution in $\mathcal{M}$.
If there exists exactly one distribution in $\mathcal{M}$, then it is trivially M-convex.
For the rest of the proof, suppose that $\mathcal{M}$ has at least two distributions.

Let $\xi, \tilde \xi \in \mathcal{M}$ be distinct. Without loss of generality assume
that $\norm{\xi} \ge \norm{\tilde \xi}$.

If $\norm{\xi} > \norm{\tilde \xi}$, then, by Lemma \ref{lem:natalt}, there exists
$(c,t) \in \calc \times \calt$ with $\xi_c^t>\tilde{\xi}_c^t$ such that
$\xi-\chi_{c,t} \in \Xi$ and $\tilde{\xi}+\chi_{c,t} \in \Xi$. However, $\tilde{\xi}+\chi_{c,t} \in \Xi$ contradicts the assumption that $\tilde{\xi}$ is
maximal in $\Xi$. Therefore, we must have $\norm{\xi} = \norm{\tilde \xi}$, which
implies that every distribution in $\mathcal{M}$ has the same sum of coordinates.
Furthermore, every distribution in $\Xi$ that has the same sum of coordinates also has
to be maximal.

By Lemma \ref{lem:natalt}, for each
$(c,t) \in \calc \times \calt$ with $\xi_c^t>\tilde{\xi}_c^t$, there exists
$(c',t') \in \calc \times \calt$ with  $\xi_{c'}^{t'}<\tilde{\xi}_{c'}^{t'}$ such that
\[\xi-\chi_{c,t}+\chi_{c',t'}\in \Xi \; \mbox{ and } \; \tilde{\xi}+\chi_{c,t}-\chi_{c',t'} \in \Xi.\]
The equations above imply
that both distributions are also maximal in $\Xi$ because
$\norm{\xi-\chi_{c,t}+\chi_{c',t'}}=\norm{\xi}$ and
$\norm{\tilde{\xi}+\chi_{c,t}-\chi_{c',t'}}=\norm{\tilde \xi}$. Therefore, we
get $\xi-\chi_{c,t}+\chi_{c',t'}\in \mathcal{M}$ and
$\tilde{\xi}+\chi_{c,t}-\chi_{c',t'} \in \mathcal{M}$, which establishes that $\mathcal{M}$
is an M-convex set.
\qed

\bigskip

\subsection*{Proof of the Statement in Example \ref{ex:ladfail}}
We prove the statement that the distributional objective defined in Example \ref{ex:ladfail}
satisfies ordinal concavity. We consider several cases depending on the value of $\xi$ used in the definition of ordinal concavity.


\medskip
\noindent
\emph{Case 1:} $\xi=\xi(\{x,y\})$. Let $t\in \calt$ be the type of the student
associated with contract $x$ and $t'\in \calt$ be the type of the student
associated with contract $z$.
If $\tilde{\xi}_{c}^{t'}=0$, then $\tilde{\xi}=\xi(\emptyset)$
or $\tilde{\xi}=\xi(\{y\})$. For $\tilde \xi=\xi(\emptyset)$, we have
$f(\tilde \xi + \xi(\{x\}))>f(\tilde \xi)$. Therefore, condition (ii) in the definition of ordinal concavity is satisfied. For $\tilde \xi=\xi(\{y\})$, we have $f(\xi-\chi_{c,t})=f(\xi)$ and $f(\tilde \xi+\chi_{c,t})=f(\tilde \xi)$. Therefore, condition (iii) in the definition of ordinal concavity is satisfied. However, if $\tilde{\xi}_{c}^{t'}=1$,
then $\tilde{\xi}=\xi(\{z\})$ or $\tilde{\xi}=\xi(\{y,z\})$. For both values of
$\tilde \xi$, $f(\xi-\chi_{c,t}+\chi_{c,t'})>f(\xi)$, which means that
condition (i) in the definition of ordinal concavity is satisfied.

\medskip
\noindent
\emph{Case 2:} $\xi=\xi(\{y,z\})$. If $\chi_{c,t}=\xi(\{y\})$ and $n>5$,
then we have $f(\xi-\chi_{c,t})>f(\xi)$, so condition (i) in the
definition of ordinal concavity is satisfied.

If $\chi_{c,t}=\xi(\{y\})$ and $n=5$, then, for $\tilde \xi = \xi(\emptyset)$, we have
$f(\tilde \xi+\chi_{c,t})>f(\tilde \xi)$, so condition (ii) in the definition
of ordinal concavity is satisfied. For $\tilde \xi = \xi(\{x\})$, we have
$f(\xi-\chi_{c,t})=f(\xi)$ and $f(\tilde \xi+\chi_{c,t})=f(\tilde \xi)$, so
condition (iii) in the definition of ordinal concavity is satisfied
For $\tilde \xi = \xi(\{z\})$, we have
$f(\tilde \xi +\chi_{c,t})=f(\tilde \xi)$ and $f(\xi-\chi_{c,t})=f(\xi)$, so
condition (iii) in the definition of ordinal concavity is satisfied. Finally, if
$\tilde \xi = \xi(\{x,z\})$, let $t'\in \calt$ be such that $\chi_{c,t'}=\xi(\{z\})$.
Then $f(\tilde \xi +\chi_{c,t}-\chi_{c,t'})=f(\tilde \xi)$ and
$f(\xi-\chi_{c,t'}+\chi_{c,t})=f(\xi)$, so condition (iii) in the definition of
ordinal concavity is satisfied.

However, if $\chi_{c,t}=\xi(\{z\})$, then $\tilde{\xi}_c^t=0$,
and for all such $\tilde \xi$ except $\xi(\{x,y\})$, we have
$f(\tilde \xi+\chi_{c,t})>f(\tilde{\xi})$. Therefore, condition (ii)  in the definition of ordinal concavity is satisfied. For $\tilde \xi = \xi(\{x,y\})$, let
$\chi_{c,t'} = \xi(\{x\})$. Then $f(\tilde \xi +\chi_{c,t}-\chi_{c,t'})=f(\tilde \xi)$
and $f(\xi-\chi_{c,t'}+\chi_{c,t})=f(\xi)$, so condition (iii) in the
definition of ordinal concavity is satisfied.

\medskip
\noindent
\emph{Case 3:} $\xi=\xi(\{x,z\})$. Since the distributional objective $f$ is symmetric
with respect to $x$ and $y$, this case is analogous to Case 2 above.

\medskip
\noindent
\emph{Case 4:} $\xi=\xi(\{z\})$. In this case, we have $\chi_{c,t}=\xi(\{z\})$ and
$\tilde{\xi}_c^t=0$. For all such $\tilde{\xi}$ except $\xi(\{x,y\})$, we have
$f(\tilde{\xi}+\chi_{c,t})>f(\tilde{\xi})$, so condition (ii) is
satisfied in the definition of ordinal concavity. For $\tilde{\xi}=\xi(\{x,y\})$,
if we let $\chi_{c,t'}=\xi(\{x\})$ where $t'\in \calt$ is the type of student
associated with contract $x$, then $f(\tilde \xi+\chi_{c,t}-\chi_{c,t'}) >
f(\tilde \xi)$, so condition (ii) in the definition of ordinal concavity is
satisfied.

\medskip
\noindent
\emph{Case 5:} $\xi=\xi(\{x\})$. In this case, we have $\chi_{c,t}=\xi(\{x\})$ and
$\tilde{\xi}_c^t=0$. If $\tilde \xi = \xi(\emptyset)$, then
$f(\tilde \xi+\chi_{c,t})>f(\tilde \xi)$.
Therefore, condition (ii) in the definition of ordinal concavity is satisfied.
If $\tilde \xi = \xi(\{y\})$, let $t'\in \calt$ be such that $\chi_{c,t'}=\xi(\{y\})$.
Then $f(\tilde{\xi}+\chi_{c,t}-\chi_{c,t'})=f(\tilde{\xi})$
and $f(\xi-\chi_{c,t}+\chi_{c,t'})=f(\xi)$, so condition (iii) in the definition of
ordinal concavity is satisfied.
If $\tilde{\xi} \in \{\xi(\{z\}),\xi(\{y,z\})\}$, then
let $t''\in \calt$ be such that $\chi_{c,t''}=\xi(\{z\})$.
Then $f(\xi-\chi_{c,t}+\chi_{c,t''})>f(\xi)$,
which means that condition (i) in the definition of ordinal
concavity is satisfied.

\medskip
\noindent
\emph{Case 6:} $\xi=\xi(\{y\})$. Since the distributional objective $f$ is symmetric
with respect to $x$ and $y$, this case is analogous to Case 5 above.

\qed

\bigskip

\subsection*{Proof of Claim \ref{claim:ex1-pMC}}
Let $\xi, \tilde \xi\in \Xi^0$ and $(c,t)$ with $\xi_c^t>\tilde \xi_c^t$. If $\tilde \xi=\xi(\emptyset)$, then
\begin{align*}
f(\tilde \xi)<f(\xi) \text{ and } f(\tilde \xi)<f(\tilde \xi+\chi_{c,t}).
\end{align*}
Together with the fact that $f(\xi(\emptyset))=0$ is the minimum function value, pseudo M$^\natural$-concavity$^+$ is satisfied. In the remaining part, suppose that $\tilde \xi\neq \xi(\emptyset)$. If $||\xi||=||\tilde \xi||=1$ or $\xi\geq \tilde \xi$, then pseudo M$^\natural$-concavity$^+$ trivially holds. In what follows, we consider the remaining three cases.

\smallskip
\noindent
\emph{Case 1:}
Suppose $\{\xi, \tilde \xi\}\subseteq \{\xi(\{x\}), \xi(\{y,z\})\}$.

\smallskip
\noindent
\emph{Subcase 1-1:}
Suppose $\xi=\xi(\{x\})$, which implies $\chi_{c,t}=\xi(\{x\})$. For $(c',t')$ with $\chi_{c',t'}=\xi(\{z\})$,
\begin{align*}
&f(\xi(\{x\})-\chi_{c,t}+\chi_{c',t'})=f(\xi(\{z\}))=n>1=f(\xi(\{x\})), \\
&f(\xi\{y,z\})+\chi_{c,t}-\chi_{c',t'})=f(\xi(\{x,y\}))=1.
\end{align*}
Together with $f(\xi(\{x\}))=1<5=f(\xi(\{y,z\})$, pseudo M$^\natural$-concavity$^+$ holds.

\smallskip
\noindent
\emph{Subcase 1-2:}
Suppose $\xi=\xi(\{y,z\})$ and $\chi_{c,t}=\xi(\{y\})$. For $(c',t')$ with $\chi_{c',t'}=\xi(\{x\})$,
\begin{align*}
&f(\xi(\{y,z\})-\chi_{c,t}+\chi_{c',t'})=f(\xi(\{x,z\}))=5=f(\xi(\{y,z\})),   \\
&f(\xi(\{x\})+\chi_{c,t}-\chi_{c',t'})=f(\xi(\{y\}))=1=f(\xi(\{x\})).
\end{align*}
Hence, pseudo M$^\natural$-concavity$^+$ holds.

\smallskip
\noindent
\emph{Subcase 1-3:}
Suppose $\xi=\xi(\{y,z\})$ and $\chi_{c,t}=\xi(\{z\})$. For $(c',t')$ with $\chi_{c',t'}=\xi(\{x\})$,
\begin{align*}
&f(\xi(\{x\})+\chi_{c,t}-\chi_{c',t'})=f(\xi(\{z\}))=5>1=f(\xi(\{x\})), \\
&f(\xi(\{y,z\})-\chi_{c,t}+\chi_{c',t'})=f(\xi(\{x,y\}))=1.
\end{align*}
Together with $f(\xi(\{x\}))=1<5=f(\xi(\{y,z\})$ pseudo M$^\natural$-concavity$^+$ holds.

\smallskip
\noindent
\emph{Case 2:}
Suppose $\{\xi, \tilde \xi\}\subseteq  \{\xi(\{y\}), \xi(\{x,z\})\}$. Since contracts $x$ and $y$ are symmetric, the proof of this case is similar to that for Case 1.

\smallskip
\noindent
\emph{Case 3:}
Suppose $\{\xi, \tilde \xi\}\subseteq \{\xi(\{z\}), \xi(\{x,y\})\}$.

\smallskip
\noindent
\emph{Subcase 3-1:}
Suppose $\xi=\xi(\{z\})$, which implies $\chi_{c,t}=\xi(\{z\})$. For $(c',t')$ with $\chi_{c',t'}=\xi(\{x\})$,
\begin{align*}
&f(\xi(\{x,y\})+\chi_{c,t}-\chi_{c',t'})=f(\xi(\{y,z\}))=5>1=f(\xi(\{x,y\})), \\
&f(\xi(\{z\})-\chi_{c,t}+\chi_{c',t'})=f(\xi(\{x\}))=1.
\end{align*}
Together with $f(\xi(\{x,y\}))=1<n=f(\xi(\{z\}))$, pseudo M$^\natural$-concavity$^+$ holds.

\smallskip
\noindent
\emph{Subcase 3-2:}
Suppose $\xi=\xi(\{x,y\})$ and $\chi_{c,t}=\xi(\{x\})$. For $(c',t')$ with $\chi_{c',t'}=\xi(\{z\})$,
\begin{align*}
&f(\xi(\{x,y\})-\chi_{c,t}+\chi_{c',t'})=f(\xi(\{y,z\}))=5>f(\xi(\{x,y\})),   \\
&f(\xi(\{z\})+\chi_{c,t}-\chi_{c',t'})=f(\xi(\{x\}))=1.
\end{align*}
Together with $f(\xi(\{x,y\}))=1<n=f(\xi(\{z\}))$, pseudo M$^\natural$-concavity$^+$ holds.

\smallskip
\noindent
\emph{Subcase 3-3:}
Suppose $\xi=\xi(\{x,y\})$ and $\chi_{c,t}=\xi(\{y\})$. Since contracts $x$ and $y$ are symmetric, the proof for this case is similar to that for Subcase 3-2.
\qed


\bigskip

\subsection*{Proof of Claim \ref{claim:ex2-pMC}}
Suppose that $\Xi^0=\{\xi\in \mathbb{Z}^{|C|\times |T|}_+ \mid \sum_{t \in \mathcal{T}}\xi_c^t\leq q\}$ for some $q\in \mathbb{Z}_+$.
Let $\xi, \tilde{\xi} \in \Xi^0$ and $(c, t) \in \mathcal{C} \times \mathcal{T}$ with $\xi_c^t>\tilde{\xi}_{c}^t$.
We consider three cases. In each case discussed below, the if-clauses of (A) and (B) in the definition of pseudo M$^\natural$-concavity$^+$ do not hold. Therefore, it suffices to show that the weak inequality in the definition holds. Recall that, for each $\xi \in \Xi^0$,
\begin{align*}
f^s(\xi)=\sum_{(c,t) \in \mathcal{C}\times \mathcal{T}} \min\{ \xi^t_c, r^t_c\}.
\end{align*}

\smallskip
\noindent
\emph{Case 1:}
Suppose $f^s(\xi)<f^s(\tilde \xi)$.

\smallskip
\noindent
\emph{Subcase 1-1:}
Suppose $r_c^t\geq \xi_c^t$. Then,
\begin{align*}
r_c^t\geq \min\{ \xi_c^t, r_c^t\}=\xi_c^t>\tilde \xi_c^t=\min\{ \tilde \xi_c^t, r_c^t\}.
\end{align*}
Together with $f^s(\xi)<f^s(\tilde \xi)$, there exists $(\corigin, t')\in \calc\times \calt$ such that
\begin{align*}
\min\{ \xi_{\corigin}^{t'}, r_{\corigin}^{t'}\}<\min\{ \tilde \xi_{\corigin}^{t'}, r_{\corigin}^{t'}\}\leq r_{\corigin}^{t'}.
\end{align*}
By the above two inequalities,
\begin{align*}
f^s(\xi)&=f^s(\xi-\chi_{c,t})+1=f^s(\xi-\chi_{c,t}+\chi_{\corigin,t'}), \text{ and } \\
f^s(\tilde \xi)&\leq f^s(\tilde \xi-\chi_{\corigin, t'})+1=f^s(\tilde \xi+\chi_{c, t}-\chi_{\corigin, t'}).
\end{align*}
It follows that pseudo M$^\natural$-concavity$^+$ is satisfied.

\smallskip
\noindent
\emph{Subcase 1-2:}
Suppose $r_c^t<\xi_c^t$.

\smallskip
\noindent
\emph{Subcase 1-2-1:}
Suppose $\sum_{(\tilde c, \tilde t)\in \mathcal{C}\times \mathcal{T}}\tilde \xi_{c}^{\tilde t}<q$, which implies $\tilde \xi+\chi_{c,t}\in \Xi^0$.
By $r_c^t<\xi_c^t$,
\begin{align*}
f^s(\xi)=f^s(\xi-\chi_{c, t}).
\end{align*}
By the definition of $f^s(\cdot)$,
\begin{align*}
f^s(\tilde \xi)\leq f^s(\tilde \xi+\chi_{c,t}).
\end{align*}
It follows that pseudo M$^\natural$-concavity$^+$ is satisfied.

\smallskip
\noindent
\emph{Subcase 1-2-2:}
Suppose $\sum_{(\tilde c, \tilde t)\in \mathcal{C}\times \mathcal{T}}\tilde \xi_{\tilde c}^{\tilde t}=q$.
By $\xi_c^t>\tilde \xi_c^t$ and $\sum_{(\tilde c,\tilde t)\in \mathcal{C}\times \mathcal{T}}\xi_{\tilde c}^{\tilde t}\leq q=\sum_{(\tilde c, \tilde t)\in \mathcal{C}\times \mathcal{T}}\tilde \xi_{\tilde c}^{\tilde t}$, there exists $(\corigin,t')\in \mathcal{C}\times \mathcal{T}$ such that $\xi_{\corigin}^{t'}<\tilde \xi_{\corigin}^{t'}$. If $r_{\corigin}^{t'}<\tilde \xi_{\corigin}^{t'}$, then together with $r_c^t<\xi_c^t$,
\begin{align*}
&f^s(\xi)=f^s(\xi-\chi_{c,t})\leq f^s(\xi-\chi_{c,t}+\chi_{\corigin,t'}), \text{ and } \\
&f^s(\tilde \xi)=f^s(\tilde \xi-\chi_{\corigin,t'})\leq f^s(\tilde \xi+\chi_{c,t}-\chi_{\corigin,t'}).
\end{align*}
It follows that pseudo M$^\natural$-concavity$^+$ is satisfied. If $r_{\corigin}^{t'}\geq \tilde \xi_{\corigin}^{t'}(>\xi_{\corigin}^{t'})$, then together with $r_c^t<\xi_c^t$,
\begin{align*}
&f^s(\xi)=f^s(\xi-\chi_{c,t})< f^s(\xi-\chi_{c,t}+\chi_{\corigin,t'}), \text{ and } \\
&f^s(\tilde \xi)-1=f^s(\tilde \xi-\chi_{\corigin,t'})\leq f^s(\tilde \xi+\chi_{c,t}-\chi_{\corigin,t'}).
\end{align*}
By the assumption of Case 1, $\min\{f^s(\xi), f^s(\tilde \xi)-1\}=\min\{f^s(\xi), f^s(\tilde \xi)\}$. Together with the above two inequalities, pseudo M$^\natural$-concavity$^+$ is satisfied.

\smallskip
\noindent
\emph{Case 2:}
Suppose $f^s(\xi)=f^s(\tilde \xi)$.\footnote{The proofs for Subcases 2-1 and 2-2-1 are similar to those for Subcases 1-1 and 1-2-1, respectively.}

\smallskip
\noindent
\emph{Subcase 2-1:}
Suppose $r_c^t\geq \xi_c^t$. Then,
\begin{align*}
r_c^t\geq \min\{ \xi_c^t, r_c^t\}=\xi_c^t>\tilde \xi_c^t=\min\{ \tilde \xi_c^t, r_c^t\}.
\end{align*}
Together with $f^s(\xi)=f^s(\tilde \xi)$, there exists $(\corigin,t')\in \calc\times \calt$ such that
\begin{align*}
\min\{ \xi_{\corigin}^{t'}, r_{\corigin}^{t'}\}<\min\{ \tilde \xi_{\corigin}^{t'}, r_{\corigin}^{t'}\}\leq r_{\corigin}^{t'}.
\end{align*}
By the above two inequalities,
\begin{align*}
&f^s(\xi)=f^s(\xi-\chi_{c,t})+1=f^s(\xi-\chi_{c,t}+\chi_{\corigin, t'}), \text{ and } \\
&f^s(\tilde \xi)\leq f^s(\tilde \xi-\chi_{\corigin, t'})+1=f^s(\tilde \xi+\chi_{c,t}-\chi_{\corigin, t'}).
\end{align*}
It follows that pseudo M$^\natural$-concavity$^+$ is satisfied.

\smallskip
\noindent
\emph{Subcase 2-2:}
Suppose $r_c^t<\xi_c^t$.

\smallskip
\noindent
\emph{Subcase 2-2-1:}
Suppose $\sum_{(\tilde c,\tilde t)\in \mathcal{C}\times \mathcal{T}}\tilde \xi_{\tilde c}^{\tilde t}<q$, which implies $\tilde \xi+\chi_{c,t}\in \Xi^0$. By $r_c^t<\xi_c^t$,
\begin{align*}
f^s(\xi)=f^s(\xi-\chi_{c,t}).
\end{align*}
By the definition of $f^s(\cdot)$,
\begin{align*}
f^s(\tilde \xi)\leq f^s(\tilde \xi+\chi_{c,t}).
\end{align*}
It follows that pseudo M$^\natural$-concavity$^+$ is satisfied.

\smallskip
\noindent
\emph{Subcase 2-2-2:}
Suppose $\sum_{(\tilde c,\tilde t)\in \mathcal{C}\times \mathcal{T}}\tilde \xi_{\tilde c}^{\tilde t}=q$. Let $\Phi=\{(\tilde c, \tilde t)\in \mathcal{C}\times \mathcal{T} \mid \xi_{\tilde c}^{\tilde t}<\tilde \xi_{\tilde c}^{\tilde t}\}$. By $\xi_c^t>\tilde \xi_c^t$ and $\sum_{(\tilde c,\tilde t)\in \mathcal{C}\times \mathcal{T}}\xi_{\tilde c}^{\tilde t}\leq q=\sum_{(\tilde c,\tilde t)\in \mathcal{C}\times \mathcal{T}}\tilde \xi_{\tilde c}^{\tilde t}$, we have $\Phi \neq \emptyset$.

Suppose, for contradiction, that $r_{\tilde c}^{\tilde t}\geq \tilde \xi_{\tilde c}^{\tilde t}$ for each $(\tilde c, \tilde t)\in \Phi$.
Then,
\begin{align*}
f^s(\xi)-f^s(\tilde \xi)&=\sum_{(\tilde c, \tilde t)\in \mathcal{C}\times \mathcal{T}}\min\{ \xi_{\tilde c}^{\tilde t}, r_{\tilde c}^{\tilde t}\} -\sum_{(\tilde c, \tilde t)\in \mathcal{C}\times \mathcal{T}}\min\{ \tilde \xi_{\tilde c}^{\tilde t}, r_{\tilde c}^{\tilde t}\} \\
&=\sum_{(\tilde c, \tilde t)\in \Phi}\min\{ \xi_{\tilde c}^{\tilde t}, r_{\corigin}^{\tilde t}\} -\sum_{(\tilde c, \tilde t)\in \Phi}\min\{ \tilde \xi_{\tilde c}^{\tilde t}, r_{\tilde c}^{\tilde t}\} \\
&+\sum_{(\tilde c, \tilde t)\in (\mathcal{C}\times \mathcal{T})\backslash \Phi}\min\{ \xi_{\tilde c}^{\tilde t}, r_{\tilde c}^{\tilde t}\} -\sum_{(\tilde c, \tilde t)\in (\mathcal{C}\times \mathcal{T})\backslash \Phi}\min\{ \tilde \xi_{\tilde c}^{\tilde t}, r_{\tilde c}^{\tilde t}\} \\
&=\sum_{(\tilde c, \tilde t)\in \Phi} \xi_{\tilde c}^{\tilde t} -\sum_{(\tilde c, \tilde t)\in \Phi}\tilde \xi_{\tilde c}^{\tilde t} \\
&+\sum_{(\tilde c, \tilde t)\in (\mathcal{C}\times \mathcal{T})\backslash \Phi}\Bigl(\min\{ \xi_{\tilde c}^{\tilde t}, r_{\tilde c}^{\tilde t}\} -\min\{ \tilde \xi_{\tilde c}^{\tilde t}, r_{\tilde c}^{\tilde t}\}\Bigr) \\
&<\sum_{(\tilde c, \tilde t)\in \Phi} \xi_{\tilde c}^{\tilde t} -\sum_{(\tilde c, \tilde t)\in \Phi}\tilde \xi_{\tilde c}^{\tilde t} \\
&+\sum_{(\tilde c, \tilde t)\in (\mathcal{C}\times \mathcal{T})\backslash \Phi} \Bigl(\xi_{\tilde c}^{\tilde t}-\tilde \xi_{\tilde c}^{\tilde t}\Bigr)  \\
&\leq 0,
\end{align*}
where the third equality follows from the assumption made for contradiction and the definition of $\Phi$, the strict inequality follows from $\min\{ \xi_{\tilde c}^{\tilde t}, r_{\tilde c}^{\tilde t}\} -\min\{ \tilde \xi_{\tilde c}^{\tilde t}, r_{\tilde c}^{\tilde t}\}\leq \xi_{\tilde c}^{\tilde t}-\tilde \xi_{\tilde c}^{\tilde t}$ for
each $(\tilde c, \tilde t)\in (\mathcal{C}\times \mathcal{T})\backslash \Phi$ and $\min\{ \xi_{c}^{t}, r_{c}^{t}\} -\min\{ \tilde \xi_{c}^{t}, r_{c}^{t}\}<\xi_{c}^{t}-\tilde \xi_{c}^{t}$ for
$(c,t)\in (\mathcal{C}\times \mathcal{T})\backslash \Phi$ (where the latter strict inequality follows from $\xi_c^t>\tilde \xi_c^t$ and $r_c^t<\xi_c^t$), and the last inequality follows from
the assumption of Subcase 2-2-2. We obtain a contradiction to the assumption of Case 2.

It follows that there exists $(\corigin,t')\in \Phi$ with $r_{\corigin}^{t'}<\tilde \xi_{\corigin}^{t'}$.
Together with $r_c^t<\xi_c^t$,
\begin{align*}
&f^s(\xi)=f^s(\xi-\chi_{c,t})\leq f^s(\xi-\chi_{c,t}+\chi_{\corigin,t'}), \text{ and } \\
&f^s(\tilde \xi)=f^s(\tilde \xi-\chi_{\corigin,t'})\leq f^s(\tilde \xi+\chi_{c,t}-\chi_{\corigin,t'}).
\end{align*}
It follows that pseudo M$^\natural$-concavity$^+$ is satisfied.

\smallskip
\noindent
\emph{Case 3:}
Suppose $f^s(\xi)>f^s(\tilde \xi)$.

\smallskip
\noindent
\emph{Subcase 3-1:}
Suppose $r_c^t\geq \xi_c^t$.

\smallskip
\noindent
\emph{Subcase 3-1-1:}
Suppose $\sum_{(\tilde c,\tilde t)\in \mathcal{C}\times \mathcal{T}}\tilde \xi_{\tilde c}^{\tilde t}<q$, which implies $\tilde \xi+\chi_{c,t}\in \Xi^0$.
By $r_c^t\geq \xi_c^t$,
\begin{align*}
f^s(\xi)-1=f^s(\xi-\chi_{c,t}).
\end{align*}
By $r_c^t\geq \xi_c^t>\tilde \xi_c^t$,
\begin{align*}
f^s(\tilde \xi)<f^s(\tilde \xi+\chi_{c,t}).
\end{align*}
By the assumption of Case 3, $\min\{f^s(\xi)-1, f^s(\tilde \xi)\}=\min\{f^s(\xi), f^s(\tilde \xi)\}$. Together with the above displayed equality and inequality, pseudo M$^\natural$-concavity$^+$ is satisfied.

\smallskip
\noindent
\emph{Subcase 3-1-2:}
Suppose $\sum_{(\tilde c,\tilde t)\in \mathcal{C}\times \mathcal{T}}\tilde \xi_{\tilde c}^{\tilde t}=q$. By $\xi_c^t>\tilde \xi_c^t$ and $\sum_{(\tilde c,\tilde t)\in \mathcal{C}\times \mathcal{T}}\xi_{\tilde c}^{\tilde t} \leq q=\sum_{(\tilde c, \tilde t)\in \mathcal{C}\times \mathcal{T}}\tilde \xi_{\tilde c}^{\tilde t}$, there exists $(\corigin,t')\in \mathcal{C}\times \mathcal{T}$ such that $\xi_{\corigin}^{t'}<\tilde \xi_{\corigin}^{t'}$. If
$r_{\corigin}^{t'}<\tilde \xi_{\corigin}^{t'}$, then together with $r_c^t\geq \xi_c^t>\tilde \xi_c^t$,
\begin{align*}
&f^s(\xi)-1=f^s(\xi-\chi_{c,t})\leq f^s(\xi-\chi_{c,t}+\chi_{\corigin,t'}), \text{ and } \\
&f^s(\tilde \xi)=f^s(\tilde \xi-\chi_{\corigin,t'})<f^s(\tilde \xi+\chi_{c,t}-\chi_{\corigin,t'}).
\end{align*}
By the assumption of Case 3, $\min\{f^s(\xi)-1, f^s(\tilde \xi)\}=\min\{f^s(\xi), f^s(\tilde \xi)\}$. Together with the above inequalities, pseudo M$^\natural$-concavity$^+$ is satisfied.
If $r_{\corigin}^{t'}\geq \tilde \xi_{\corigin}^{t'}(>\xi_{\corigin}^{t'})$, together with $r_c^t\geq \xi_c^t>\tilde \xi_c^t$,
\begin{align*}
&f^s(\xi)=f^s(\xi-\chi_{c,t})+1=f^s(\xi-\chi_{c,t}+\chi_{\corigin,t'}), \text{ and }\\
&f^s(\tilde \xi)=f^s(\tilde \xi-\chi_{\corigin,t'})+1=f^s(\tilde \xi+\chi_{c,t}-\chi_{\corigin,t'}).
\end{align*}
It follows that pseudo M$^\natural$-concavity$^+$ is satisfied.

\smallskip
\noindent
\emph{Subcase 3-2:}
Suppose $r_c^t<\xi_c^t$.

\smallskip
\noindent
\emph{Subcase 3-2-1:}
Suppose $\sum_{(\tilde c,\tilde t)\in \mathcal{C}\times \mathcal{T}}\tilde \xi_{\tilde c}^{\tilde t}<q$, which implies $\tilde \xi+\chi_{c,t}\in \Xi^0$. By $r_c^t<\xi_c^t$,
\begin{align*}
f^s(\xi)=f^s(\xi-\chi_{c,t}).
\end{align*}
By the definition of $f^s$,
\begin{align*}
f^s(\tilde \xi)\leq f^s(\tilde \xi+\chi_{c,t}).
\end{align*}
It follows that pseudo M$^\natural$-concavity$^+$ is satisfied.

\smallskip
\noindent
\emph{Subcase 3-2-2:}
Suppose $\sum_{(\tilde c,\tilde t)\in \mathcal{C}\times \mathcal{T}}\tilde \xi_{\tilde c}^{\tilde t}=q$. Let $\Phi=\{(\tilde c, \tilde t)\in \mathcal{C}\times \mathcal{T} \mid \xi_{\tilde c}^{\tilde t}<\tilde \xi_{\tilde c}^{\tilde t}\}$. By $\xi_c^t>\tilde \xi_c^t$ and $\sum_{(\tilde c,\tilde t)\in \mathcal{C}\times \mathcal{T}}\xi_{\tilde c}^{\tilde t}\leq q=\sum_{(\tilde c,\tilde t)\in \mathcal{C}\times \mathcal{T}}\tilde \xi_{\tilde c}^{\tilde t}$, we have $\Phi \neq \emptyset$.

Suppose, for contradiction, that $r_{\tilde c}^{\tilde t}\geq \tilde \xi_{\tilde c}^{\tilde t}$ for each
$(\tilde c, \tilde t)\in \Phi$. Then, by following the same line of argument as in Subcase 2-2-2, we obtain $f^s(\xi)<f^s(\tilde \xi)$, which is a contradiction to the assumption of Subcase 3. It follows that there exists $(\corigin,t')\in \Phi$ with $r_{\corigin}^{t'}<\tilde \xi_{\corigin}^{t'}$. Together with $r_c^t<\xi_c^t$,
\begin{align*}
&f^s(\xi)=f^s(\xi-\chi_{c,t})\leq f^s(\xi-\chi_{c,t}+\chi_{\corigin,t'}), \text{ and } \\
&f^s(\tilde \xi)=f^s(\tilde \xi-\chi_{\corigin,t'})\leq f^s(\tilde \xi+\chi_{c,t}-\chi_{\corigin,t'}).
\end{align*}
We conclude that pseudo M$^\natural$-concavity$^+$ is satisfied.
\qed

\medskip
\noindent
\emph{Counterexample to Claim \ref{claim:ex2-pMC} when $\Xi^0$ is not given as in the statement:}
Let $\mathcal{C}=\{c,c'\}$ and $\mathcal{T}=\{t,t'\}$. Suppose that each school's capacity is given by $q_c=2$ and $q_{c'}=1$, i.e.,
\begin{align*}
\Xi^0=\Bigl\{\xi\in \mathbb{Z}^{|\mathcal{C}|\times |\mathcal{T}|}_+ \mid \sum_{t\in T}\xi_c^t\leq 2, \sum_{t\in T}\xi_{c'}^t\leq 1\Bigr\}.
\end{align*}
The number of reserved seats is given by $r_c^t=1$, $r_c^{t'}=1$, $r_{c'}^t=1$, and $r_{c'}^{t'}=0$. Let $\xi, \tilde \xi\in \Xi^0$ be such that
\begin{align*}
&\xi_c^t=2, \: \xi_c^{t'}=0, \: \xi_{c'}^t=1, \: \xi_{c'}^{t'}=0, \\
&\tilde \xi_c^t=1, \: \tilde \xi_c^{t'}=1, \: \tilde \xi_{c'}^t=0, \: \tilde \xi_{c'}^{t'}=0.
\end{align*}
For $(c,t)$, the only candidate of $(c'',t'')\in (\mathcal{C}\times \mathcal{T})\cup\{\emptyset\}$ with $\tilde \xi+\chi_{c,t}-\chi_{c'',t''}\in \Xi^0$ is $(c'',t'')=(c,t')$ (otherwise the capacity constraint for school $c$ is violated at $\tilde \xi+\chi_{c,t}-\chi_{c'',t''}$). Then,
\begin{align*}
&\min\{f^s(\xi), f^s(\tilde \xi)\}=\min\{2,2\}=2, \text{ and } \\
&f^s(\tilde \xi)=2>1=f^s(\tilde \xi+\chi_{c,t}-\chi_{c,t'}).
\end{align*}
It follows that $f^s$ violates pseudo M$^\natural$-concavity, and hence violates pseudo M$^\natural$-concavity$^+$.

\bigskip

\begin{center}
\textbf{Omitted Examples}
\end{center}

\begin{example}\label{ex:oc-truncation-fail}
In this example we show that ordinal concavity of $f$ does not necessarily imply ordinal concavity of $f_\lambda$.
Let $\mathcal{C}=\{c\}$, $\mathcal T=\{t,t'\}$, $\Xi^0=\{0,1\}^2 \subseteq \mathbb{Z}^2_+, \xi(\{x\})=(1,0)$, and $\xi(\{y\})=(0,1)$.
Let $\calx=\{x,y\}$ and the distributional objective $f$ be defined as follows:
\[f(\xi(\emptyset))=1, f(\xi(\{x\}))=0, f(\xi(\{y\}))=2, \mbox{ and } f(\xi(\{x,y\}))=1.\]
It is easy to see that $f$ is ordinally concave.
For $\lambda=1$,
\[f_\lambda(\xi(\emptyset))=1, f_\lambda(\xi(\{x\}))=0, f_\lambda(\xi(\{y\}))=1, \mbox{ and } f_\lambda(\xi(\{x,y\}))=1.\]
Consider $\xi(\{x,y\})$, $\xi(\emptyset)$, and $t \in \mathcal{T}$ with $\chi_{c,t}=\xi(\{x\})$. Since
\begin{align*}
&f_\lambda(\xi(\{x,y\}))=1=f_\lambda(\xi(\{y\})) \text{ and } \\
&f_\lambda(\xi(\emptyset))=1>0=f_\lambda(\xi(\{x\})),
\end{align*}
$f_\lambda$ violates ordinal concavity.
\end{example}

\begin{example}\label{ex:truncation-sufficient-fail}
In this example we show that the converse of Proposition \ref{prop:truncation-necessary} is false.
Let $\mathcal{C}=\{c\}$, $\mathcal{T}=\{t\}$, and $\Xi^0=\{0,1,2\}\subseteq \mathbb{Z}_+$. We identify $\mathbb{Z}^{|\mathcal{C}|\times |\mathcal{T}|}_+$ with $\mathbb{Z}_+$. Define
$f:\Xi^0\rightarrow \mathbb{R}$ as
\begin{align*}
f(0)=0, \: f(1)=0, \: \text{ and } f(2)=1.
\end{align*}
It is easy to see that $f$ satisfies pseudo M$^\natural$-concavity. However, $f_\lambda$ violates ordinal concavity whenever $\lambda \geq 1$ (in which case $f_\lambda=f$).\footnote{Therefore,
this example in fact shows that pseudo M$^\natural$-concavity of $f$ does not imply ordinal concavity of $f$.} To see this point, let $\xi=2$ and $\tilde \xi=0$. Since
\begin{align*}
&1=f_\lambda(\xi)>f_\lambda(\xi-\chi_{c,t})=f_\lambda(1)=0 \text{ and } \\
&0=f_\lambda(\tilde \xi)=f_\lambda(\tilde \xi+\chi_{c,t})=f_\lambda(1)=0,
\end{align*}
$f_\lambda$ violates ordinal concavity.
\end{example}

\begin{example}\label{ex:frontier}
In this example we illustrate the trace algorithm in Section \ref{sec:choice-target}.
Consider the setting in Example \ref{ex:ladfail} and suppose that $n\geq 6$.
Suppose that the university is considering
the set of applications $X=\{x,y,z\}$ and the merit ranking of contracts is
$x\succ y \succ z$. Note that the distribution-conscious choice rule outcome is
$C^d(X)=\{z\}$.

At the beginning of the algorithm $k=0$,
$\lambda_0=0$, and $\mathcal{X}_0=\emptyset$.
Therefore, we need to calculate $C^d_{\lambda_0}(X)$. For $\lambda_0=0$, $f_{\lambda_0}$ assigns zero to all sets.
Hence, the set of maximal distributions in the
set of maximizers of $f_{\lambda_0}$ is $\{ \xi(\{x,y\}), \xi(\{x,z\}), \xi(\{y,z\})\}$,
and thus, $C^d_{\lambda_0}(X)=\{x,y\}$. Since $C^d_{\lambda_0}(X)\neq C^d(X)=\{z\}$, we set
$\mathcal{X}_1=\mathcal{X}_0 \cup \{C^d_{\lambda_0}(X)\}=\{\{x,y\}\}$ and $\lambda_1=f(\xi(\{x,y\}))+1=2$.

In the second iteration we have $k=1$, $\lambda_1=2$, and $\mathcal{X}_1=\{\{x,y\}\}$.
Hence, we need to find $C^d_{\lambda_1}(X)$. For $\lambda_1=2$, $f_{\lambda_1}$ assigns two to
all sets with a distributional objective (with respect to $f$) of at least two. Therefore, the set of maximal
distributions in the set of maximizers for the distributional objective $f_{\lambda_1}$ is
$ \{ \xi(\{x,z\}), \xi(\{y,z\})\}$,
and thus $C^d_{\lambda_1}(X)=\{x,z\}$. Since $C^d_{\lambda_1}(X)\neq C^d(X)=\{z\}$, we set
$\mathcal{X}_2=\mathcal{X}_1 \cup \{C^d_{\lambda_1}(X)\}=\{\{x,y\},\{x,z\}\}$ and $\lambda_2=f(\xi(\{x,z\}))+1=6$.

In the third iteration we have $k=2$, $\lambda_2=6$, and $\mathcal{X}_2=\{\{x,y\},\{x,z\}\}$.
Hence, we need to construct $C^d_{\lambda_2}(X)$. For $\lambda_2=6$, $f_{\lambda_2}$
assigns six to all sets with a distributional objective (with respect to $f$) of at least
six. Therefore, the set of maximal distributions in the set of maximizers for the
distributional objective $f_{\lambda_2}$ is $\{ \xi(\{z\})\}$, which implies that
$C^d_{\lambda_2}(X)=\{z\}$. Since $C^d_{\lambda_2}(X)=C^d(X)=\{z\}$, we set
$\mathcal{X}_3=\mathcal{X}_2 \cup \{C^d_{\lambda_2}(X)\}=\{\{x,y\},\{x,z\},\{z\}\}$ and return this as the outcome of the trace algorithm.
The outcome generates all the sets in the Pareto frontier of the distributional objective and merit by Theorem \ref{thm:trace}; see Figure \ref{fig:frontier}.

\begin{figure}
  \centering
\definecolor{ffffff}{rgb}{1.,1.,1.}
\begin{tikzpicture}[scale=.8]
\begin{axis}[
x=1.0cm,y=1.0cm,
axis lines=middle,
xmin=-1.5,
xmax=11.7,
ymin=-1.5,
ymax=8.0,
xtick=\empty,
ytick=\empty,]
\draw (0.38,7.72) node[anchor=north west] {$\{x,y\}$};
\draw [dashed] (5.,0.)-- (5.,5.);
\draw [dashed] (5.,5.)-- (0.,5.);
\draw (4.4,5.72) node[anchor=north west] {$\{x,z\}$};
\draw [dashed] (8.,0.)-- (8.,1.);
\draw [dashed] (8.,1.)-- (0.,1.);
\draw (7.55,1.72) node[anchor=north west] {$\{z\}$};
\draw [dashed] (1.,0.)-- (1.,7.);
\draw [dashed] (1.,7.)-- (0.,7.);
\draw (4.75,-0.1) node[anchor=north west] {5};
\draw (7.75,-0.2) node[anchor=north west] {$n$};
\draw [dashed] (1.,4.)-- (1.,0.);
\draw [dashed] (1.,4.)-- (0.,4.);
\draw [dashed] (5.,3.)-- (0.,3.);
\draw (.75,-0.1) node[anchor=north west] {1};
\draw (4.4,4.70) node[anchor=north west] {$\{y,z\}$};
\draw (0.58,4.70) node[anchor=north west] {$\{x\}$};
\draw (0.58,3.75) node[anchor=north west] {$\{y\}$};
\draw (9.20, -0.1) node[anchor=north west] {distributional};
\draw (9.20, -0.6) node[anchor=north west] {objective};
\draw (-1.3, 8.0) node[anchor=north west] {merit};
\begin{scriptsize}
\draw [fill=black] (5.,5.) circle (2.5pt);
\draw [fill=black] (8.,1.) circle (2.5pt);
\draw [fill=black] (1.,7.) circle (2.5pt);
\draw [fill=ffffff] (5.,4.) circle (2.5pt);
\draw [fill=ffffff] (1.,4.) circle (2.5pt);
\draw [fill=ffffff] (1.,3.) circle (2.5pt);
\end{scriptsize}
\end{axis}
\end{tikzpicture}
\caption{The filled nodes form the Pareto frontier of the distributional objective and merit in Example \ref{ex:frontier} when $n\geq 6$.}
\label{fig:frontier}
\end{figure}

\end{example}


\end{document}